\documentclass[11pt]{article}
\usepackage{setspace}
\usepackage{authblk}
\usepackage{natbib}
\usepackage[short,nodayofweek,12hr]{datetime}
\usepackage{amsmath}
\usepackage{float}
\usepackage{xcolor}
\usepackage{amsfonts}
\usepackage{amsmath}
\usepackage{bbm}
\usepackage{adjustbox}
\usepackage[flushleft]{threeparttable}
\usepackage{graphics}
\usepackage{hyperref}
\usepackage{booktabs}
\usepackage{graphicx}
\usepackage{comment}
\usepackage[utf8]{inputenc}
\usepackage{sectsty}
\usepackage[textsize=footnotesize]{todonotes}

\hypersetup{
  colorlinks=true,
  urlbordercolor={1 1 1},
  urlcolor={blue},
  linkcolor={blue},
  citecolor={blue},
  raiselinks=false
}

\subsectionfont{\normalfont\itshape}
\subsubsectionfont{\normalfont\itshape}

\usdate

\title{\textbf{Generalized Autoregressive Score \\Trees and Forests}\footnote{We thank Saketh Aleti, Tim Bollerslev, Anna Bykhovskaya, Anna Cieslak, Peter Reinhard Hansen and Haozhe Zhang for valuable comments. Email: \textcolor{blue}{\href{mailto:andrew.patton@duke.edu}{andrew.patton@duke.edu}}.}}
\author[a]{Andrew J. Patton}
\author[a]{Yasin Simsek}
\affil[a]{\textit{\small Department of Economics, Duke University}}
\date{This version: \today \\ \vspace{1cm}}

\begin{document}
\maketitle

\begin{abstract}
\doublespacing
We propose methods to improve the forecasts from generalized autoregressive score (GAS) models (\citealp{creal2013generalized}; \citealp{harvey2013dynamic}) by localizing their parameters using decision trees and random forests. These methods avoid the curse of dimensionality faced by kernel-based approaches, and allow one to draw on information from multiple state variables simultaneously. We apply the new models to four distinct empirical analyses, and in all applications the proposed new methods significantly outperform the baseline GAS model. In our applications to stock return volatility and density prediction, the optimal GAS tree model reveals a leverage effect and a variance risk premium effect. Our study of stock-bond dependence finds evidence of a flight-to-quality effect in the optimal GAS forest forecasts, while our analysis of high-frequency trade durations uncovers a volume-volatility effect. 

\end{abstract}
 
\doublespacing
 

\textbf{J.E.L. Codes:} C22, C32, C53.


\newpage
\section{Introduction}

Models for economic time series data that can capture time variation in features of the predictive density are widely used for policy making, investment decisions, risk management, and in many other applications. Such models include the autoregressive-moving average model of \cite{box1970time}, the ARCH/GARCH models of \cite{engle1982autoregressive} and \cite{bollerslev1986generalized}, and many others. The family of ``generalized autoregressive score'' (GAS) 
models, proposed by \cite{creal2013generalized} and \cite{harvey2013dynamic}, nests these time series models and others, and has been applied to a wide range of problems. \cite{artemova2022scoreTheory,artemova2022scoreApplications} and \cite{harvey2022score} provide recent surveys of this large and growing literature.

Despite their success, score-driven models are inevitably only approximations to the true data generating process. We propose to use data mining methods from the machine learning literature to improve the performance of these models. Specifically, we propose a ``GAS tree,'' that combines the parsimonious structure of the GAS model with the flexible, data-driven learning of decision trees \cite{breiman1984classification,breiman2017classification}. A GAS tree allows the parameters of the model to vary across ``branches'' of the tree, which are formed using a possibly large collection of state variables. This leads to a model that can incorporate information from outside the GAS model, and that allows for potentially complicated nonlinearities and interactions. We further propose ``GAS forests,'' analogous to the ``random forests'' of \cite{breiman2001random} for linear regression, where we create many GAS trees using bootstrap samples of the original data and then average the forecasts from these trees. In many applications random forests have been found to improve upon regression trees due to the reduction in variance obtained via averaging, see e.g. \cite{hastie2009elements}.

The estimation of GAS trees and GAS forests is computationally demanding. It involves finding the optimal state variables and thresholds from the set of candidate variables, as well as estimating the parameters of the GAS model. We use cluster computing and a ``greedy'' estimation algorithm related to that of \cite{breiman1984classification} for regression and \cite{audrino2001tree} for GARCH trees. This algorithm finds a near-optimal solution and converges quickly. A key hyper-parameter in tree and forest models is the maximum depth of the tree (essentially, how many subsamples of the data will be considered) and we tune this parameter using a validation sample, separate from our forecast evaluation sample.

We apply the proposed GAS tree and GAS forest models in four empirically relevant problems: forecasting stock return volatility, the distribution of stock returns, the joint distribution of stock and bond returns, and high-frequency trade durations. As baseline models for these applications we use the GARCH model of \cite{bollerslev1986generalized}, the t-GAS model of \cite{creal2011dynamic}, a joint distribution model with Student's $t$ margins and a Student's $t$ copula, as in \cite{janus2014long}, and the ACD model of \citet{engle1998autoregressive}. We then consider tree and forest extensions of these models, and in all four cases we find that the baseline model is significantly out-performed. For the two stock return applications, we find that the GAS tree provides the best out-of-sample forecasts. The estimated tree structures provide significantly better forecasts, and turn out to be relatively simple: we find evidence of a leverage effect, where the GAS model parameters differ depending on whether the lagged stock return was positive or negative, and a variance risk premium effect, where the model parameters differ depending on whether the difference between option-implied and historical volatilities is large or small. 

In our study of the joint predictive distribution of stock and bond returns, we find that the GAS forest produces the best out-of-sample forecasts. Variable importance analyses indicate that the most important variables for the GAS forest are the lagged stock and bond returns themselves, indicating omitted nonlinearity in the baseline GAS model. We find evidence of a flight-to-quality effect, where higher bond returns or lower stock returns are associated with even more negative long-run correlations between the stock and bond markets. In our analysis of trade durations, defined as the time taken for 10,000 shares of the S\&P 500 exchange traded fund, SPY, to be transacted, we again find the forest-based extension to be the preferred model. In this application the most important state variables are both measures of volatility, consistent with the well-known volume-volatility relationship (see, e.g., \citealp{Karpoff87}).  

This paper is part of the fast-growing literature using tools from machine learning in econometrics, see \cite{varian2014big} and \cite{athey2019machine} for recent surveys. 
Various studies have found that machine learning techniques  bring significant gains over traditional econometric methods for forecasting applications. For example, \cite{medeiros2021forecasting}, \cite{goulet2020macroeconomy} and \cite{huber2020nowcasting} show that tree-based methods, including random forests, can produce more accurate forecasts of important macroeconomic variables like unemployment and inflation. \cite{gu2020empirical} and \cite{bianchi2021bond} show how machine learning methods can improve forecasts of stock and bond returns.

In addition to macroeconomic and financial forecasting, some recent papers have found success applying machine learning methods to volatility models such as the GARCH model of \cite{bollerslev1986generalized} and the HAR model of \cite{corsi2009simple}. For instance, \cite{christensen2021machine} shows that neural networks and random forests significantly improve over HAR model, and \cite{nguyen2022statistical,nguyen2022recurrent} create hybrid stochastic volatility and GARCH models with recurrent neural networks. \cite{reisenhofer2022harnet} and \cite{tetereva2022forest} use convolutional neural networks and random forests, respectively, combined with the HAR model to obtain improved out-of-sample forecasts. 

This study also relates to a broadly defined ``local estimation'' literature.  \cite{tibshirani1987local}, \cite{fan1998local} and \cite{fan2009local} use kernel-based methods to localize (quasi-) maximum likelihood models. A more recent strand of this literature includes \cite{breiman2001random}, \cite{schlosser2019distributional} and \cite{athey2019generalized}, who use decision trees and random forests to localize regressions, parametric distributions, and GMM models respectively. Our paper is related to \cite{oh2021better}, which is part of the first strand of this literature. That paper's approach suffers from the curse of dimensionality, due to its use of kernel-based methods, and it additionally requires that all (or none) parameters of the baseline model are localized. In the next section, we show that our proposed approach can deal with a large number of state variables and permits a subset of parameters to be localized, allowing the researcher to impose more or less structure on the model as needed.

The remainder of the paper is structured as follows. In Section \ref{sec:models} we review the class of generalized autoregressive score models of \cite{creal2013generalized} and \cite{harvey2013dynamic} and introduce our new GAS tree and GAS forest models. Section \ref{sec:models} also includes computational details on the implementation of these models. Section \ref{sec:emp_app} presents four empirical analyses, applying the new methods to forecasting volatility, correlation, and univariate and bivariate distributions. Section \ref{sec:conclusion} concludes, and the appendix presents details on the derivations for the third application. A supplemental appendix contains additional results.

\section{GAS Trees and Forests} \label{sec:models}
The class of generalized autoregressive score (GAS) models of \cite{creal2013generalized} and \cite{harvey2013dynamic} provide a parsimonious and powerful way to capture time variation in the parameter(s) of a given probability density function. We describe this model below, and in Sections \ref{sec:GAStree} and \ref{sec:GASforest} we introduce tree- and forest-based extensions of this class of models.

\subsection{GAS models}
Let the dependent variable be denoted $\mathbf{y}_t \in \mathbb{R}^K$. Conditional on the information set $\mathcal{F}_t$, this variable is assumed to have a parametric predictive density $p$, with $d$-dimensional time-varying parameter $f_t$, and potentially a static parameter $\nu$. The GAS($p,q$) model specifies the evolution of $f_t$ as:
\begin{eqnarray} \label{eq:tvp}
	f_{t} &=& \omega +\sum_{j=1}^{q} B_{j} f_{t-j} +\sum_{i=1}^{p} A_{i} s_{t-i}\\
 \text{where~~~~~~} s_t &=& S_t \cdot \nabla_t  \nonumber \\
 \nabla_t &=& \frac{\partial \log p(\mathbf{y}_t;f_t,\nu)}{\partial f'_t}   \nonumber \\
 S_t &=& \mathbb{E}_{t-1}[\nabla_t \nabla'_t]^{-1} \nonumber
\end{eqnarray}
It is the appearance of the score, $\nabla_t$, in the evolution equation for $f_t$ that gives this class of models its name.\footnote{We follow \cite{creal2011dynamic} and use the inverse information matrix to scale the score in all of our applications, though other choices for this matrix are possible, such as the square root of this matrix, or simply the identity matrix.} Similar to the well-known Newton-Raphson algorithm for numerical optimization, at each date $t$, $f_t$ moves in the direction that most improves the model fit.

Let $\theta =(\omega, \mathrm{vec}(B_1), ..., \mathrm{vec}(B_q), \mathrm{vec}(A_1), ..., \mathrm{vec}(A_p))$ denote the vector of all GAS parameters of this model, making $(\theta,\nu)$ the full set of unknown parameters. Since GAS models are ``observation driven,'' as opposed to ``parameter-driven'', the likelihood function is available in closed form, and $(\theta,\nu)$ can be estimated by maximum likelihood with low computational cost. This feature makes it feasible to consider tree- and forest-based extensions of this class of models, which we introduce below. 


\subsection{GAS Trees} \label{sec:GAStree}
\begin{figure}[t!]
    \centering
    \caption{A decision tree example}
    \vspace{0.55cm}
    \begin{tabular}{cc}
         \includegraphics[scale=0.3]{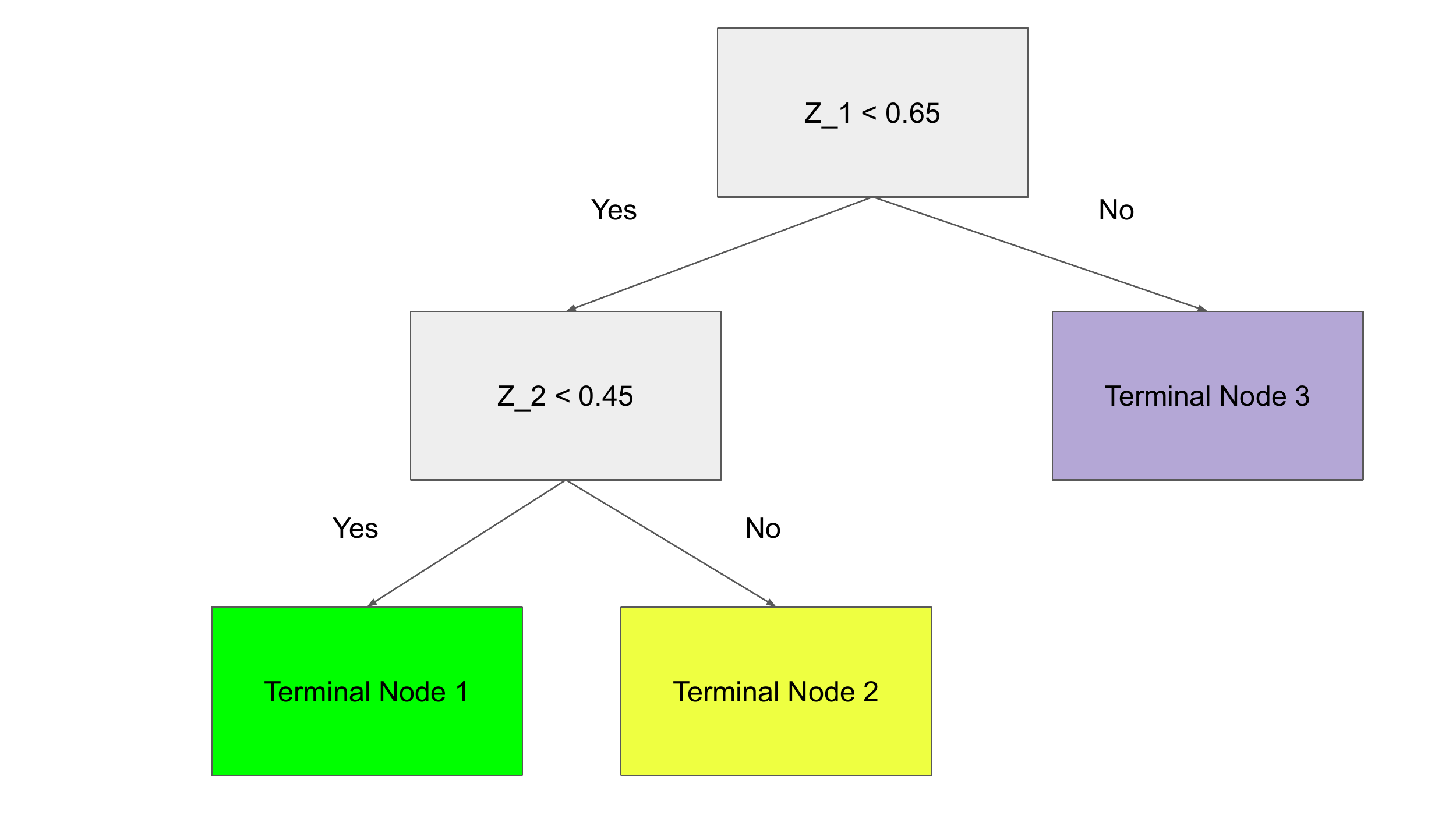} &  \includegraphics[scale=0.45]{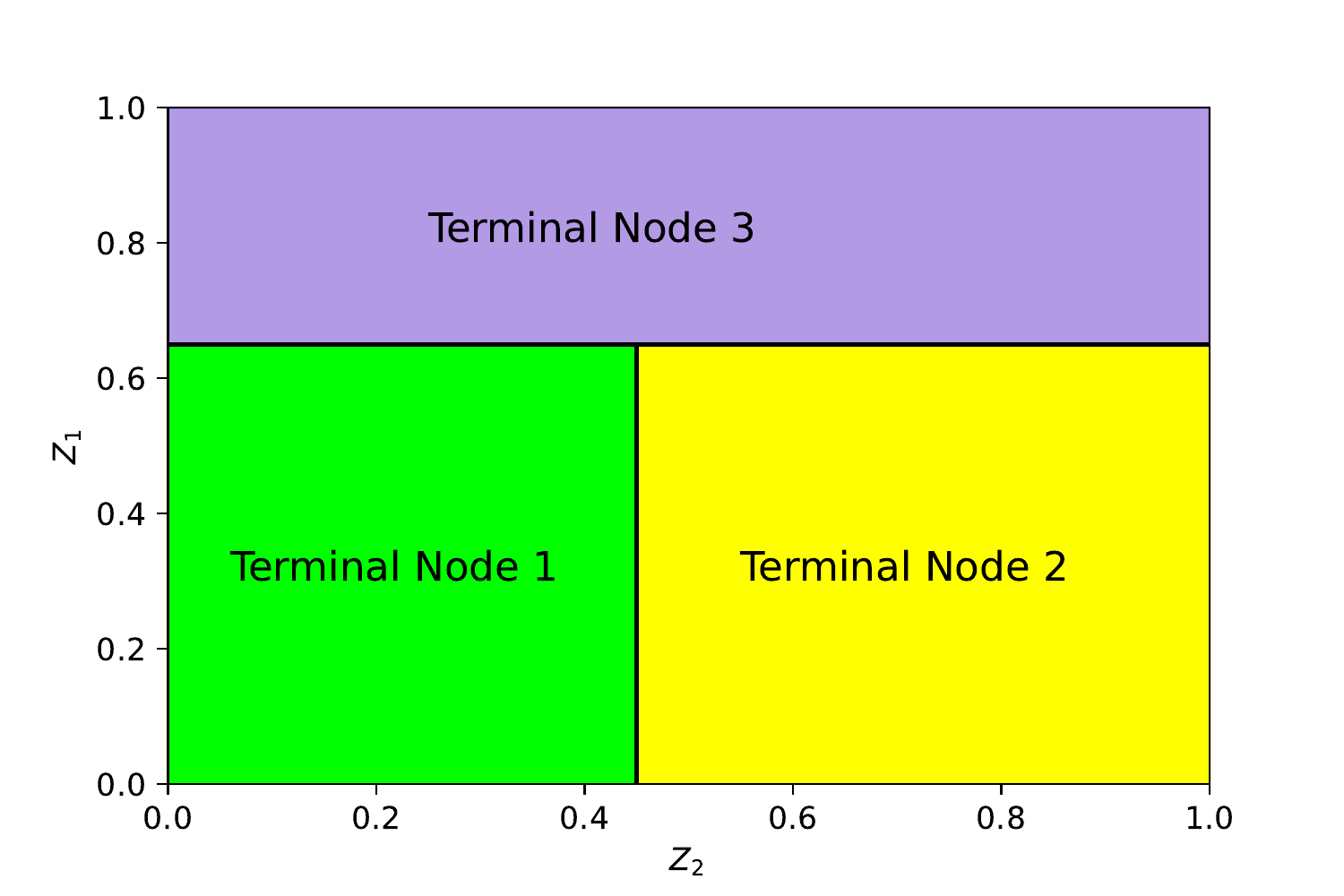} \\
    \end{tabular}
    \vspace{0.5cm}
    \label{fig:tree_ex}
\end{figure}

Regression trees (\citealp{breiman1984classification}; \citealp{breiman2017classification}) 
are a type of nonparametric regression based on sequentially splitting the available data into partitions. The partitions are formed using one or more state variables, $Z_t$, and estimated threshold value(s), $c$. Figure \ref{fig:tree_ex} illustrates a simple tree structure. The left panel shows a tree with two state variables and specific thresholds, and the right panel shows the corresponding partition of the support of state variables. This hypothetical tree has three ``terminal nodes'' and implies a specific partition of the data, denoted $\boldsymbol{\mathcal{P}}=\{\mathcal{P}_1,\mathcal{P}_2,\mathcal{P}_3\}$. 
Given a tree structure, a ``regression tree'' is obtained by estimating a linear regression separately for each of the terminal nodes in the tree. In so doing, regression trees allow for nonlinearities and multi-way interactions, greatly generalizing the baseline regression model. Naturally this flexibility makes trees prone to overfit the training data, and therefore, trees must be regularized, or ``pruned.'' We describe the estimation and regularization methods we use for GAS trees and forests in Section \ref{sec:estimation}. 

We adapt the idea of regression trees for application to generalized autoregressive score (GAS) models. For a given tree structure with $J$ terminal nodes, $\boldsymbol{\mathcal{P}}=\{\mathcal{P}_1,..., \mathcal{P}_J\}$, the GAS(1,1) tree is based on the evolution equation:
\begin{eqnarray} \label{eq:GASforecast}
    f_t &=& \omega(\mathbf{Z_t}) + \beta(\mathbf{Z_t}) f_{t-1} + \alpha(\mathbf{Z_t}) s_{t-1} \\
    \text{where~~~~~} \theta(\mathbf{Z_t}) &=& \sum_{j=1}^J \theta_j \mathbbm{1}(\mathbf{Z_t}\in \mathcal{P}_j) \nonumber   
\end{eqnarray}
where $\theta_j \equiv [\omega_j,\beta_j,\alpha_j]$ are the GAS parameters for partition $j$, and $s_t$ is as in equation (\ref{eq:tvp}). By allowing the parameters of the GAS model to vary across partitions we greatly increase the flexibility of this class of models to fit the data. Furthermore, by retaining the GAS structure for each partition, we can more easily interpret \textit{how} the tree structure improves the fit of the model, in contrast with more ``black box'' machine learning algorithms.

The parameters of the predictive density that are assumed constant in the baseline GAS model, denoted $\nu$ above, can either be held constant across partitions or can be allowed to vary.\footnote{In the kernel-based local M-estimation approach of \cite{oh2021better} it is not possible to allow only a subset of parameters to vary with the state variable(s); the framework adopted in that paper requires an ``all-or-nothing'' assumption on parameter variation.} In our description below we impose they are fixed across partitions.

\subsection{GAS Forests} \label{sec:GASforest}

``Random forests'' (\citealp{breiman2001random}) are an extension of regression trees designed to reduce the estimation error in predictions, while retaining the information contained in the tree-based forecast, see \cite{hastie2009elements} for example. Similar to bootstrap aggregation, or ``bagging,'' a random forest is populated by trees that are each estimated on a bootstrap sample of the original data. In addition, each tree uses only a randomly-selected subset of the original state variables. The predictions from each of these trees are then averaged to obtain the random forest forecast. 


If we denote trees in the random forest as $\boldsymbol{\mathcal{P}_b}$ for $b=1,...,B$, and the  forecast from each tree for a given value of the vector of state variables, $\mathbf{Z_t}$ as $f_t^{(b)}(\mathbf{Z_t})$, obtained using equation (\ref{eq:GASforecast}), then the GAS forest forecast is obtained simply as
\begin{eqnarray}
    f_t(\mathbf{Z_t}) &=& \frac{1}{B} \sum_{b=1}^B f_t^{(b)}(\mathbf{Z_t}) \label{eq:GASforest}
\end{eqnarray}


We next turn to the estimation of the tree structure used in GAS trees and forests.

\subsection{Estimating GAS Trees and Forests} \label{sec:estimation}

The estimation of a GAS tree requires finding the optimal state variables and thresholds from the set of candidate variables, as well as estimating the parameters of the GAS model. Finding the global optimum of this optimization problem is computationally infeasible in even moderately-sized regression tree applications, and to reduce the computational burden \cite{breiman1984classification} proposed a greedy estimation algorithm that finds a near-optimal solution and converges quickly. The algorithm finds a state variable and a threshold to locally minimize the prediction error at each splitting step, continuing until a stopping criteria is satisfied. 

Standard regression tree estimation involves estimating a regression separately for each terminal node in the tree, but given the autoregressive nature of GAS models, this is not possible for our application. We propose a modified estimation algorithm similar to \cite{audrino2001tree} that uses a tree structure for the GAS model parameters and retains the autoregressive structure for $f_t$.

In the case that the number of terminal nodes, $J$, is one, there is no tree structure and the original GAS model is estimated via maximum likelihood:
\begin{eqnarray}
    (\hat{\theta}_T,\hat{\nu}_T) &=& \underset{\theta,\nu}{\operatorname{argmax}} ~ \frac{1}{T}\sum_{t=1}^T \log p(y_t; f_t(\theta), \nu)  \label{eq:mle}
\end{eqnarray}
For $J\geq 2$ we use the following estimation algorithm to estimate the tree structure or, equivalently, to find the optimal partition $\boldsymbol{\mathcal{P}}$. Estimation of the GAS tree involves Steps 1--5 below, and the GAS forest additionally uses Step 6. 

\bigskip

\noindent \textit{Step 1:} Denote the entire sample as the trivial partition $\boldsymbol{\mathcal{P}}^{(0)}$. Estimate the parameters of the model as in equation (\ref{eq:mle}), and denote these as $(\hat{\theta}^0,\hat{\nu}^0)$.

\bigskip

\noindent \textit{Step 2:} Define a new partition: $\boldsymbol{\mathcal{P}}_{j,k}^{(m+1)}=\boldsymbol{\mathcal{P}}^{(m)}_{-j} \cup \{\mathcal{P}_{j,k,L}^{(m)},\mathcal{P}_{j,k,R}^{(m)}\}$ where $\boldsymbol{\mathcal{P}}^{(m)}_{-j}=\boldsymbol{\mathcal{P}}^{(m)} / \mathcal{P}_j$ contains all the partitions of $\boldsymbol{\mathcal{P}}^{(m)}$ except for the $j^{th}$, and the $j^{th}$ partition is split into ``left'' and ``right'' subpartitions based on the $k^{th}$ state variable and a threshold $c$
\begin{eqnarray}    
     \mathcal{P}_{j,k,L}^{(m)} & = & \{\mathbf{Z_t}: \mathbf{Z_t}\in         \mathcal{P}_j^{(m)} \quad \text{and} \quad Z_{t,k} \leq  c\} \\
    \mathcal{P}_{j,k,R}^{(m)} & = & \{\mathbf{Z_t}: \mathbf{Z_t}\in         \mathcal{P}_j^{(m)} \quad \text{and} \quad Z_{t,k} > c\}  \nonumber
\end{eqnarray}

\bigskip

\noindent \textit{Step 3:} Estimate the parameters for new subpartitions, taking the parameters of the other partitions, $\hat{\theta}^{(m)}_{-j}$, as fixed:\footnote{Optimizing the split is the most demanding step in  the entire algorithm. This assumption significantly reduces the computational burden without hurting the results. Similar ideas are also implemented in the literature, for example \cite{athey2019generalized} uses a gradient approximation in the split selection step.}
\begin{eqnarray}
    (\hat{\theta}_{j,k,L}^{(m+1)}, \hat{\theta}_{j,k,R}^{(m+1)}) &=& \underset{\theta_L,\theta_R}{\operatorname{argmax}} ~ \frac{1}{T}\sum_{t=1}^T \log p(y_t; f_t(\hat{\theta}^{(m)}_{-j},\theta_L, \theta_R), \hat{\nu}^{(m)})
\end{eqnarray}
The compute the log-likelihood value at estimated parameter values. 
\begin{equation} \label{eq:split_criterion}
    \log p(y;\boldsymbol{\mathcal{P}}^{(m+1)}_{j,k}) = \frac{1}{T}\sum_{t=1}^T \log p(y_t; f_t(\hat{\theta}^{(m)}_{-j}, \hat{\theta}_{j,k,L}^{(m+1)}, \hat{\theta}_{j,k,R}^{(m+1)}), \hat{\nu}^{(m)})
\end{equation}

\bigskip
\noindent \textit{Step 4:} Maximize equation \eqref{eq:split_criterion} over the partition $j$, state variable $k$, and threshold $c$. Denote the optimized new partition as $\boldsymbol{\mathcal{P}}^{(m+1)}$ and estimate all the model parameters using equation (\ref{eq:mle}), denote these as $\hat{\theta}^{(m+1)}$.\footnote{In steps 3 and 4, parameter estimations are done by nonlinear solvers available in Scipy package of Python language. Following the \textit{warm start} idea in the optimization literature, we use the estimates from the previous iteration, $\hat{\theta}^{(m)}$, as starting values for optimization procedure to accelerate the algorithm.}

\bigskip

\noindent \textit{Step 5:} Repeat steps 2-4 until the depth of the tree, $m$, reaches a prespecified maximum value, $M$. The depth of the tree controls the model complexity, and we consider values of $M$ between one and six. We tune this parameter using a validation sample.

\bigskip
\noindent \textit{Step 6:} For the GAS forest, repeat steps 2-5 for $B=200$ trees.\footnote{In preliminary analyses we obtained very similar results for $B=500$.} Each tree in the forest uses bootstrap data obtained from a circular block bootstrap (see, e.g., \citealp{politis1999subsampling}), with block length of 100 observations, and a random selection of one-third of the total state variables. One-third is a common choice in the machine learning literature, see \cite{hastie2009elements} for example. The forecasts from each of the bootstrap trees are then averaged to obtain the GAS forest forecast.
\bigskip



All computations are done using the Duke Computing Cluster exploiting multiple computing nodes. We parallelize the split optimization steps, and use the Numba package to speed up the code. Estimating a single tree takes around five minutes with forty CPUs. We apply fixed-window estimation all models: we estimate the model parameters using the estimation sample and use those parameters to compute all out-of-sample forecasts.


\section{Forecast performance of GAS trees and GAS forests} \label{sec:emp_app}

We apply our new GAS tree and forest models in four out-of-sample forecasting analyses. We firstly consider forecasting S\&P 500 return volatility using the GARCH model of \cite{bollerslev1986generalized}, followed by predicting the entire conditional density of S\&P 500 returns using the ``t-GAS'' model of \cite{creal2013generalized}. In our third application we consider a flexible model for the joint distribution of S\&P 500 returns and 10-year U.S. government bond returns, motivated in part by work on the switching sign of this correlation, see \cite{guidolin2006econometric} and \cite{baele2010determinants}, using the t-GAS copula to link t-GAS models for the marginal distributions, as in \cite{janus2014long}. Finally, we consider the ``autoregressive conditional duration'' model of \cite{engle1998autoregressive}, using high frequency transaction data on the exchange traded fund tracking the S\&P 500 index, the SPY. These four applications represent a range of predictive environments, and we provide evidence of the merits of GAS trees and forests in each of them.

In addition to the baseline GAS model for each application, we consider two other benchmark models. The first is a low-dimensional GAS tree (``small GAS tree''), in which we only consider the lag of the dependent variable(s) as a state variable. This model is similar to that of \cite{audrino2001tree}, and comparing the GAS tree and forest forecasts with this benchmark reveals the benefits of using a larger set of state variables. The third benchmark model is the ``distributional random forest'' of \cite{schlosser2019distributional}. 
This model has no time series dynamics, but can provide a flexible distributional fit through the forest structure. 


We compare all models in terms of one-step-ahead predictive performance.
For the volatility forecasting application, we use the QLIKE loss function with realized volatility as the volatility proxy, see \cite{Patton2011} for details. For the remaining applications we use the negative log-likelihood, which is a consistent scoring rule for density forecasts, see \cite{Gneiting2007}. We conduct \cite{dieboldMariano1995} tests of equal predictive accuracy, using \cite{newey1987simple} standard errors based on 10 lags.


\subsection{Data description} \label{sec:data}

We consider GAS trees and forests in four empirical applications. The first three of these use daily data on S\&P 500 index returns and 10-year U.S. government bond returns from January 2000 to December 2021, a total of 5447 observations. Our fourth application uses high frequency trade durations for the S\&P 500 index tracker fund, SPY, during the calendar year 2021, and has 5,100 observations. In all applications we split the sample into three sub-periods: an estimation sample (first 30\% of observations), a validation sample for optimizing hyperparameters (next 30\%), and a test sample for out-of-sample forecast comparisons (remaining 40\%). 

We consider ten state variables for use in the applications based on daily data, and we add three high frequency state variables in the fourth application.  
We firstly include the (lagged) return on the S\&P 500 index and the 10-year U.S. government bond, to capture any nonlinearities omitted by the GAS models. We next consider three measures of volatility: 5-min subsampled realized volatility (RVOL) on the S\&P 500 index, a one-month (backward-looking) rolling average of RVOL, motivated by prominent HAR model of \cite{corsi2009simple}, and the VIX index, a measure of S\&P 500 volatility implied by options prices. We then consider three measures from the fixed income market: the federal funds rate, the difference between 10-year and 3-month bond yields, representing the level and slope of the yield curve, and the ``default spread'' defined as the difference between BAA and AAA rate corporate bond yields. For our ninth state variable we include the economic policy uncertainty index proposed by \cite{baker2016measuring}, based on newspaper coverage. This index tracks important policy related events like the failure of Lehman Brothers or presidential elections. 
We take a rolling monthly average of policy uncertainty index to eliminate the noise in the data. Our tenth state variable is time, to capture potential structural breaks, see for example \cite{coulombe2020machine} and \cite{goulet2020macroeconomy}. In our fourth application, we additionally consider three high-frequency state variables: the first lag of duration, which can capture nonlinearities missed by the benchmark model, the return on SPY over the last trade event period, which can capture leverage-type effects, and the market liquidity of \cite{amihud2002illiquidity}, which can gauge whether the ACD model parameters differ during periods of high versus low liquidity.

We use the augmented Dickey-Fuller test (\citealp{DickeyFuller1979}) for each state variable to test for the presence of a unit root. We fail to reject the null of a unit root for the federal funds rate and the difference between 10 year and 3 month yield, and we take the first difference of these two variables. 
To avoid look-ahead bias, we use a one-period lag of the state variables when forming the tree and forest forecasts. 

All of our data comes from the FRED database at the Federal Reserve Bank of St. Louis, with the following exceptions: the realized volatility data comes from the Oxford Realized Library; the high frequency data comes from the New York Stock Exchange's TAQ database; the 10 year bond return series is from \cite{liu2021reconstructing}; and the policy uncertainty series is from \cite{baker2016measuring}.\footnote{The data for the latter two variables is available at \url{https://sites.google.com/view/jingcynthiawu/yield-data} and 
\url{www.policyuncertainty.com}.}



\subsection{Forecasting stock return volatility} \label{sec:vol_for}
The GARCH model of \cite{bollerslev1986generalized} is widely used for forecasting asset return volatility, and has been shown to be difficult to beat in a range of applications, see \cite{hansen2005forecast}. Assuming a zero conditional mean, 
the model is:
\begin{equation}
\begin{array}{rcl}
     y_t  & = & \sigma_t \epsilon_t;  \quad \epsilon_t  \sim ~i.i.d.~\mathcal{N}(0,1) \\
     \sigma_t^2  & = & \omega + \beta \sigma_{t-1}^2 + \alpha y_{t-1}^2.
\end{array}
\end{equation}
\cite{creal2013generalized} show that this model can be interpreted as a GAS model for the scale parameter of the Normal distribution. Given this equivalence, the GAS tree and forest models for this case can also be labeled GARCH tree and forest models. The ``distributional random forest'' (DRF) of \cite{schlosser2019distributional} in this application sets $\beta =\alpha =0$ and allows the intercept, $\omega$ to vary with the forest structure, while the ``small GAS tree'' model of \cite{audrino2001tree} uses a decision tree with only $y_{t-1}$ as a state variable. We compare forecasts from these three models with those from the new GAS tree and GAS forest models introduced in Sections \ref{sec:GAStree} and \ref{sec:GASforest} in Table \ref{tab:garch_oos}.


We observe that each variant of tree-based GARCH model (small GAS tree, GAS tree and GAS forest) significantly outperforms the benchmark GARCH model with $t$ statistics all less than $-2.5$. Moreover, the GARCH tree and forest models significantly beat the DRF specification. Thus, in this first application, we find that tree structured models improve the out-of-sample forecast accuracy over simple GARCH, a conventional econometrics model, and DRF, a machine learning tool. Table \ref{tab:garch_oos} also shows that the ``small GAS tree'' outperformed by the GAS tree and forest models, with $t$-statistics below $-2.7$,  revealing that external variables carry important information about future volatility.

Interestingly, and in contrast with both the econometrics and the machine learning literatures which generally find ensemble methods tend to outperform forecasts from individual models, we find that the GAS tree outperforms the GAS forest, with a $t$-statistic of nearly five. We interpret this result by noting that random forests have the potential to improve forecast accuracy through variance reduction at the cost of increasing bias, see for example  \cite{hastie2009elements}. In our case, the variance reduction attained by the GAS forest cannot compensate the associated increased bias, leading to less accurate forecasts.  

\begin{table}[t!]
    \centering
    \caption{\textbf{Out-of-sample performance of GARCH models using QLIKE loss.} This table presents $t$-statistics from Diebold-Mariano tests of out-of-sample forecast performance (top four rows) and average out-of-sample QLIKE losses (bottom row). A negative $t$-statistic indicates that the model in the column had lower average loss than the model in the row, while a positive $t$-statistic indicates the opposite.  \vspace{0.5cm}}
    \begin{tabular}{lccccc}
    \toprule 
&  &  & Small & GARCH & GARCH  \\
                  & GARCH & DRF & GARCH Tree & Tree & Forest  \\
         \cmidrule(lr){2-6} 
& \phantom{GARCH} & \phantom{GARCH} & \phantom{GARCH}  & \phantom{GARCH} &  \\     
         DRF       & -1.470 &  &   &  &  \\
         Small GARCH Tree & -2.547 & -0.414  &   &  &  \\
         GARCH Tree      & -8.651 & -5.577 & -8.288   &  &  \\
         GARCH Forest    & -6.409 & -3.429 & -2.777  & 4.973 &  \\ \midrule
         &  &  &   &  &  \\
         Avg loss     & 0.393  & 0.375  & 0.367   & 0.303  & 0.343   \\
         \bottomrule
    \end{tabular}
    \vspace{0.75cm}
    \label{tab:garch_oos}
\end{table}

\begin{figure}[t!]
    \centering
    \caption{\textbf{The estimated GARCH tree model.} This figure depicts the tree structure for the GARCH model. The tree's splits are based on SPX, RVOL and VIX, which refer to the S\&P 500 return, realized volatility, and the option-implied volatility index respectively. \vspace{0.5cm}}
    \includegraphics[scale=0.60]{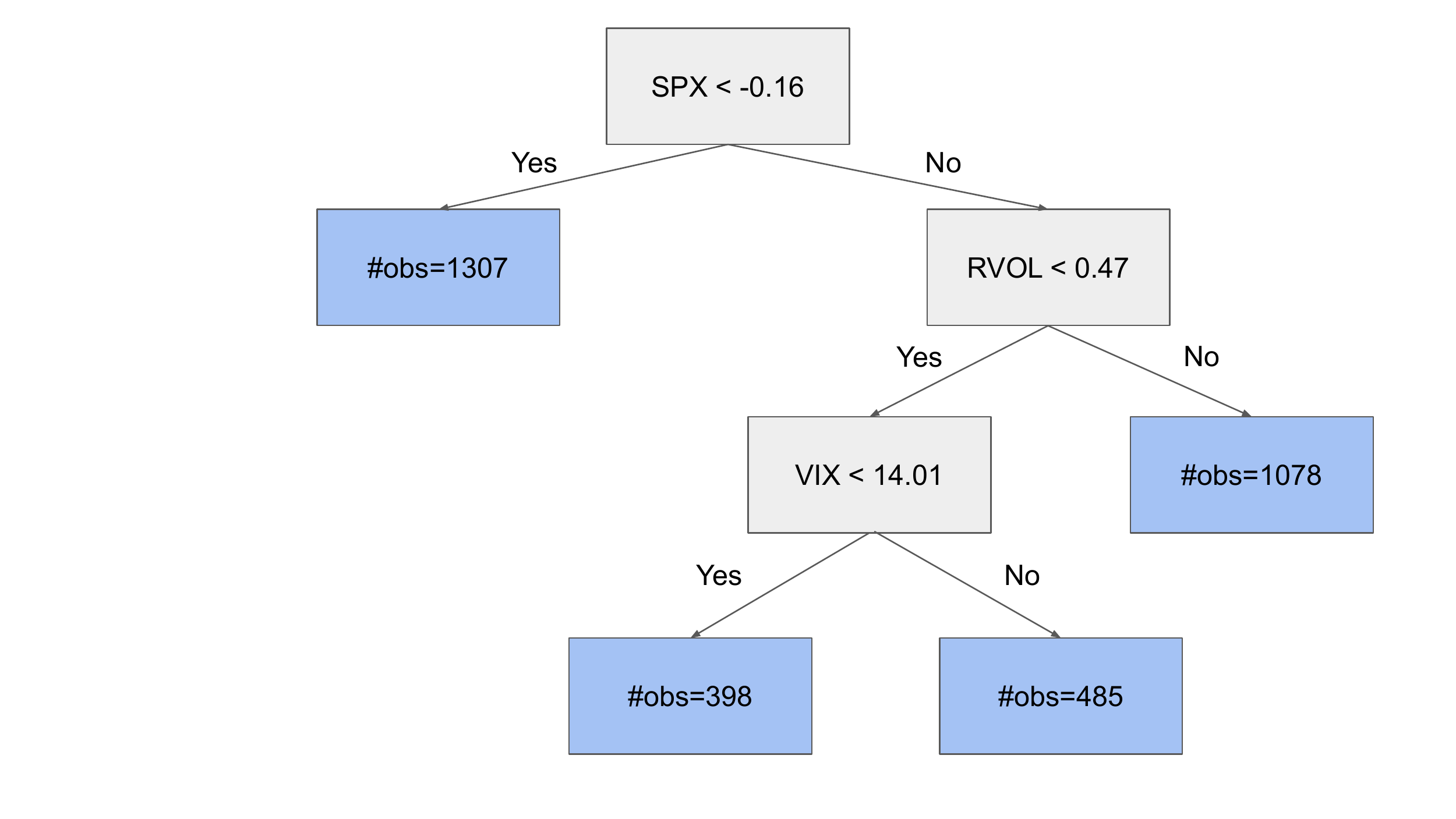}
    \label{fig:garch_tree}
    \vspace{0.75cm}
\end{figure}

To understand the source of forecast gains from the GAS tree model, Figure \ref{fig:garch_tree} presents the estimated tree structure. The optimal tree depth was found to be three, with three different splitting variables. The algorithm first chooses the S\&P 500 return with a threshold value $-0.16$ (its $40^{th}$ percentile) which approximately splits the sample using positive and negative market returns, consistent with an asymmetric reaction of future volatility to past returns, also known as a ``leverage effect'' (\citealp{black1976}). The second split in the tree is for positive returns and uses realized volatility with a threshold of 0.47 (10.9\% in annualized standard deviation form), corresponding to the $45^{th}$ percentile of RVOL (conditional on the first split), thus approximately splitting positive return days into ``high'' and ``low'' volatility days. The third and final split is for low volatility days and uses VIX with a threshold of 14.01, corresponding to its $45^{th}$ conditional percentile. Recalling that the ``variance risk premium'' (\citealp{carrWu2008}; \citealp{BollerslevVRP2009}) can be approximated as the difference between VIX$^2$ and RVOL, the four terminal nodes of the tree in Figure \ref{fig:garch_tree} can be interpreted, approximately, as those associated with (1) negative returns, (2) positive returns and high realized volatility, (3) positive returns, low realized volatility and low variance risk premium, (4) positive returns, low realized volatility and high variance risk premium.\footnote{Employing a semi-structural regime switching model, \cite{baele2010determinants} finds that variance risk premium is an important economic factor in explaining stock return volatility.}


\subsection{Forecasting the distribution of future stock returns} \label{sec:uni_for}
We next consider the problem of forecasting the entire distribution of daily returns on the S\&P 500 index. Our baseline model is the t-GAS model introduced by \cite{creal2013generalized}, which captures both excess kurtosis, through the use of the Student's $t$ distribution for the standardized residuals, and time-varying volatility, through the GAS structure for the scale parameter. Assuming a zero conditional mean, the t-GAS model is:
\begin{eqnarray}
     y_t  & = & \sigma_t \epsilon_t;  \quad \epsilon_t  \sim~i.i.d.~ t(v)  \label{eq:tGAS} \\
     \sigma_t^2 & = & \omega + \beta \sigma_{t-1}^2 + \alpha(1+3v^{-1}) \left(\frac{1+v^{-1}}{1-2 v^{-1}} \biggl\{ 1+\frac{v^{-1}}{1-2 v^{-1}} \frac{y_{t-1}^2}{\sigma_{t-1}^2}  \biggr\} ^{-1} y_{t-1}^2-\sigma^2_{t-1}\right) \nonumber
\end{eqnarray}
where $\nu$ is the degrees of freedom parameter for the $t$ distribution. As in \cite{creal2013generalized}, $\nu$ is assumed constant, while $\sigma_t^2$ varies over time. The dynamics of $\sigma_t^2$ differs from the familiar GARCH structure when $\nu < \infty$, and simplifies to the GARCH model when $\nu \rightarrow \infty$. The $\{ \cdot \}$ term in equation (\ref{eq:tGAS}) implies a more moderate reaction to a large past return than in the GARCH model, as large returns are more common under the  $t$ distribution than the Normal distribution. 

\begin{table}[t!]
    \centering
     \caption{\textbf{Out-of-sample performance of t-GAS models using negative log-likelihood loss.} This table presents $t$-statistics from Diebold-Mariano tests of out-of-sample forecast performance (top four rows) and average out-of-sample negative $\log \mathcal{L}$ losses (bottom row). A negative $t$-statistic indicates that the model in the column had lower average loss (i.e., a higher out-of-sample log-likelihood) than the model in the row, while a positive $t$-statistic indicates the opposite. \vspace{0.5cm}}
    \begin{tabular}{lccccc}
    \toprule
        &  &  & Small & GAS & GAS  \\
                  & t-GAS & DRF & GAS Tree & Tree & Forest  \\
         \cmidrule(lr){2-6}
         & \phantom{GAS Tree} & \phantom{GAS Tree} & \phantom{GAS Tree}  & \phantom{GAS Tree} &  \\
         DRF       & -5.396 &  &   &  &  \\
         Small Tree & -3.555 & 1.571 &   &  &  \\
         GAS Tree      & -6.517 & -1.240 & -4.924 &  &  \\
         GAS Forest    & -5.485 & 1.555 & -1.048  & 2.755  &  \\ \midrule
         & & & & & \\
         Avg Loss     & 1.179 & 1.141 & 1.153 & 1.132 & 1.147  \\
         \bottomrule
    \end{tabular}
    \label{tab:tgas_oos}
    \vspace{0.75cm}
\end{table}

\pagebreak
Table \ref{tab:tgas_oos} presents comparisons of the t-GAS model with the distributional random forest (DRF), the ``small GAS tree'' (which only uses the lagged return as a state variable), and the GAS tree and forest models, both of which use all ten state variables described in Section \ref{sec:data}. The first column shows that the t-GAS model is significantly out-performed by all four competing models, with Diebold-Mariano $t$-statistics less than -3.5 in all cases. We also observe that the GAS tree significantly outperforms the ``small GAS tree'' and also the GAS forest, both with $p$-values less than 0.01. The GAS tree also outperforms the DRF forecast, but the difference is not statistically significant at the 5\% level.

\begin{figure}[t!]
    \centering
    \caption{\textbf{The estimated t-GAS tree model.} This figure depicts the tree structure for the t-GAS model. The tree's splits are based on SPX and RVOL, which refer to the S\&P 500 return and realized volatility respectively. \vspace{0.0cm}}
    \includegraphics[scale=0.6]{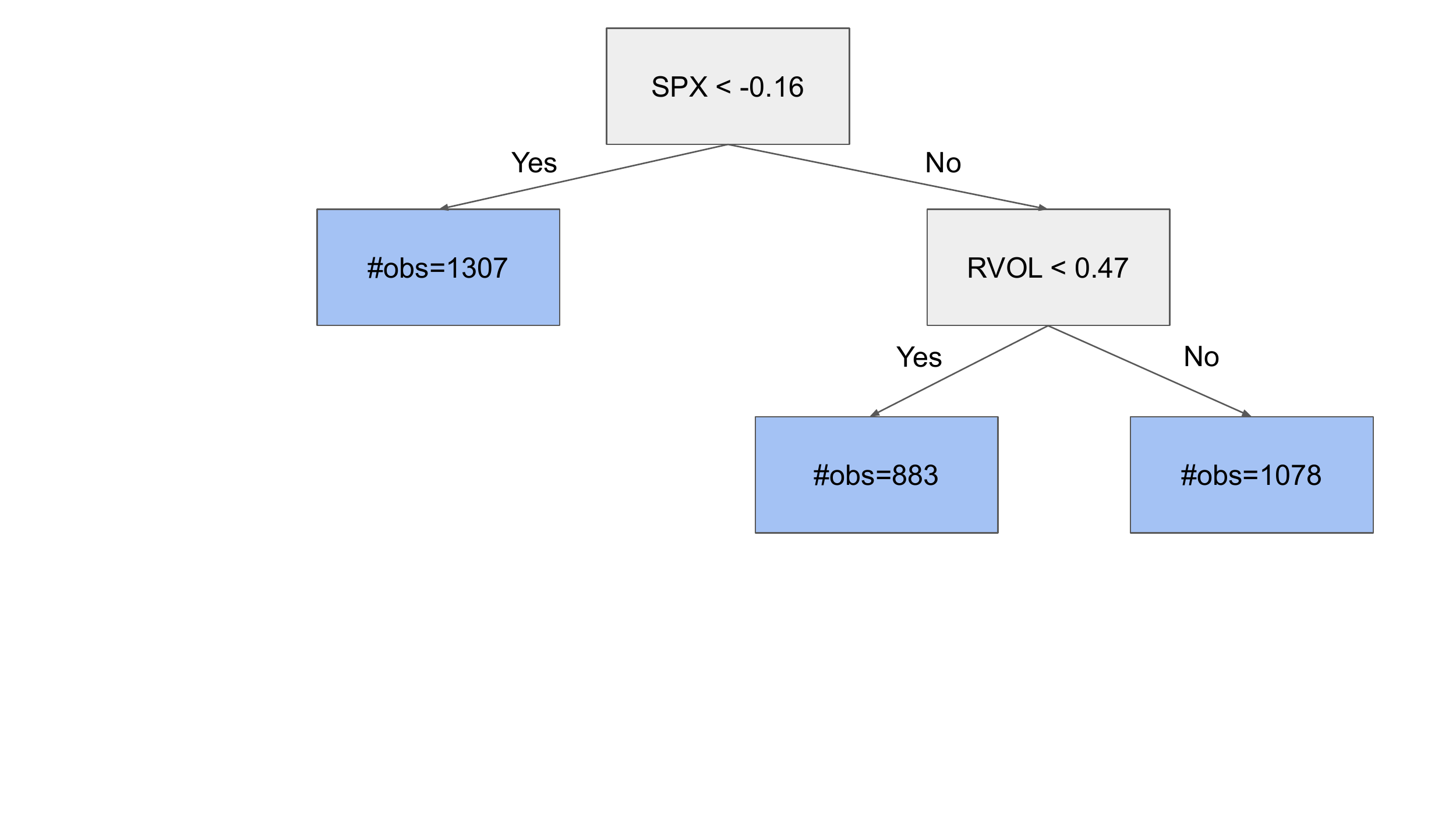}
    \label{fig:tgas_tree}
\end{figure}

Figure \ref{fig:tgas_tree} shows the estimated t-GAS tree. Unlike the structure for the GARCH tree in the previous section, this tree only has depth of two, but those first two levels are identical to those of the GARCH tree.\footnote{We use a grid of $19$ values, corresponding to the 0.05, 0.10, ..., 0.95 quantiles of the state variable, for the threshold for each state variable, making our finding of identical threshold values less surprising.} The three terminal nodes of the tree have roughly equal numbers of observations, and can be interpreted as (1) negative returns, (2) positive returns and low realized volatility, (3) positive returns and high realized volatility.




\subsection{Forecasting the joint distribution of stock and bond returns}

We now focus on forecasting the joint distribution of stock and bond returns, using the S\&P 500 index and 10-year Treasury bond for this purpose. We construct this model by combining t-GAS models for the marginal distributions, as utilized in the previous section, see equation \eqref{eq:tGAS}, with a Student's $t$ copula, as in \cite{janus2014long}. Copulas are convenient tools for capturing the dependence between variables separately from the marginal distributions of each of the variables; see \cite{PattonHBOOK2013} for a review of these methods for economic time series. The use of a $t$ copula allows for the possibility of tail dependence, or ``joint crashes and joint booms.''  Although the marginal distributions and the copula are all from the Student's $t$ family, this joint distribution is not bivariate Student's $t$ distribution unless all three degrees-of-freedom parameters are identical; instead, this joint distribution allows for varying degrees of fat tails in each marginal distribution and the joint tails. The Student's $t$ copula with GAS dynamics for the correlation parameter is:
\begin{eqnarray}
     \mathbf{u}_t & \sim  &  \mathbf{C}_{\text{Student}}(\rho_t,\nu) \\ 
  \rho_t &=& \frac{\exp{ \{ \tilde{\rho}_t} \} -1}{\exp{ \{ \tilde{\rho}_t \}}+1} \nonumber \\
     \tilde{\rho}_t & = & \omega + \beta \tilde{\rho}_{t-1} \nonumber \\ 
    && + \alpha \left( \frac{2}{1-\rho_t^2}\right) \left(\frac{1+\rho_t^2}{g + (2g-1)\rho_t^2}\right) \left(w_t(x_{1,t}x_{2,t}-\rho_t) - \frac{\rho_t}{1+\rho_t^2}(w_tx_{1,t}^2+w_tx_{2,t}^2-2)\right) \nonumber
\end{eqnarray}
where $x_{i,t} \equiv F_{\text{Student}}^{-1}(u_{i,t};\nu_i)$ uses the inverse $t$ CDF with degrees-of-freedom parameter $\nu_i$, and $g$ and $w_t$ are scalars defined in Appendix \ref{app:derivation}. As in the univariate $t$-GAS model, we impose that the copula degrees of freedom parameter, $\nu$ is constant over time.

\begin{table}[t!]
    \centering
    \caption{\textbf{Out-of-sample performance of $t$ Copula GAS models using negative log-likelihood loss.} This table presents $t$-statistics from Diebold-Mariano tests of out-of-sample forecast performance (top four rows) and average out-of-sample negative $\log \mathcal{L}$ losses (bottom row). A negative $t$-statistic indicates that the model in the column had lower average loss (i.e., a higher out-of-sample log-likelihood) than the model in the row, while a positive $t$-statistic indicates the opposite. \vspace{0.5cm}}
    \begin{tabular}{lccccc}
    \toprule
  &  &  & Small & GAS & GAS  \\
                  & GAS & DRF & GAS Tree & Tree & Forest  \\
         \cmidrule(lr){2-6}
         & \phantom{GAS Tree} & \phantom{GAS Tree} & \phantom{GAS Tree}  & \phantom{GAS Tree} &  \\
         DRF       & 2.598  &  &   &  &  \\
         Small Tree & -1.451 & -2.811  &   &  &  \\
         GAS Tree      & -1.451 & -2.811 & \textbf{---}   &  &  \\
         GAS Forest    & -3.680 & -4.092 & -0.795 & -0.795  &  \\ \midrule
         & & & & & \\
         Avg Loss     & -0.079 & -0.063 & -0.084 & -0.084 & -0.087  \\
         \bottomrule
    \end{tabular}
    \label{tab:tcopula_oos}
    \vspace{0.75cm}
\end{table}

Table \ref{tab:tcopula_oos} shows the out-of-sample performance of competing models. We see that the benchmark GAS model significantly beats the distributional random forest (DRF), unlike in the two univariate applications, but it is beaten by both the ``small tree'' and the GAS tree models. The best-performing model is the GAS forest, which significantly outperforms the GAS model, with a $t$-statistic of -3.7. The GAS forest also significantly beats the DRF, but does not significantly beat either of the tree models. Interestingly, in this application the GAS tree reduces to the ``small GAS tree'' model, in that the only state variables selected for use in the tree structure are lags of the stock and bond returns. 

\begin{figure}[t!]
    \centering
    \caption{\textbf{Leave-one-out variable importance for the Student's $t$ copula GAS forest.} This figure plots the change in out-of-sample average negative log-likelihood between using the original GAS forest and a GAS forest with a  state variable (listed on the $y$-axis) omitted. Positive values indicate a worsening of forecast performance, and thus that the omitted state variable is an important component of the original model. The horizontal lines represent 95\% confidence intervals for the difference in average log-likelihoods. \vspace{0.5cm}}
    \includegraphics[scale=0.42]{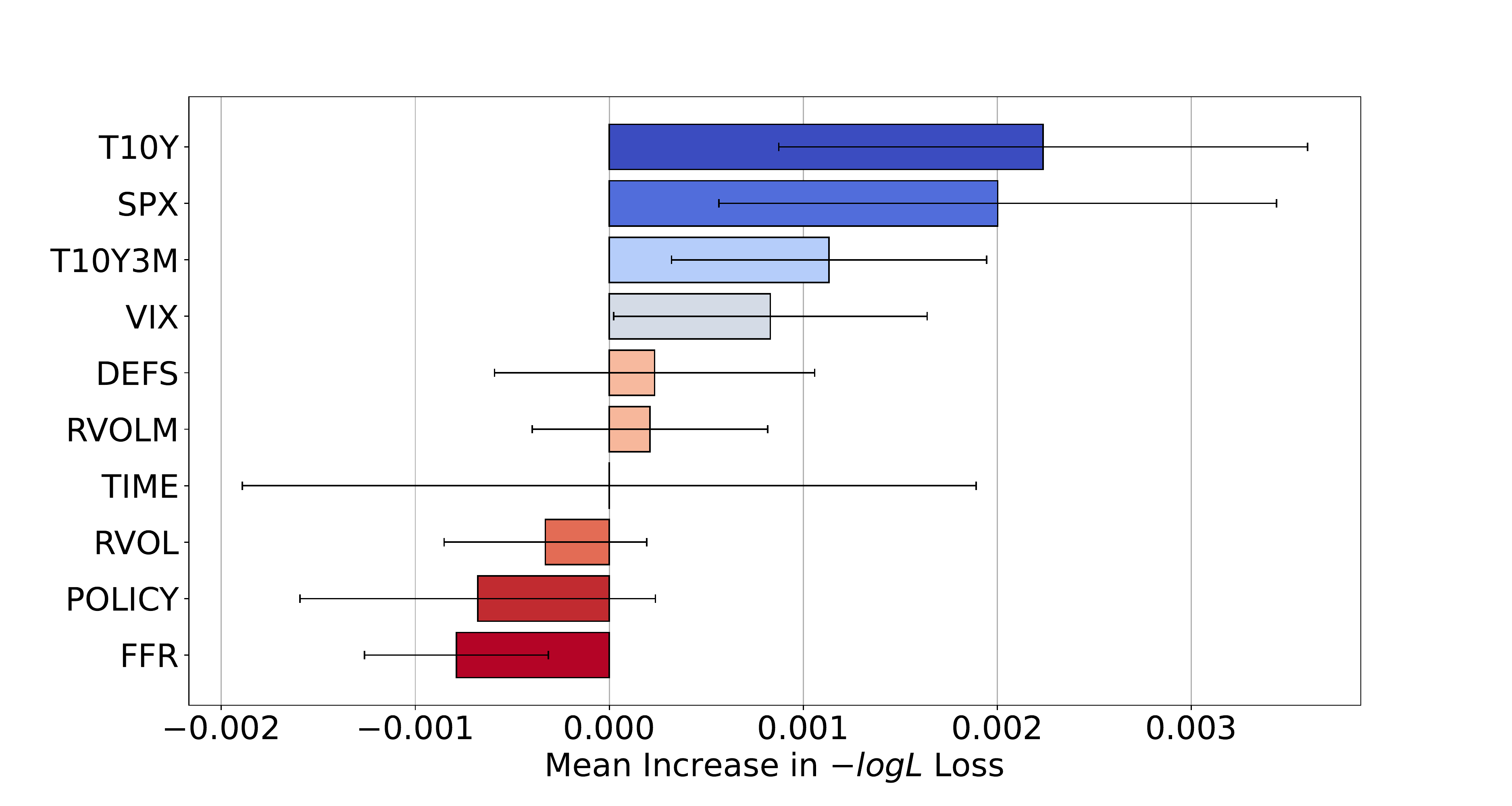}
    \label{fig:tcopula_vi} \vspace{0.5cm}
\end{figure}

In contrast with the univariate applications, the best-performing model is the GAS forest, not the GAS tree, and so we cannot present a tree diagram to better understand the structure of the best model. In its place, we consider two methods for interpreting the optimal model. Firstly, we conduct a \textit{leave-one-out} analysis to measure the importance of each state variable. Specifically, we drop each state variable from the analysis, one at a time, and re-compute the optimal GAS forest forecasts. We then compare the average out-of-sample average loss from the original GAS forest and the GAS forest using one fewer state variable. If the difference is small, then the omitted state variable is unimportant, while if the difference is positive, then the omitted variable is important for forecast performance.\footnote{As this is an out-of-sample comparison of models, it is possible that the difference is negative, meaning that the smaller GAS forest is \textit{preferred} to the original GAS forest. With a large enough sample size, including irrelevant state variables leads to no change, positive or negative, in out-of-sample loss, as such state variables will never be selected. In finite samples, however, irrelevant state variables may be mistakenly included.} We can use Diebold-Mariano tests to determine whether the change in out-of-sample loss is statistically significant. Figure \ref{fig:tcopula_vi} presents the results of this analysis, and shows that the most important state variable for the GAS forest is the lagged bond return, T10Y, followed by the lagged stock market return, SPX. Omitting either of these significantly (at the $5\%$ level) deteriorates the GAS forest forecasts. The slope of the term structure (T10Y3M) and the volatility index (VIX) are also found to be important for the quality of GAS forest forecasts. Interestingly, we observe a statistically significant \textit{improvement} in forecast performance by omitting the Federal funds rate (FFR) as a state variable, indicating that this variable is unhelpful for out-of-sample forecasting, but is selected for inclusion in the forest often enough to deteriorate the forecast.

\begin{figure}[t!]
    \centering
    \caption{\textbf{Parameter estimates as a function of S\&P 500 index returns (SPX) for the Student's $t$ copula GAS forest.} This figure plots the average, across bootstrap samples, values of $\omega/(1-\beta)$ (upper-left), $\beta$ (upper-right), and $\alpha$ (lower-left) from the GAS model in equation \eqref{eq:tvp}, as well as the average predicted correlation, $\rho_t$ (lower-right) from that model. The state variable is discretized into bins based on 1\% quantiles. The values for each of these quantities from the benchmark GAS model are plotted in horizontal dashed lines. The solid lines are local quadratic polynomials fitted to the grey dots. \vspace{0.5cm}}
    \begin{tabular}{cc}
         \includegraphics[scale=0.5]{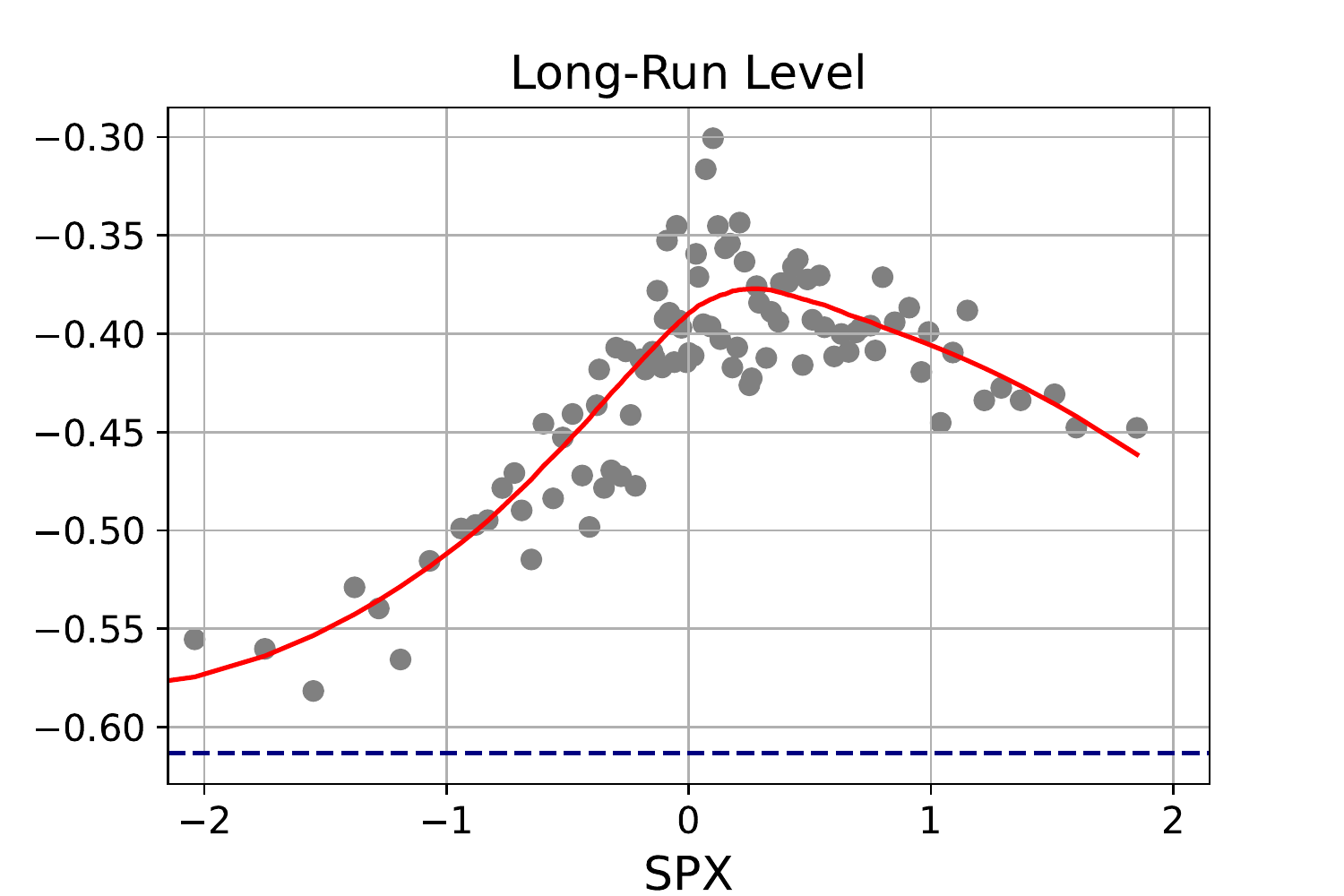} &  
         \includegraphics[scale=0.5]{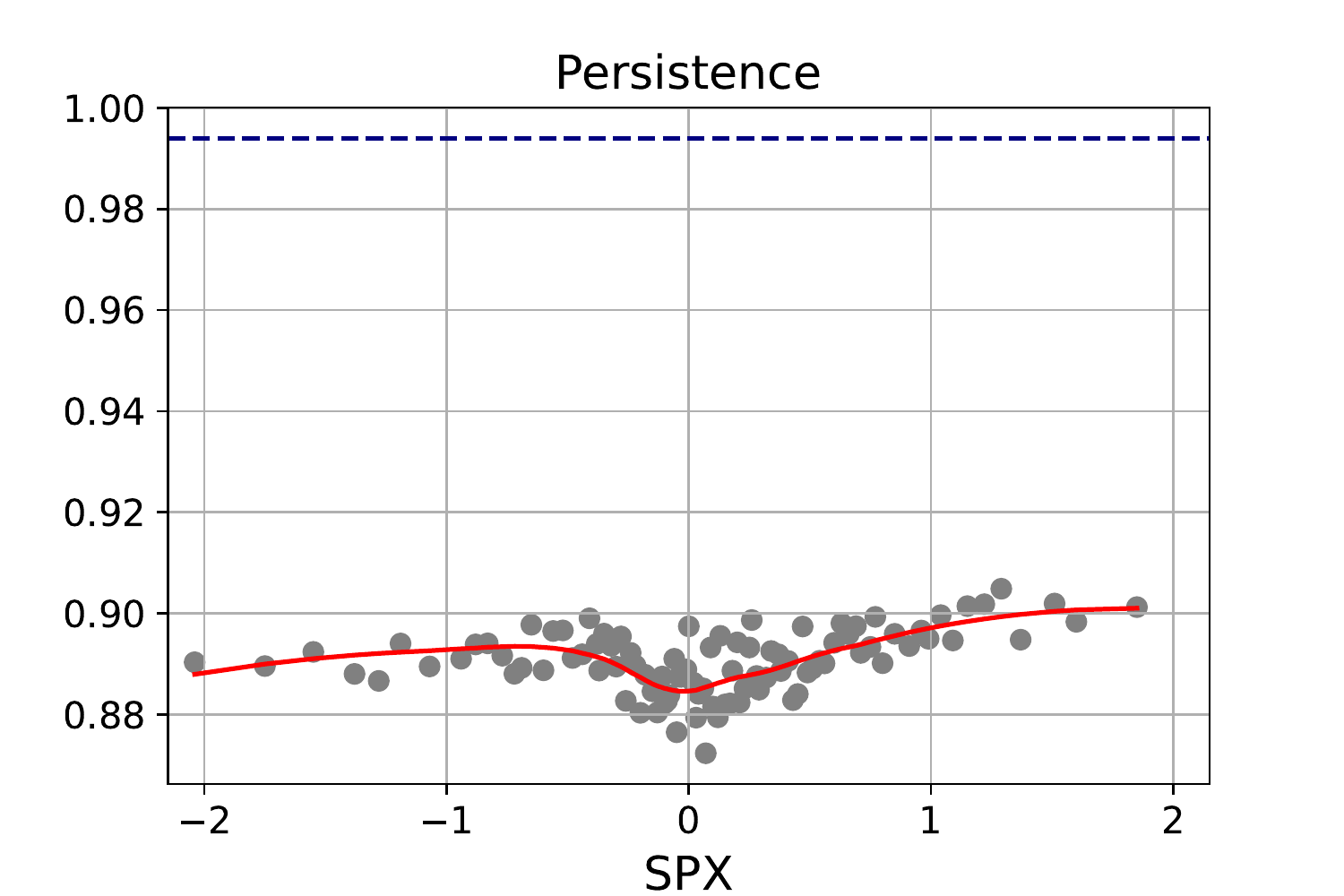} \\
         \includegraphics[scale=0.5]{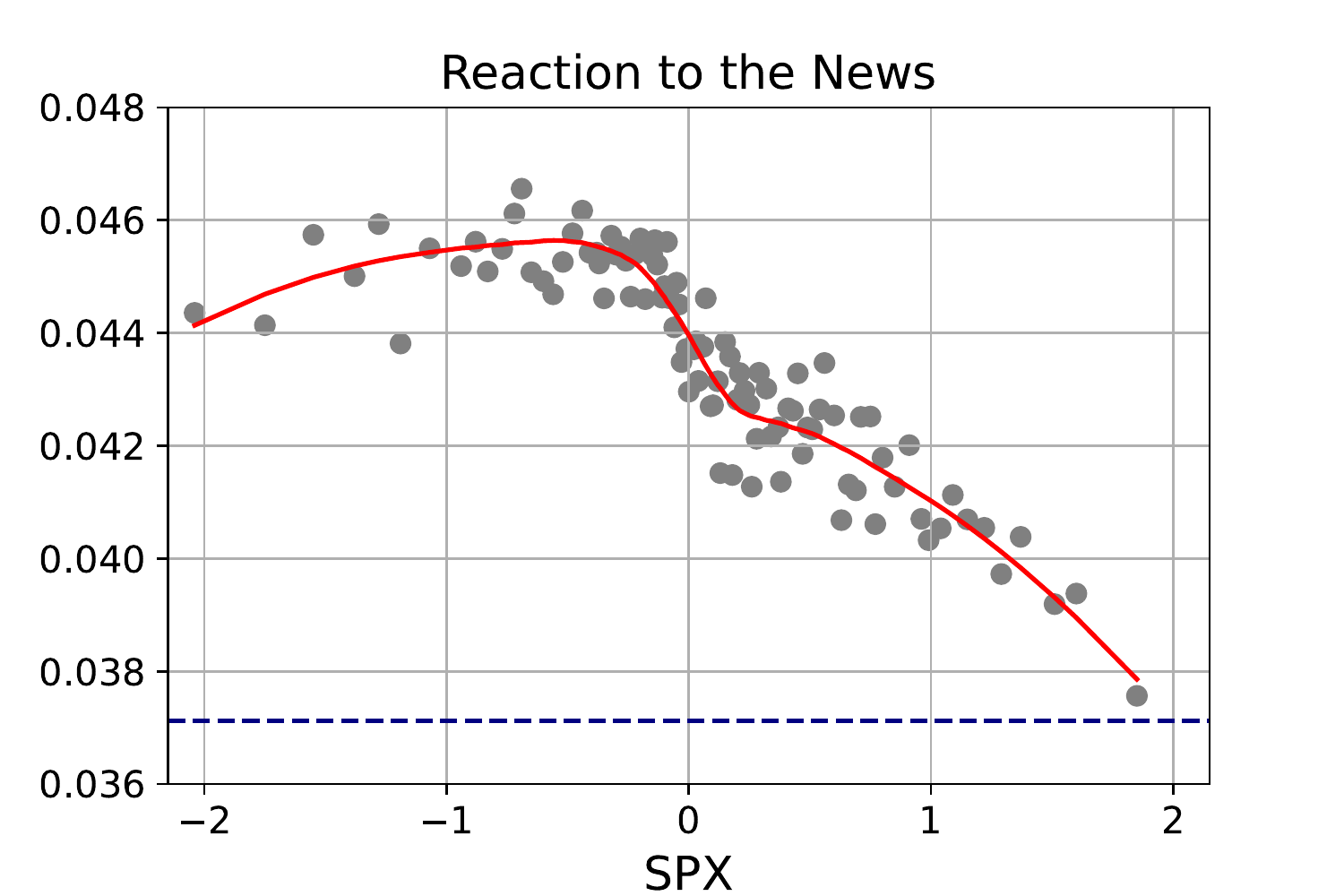} & 
         \includegraphics[scale=0.5]{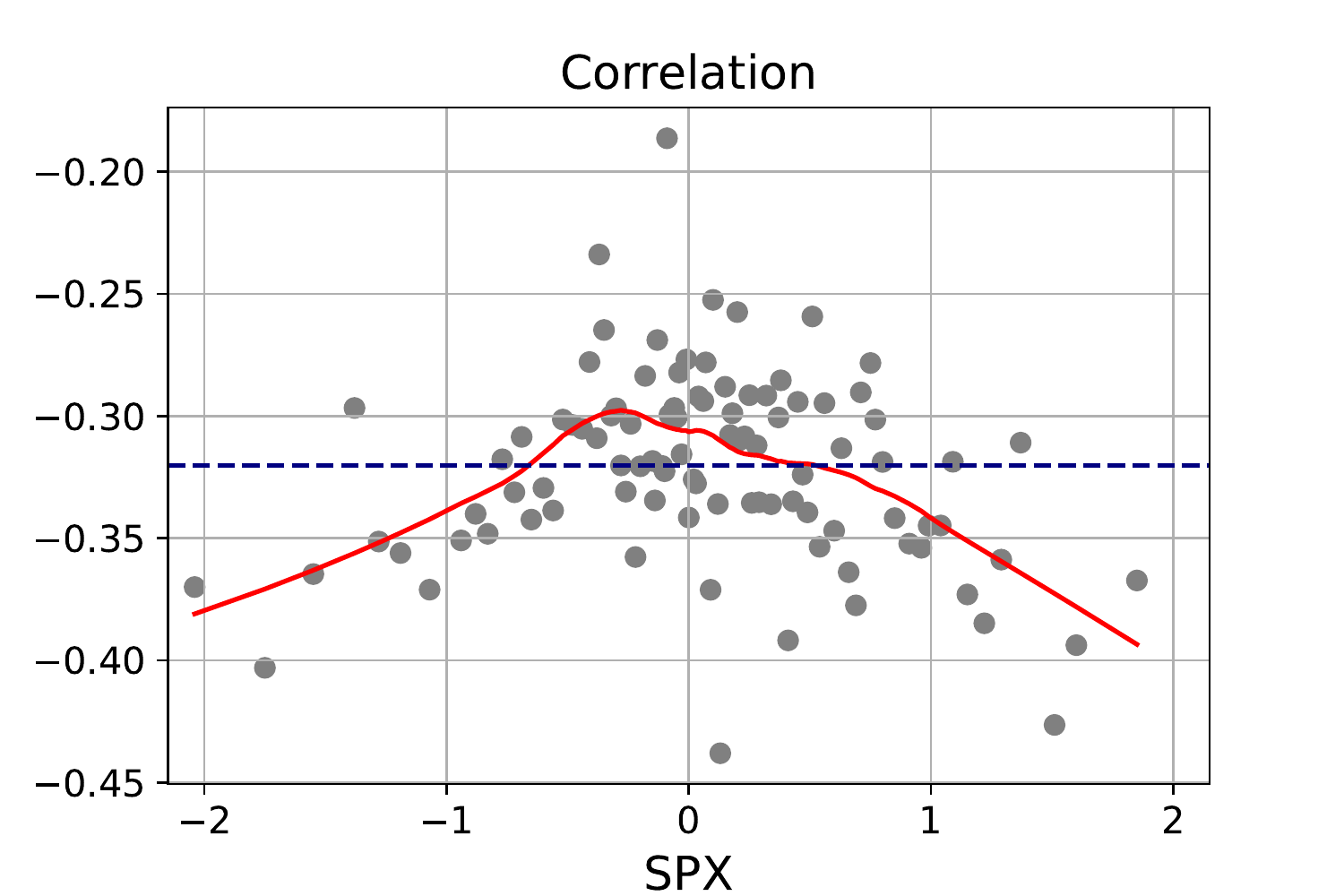}
    \end{tabular}
    \label{fig:tcopSPX} \vspace{0.5cm}
\end{figure}

We next analyze the impact of the most important state variable on the GAS forest model by plotting the parameters of the GAS model (recall equation \ref{eq:tvp}) as a function of the state variable, see Figure \ref{fig:tcopSPX}.\footnote{Similar plots for the other variables found to be significant in Figure \ref{fig:tcopula_vi} are presented in the supplemental appendix.} The parameters $\beta$ and $\alpha$ are interpretable as the persistence and reaction-to-news of the model. The intercept, $\omega$ is not directly interpretable, and we instead plot $\omega/(1-\beta)$ which is interpretable as the long-run level of the GAS process. As the GAS forest involves averaging 200 bootstrap samples, each based on a random subset of state variables, we construct this plot by averaging the GAS parameters within each 1\% quantile of the state variable. 

Figure \ref{fig:tcopSPX} presents the results for the stock return as a state variable.  The upper-left panel shows that the long-run correlation peaks when the stock market return is around zero, at about $0.35$. It declines to about $-0.45$ as the stock return increases to $2\%$, while it declines markedly to nearly $-0.6$ when the stock return is $-2\%$. This is interpretable as a ``flight-to-quality'' effect, with low stock market returns leading to more negative comovements between the stock and bond markets. Figure \ref{fig:tcop_T10Y} in the supplemental appendix plots the corresponding results with the bond return as the state variable, and that figure is also consistent with a flight-to-quality effect.

The upper-right panel of Figure \ref{fig:tcopSPX} shows that the persistence of the GAS model is roughly unrelated to the stock market return. The lower-left panel shows that the GAS model reacts about $20\%$ more strongly to news when the stock market is down versus up: the $\alpha$ parameter is $0.046$ when stocks are down, while it is $0.038$ when stocks are up. This is consistent with investors paying closer attention to bad news than good news, a finding similar to that of \cite{PattonSheppard2015} in a different context. 

The lower-right panel of Figure \ref{fig:tcopSPX} shows the predicted correlation from the GAS forest as a function of the stock market return, and reveals an inverted U-shaped pattern, though with substantial noise. Without an underlying model to guide interpretation, one might be hesitant to draw too much from this panel. With the benefit of the GAS structure underlying our forest forecast, we know that this shape is primarily coming from the long-run level, $\omega/(1-\beta)$, in the upper-left panel, and that that relationship is strong. This reveals an important benefit of combining machine learning tools with economically motivated, and/or empirically successful, econometric models.

\subsection{Forecasting market activity} \label{sec:duration_for}
Finally, we consider the problem of forecasting the time between consecutive trade events, known as a ``trade duration.''\footnote{A ``trade event'' could be a single transaction occurring, or a total of $x$ transactions occuring, or a total of $\$y$ value of transactions occuring, or a total of $z$ shares being transacted, or some other event defined as a function of characteristics of transactions.}  Trade durations are a measure of market activity, and are important for high-frequency risk management and transaction cost minimization. We take as our benchmark the ``autoregressive conditional duration'' (ACD) model of \cite{engle1998autoregressive}. Denoting $y_t$ as the time (in minutes) between consecutive trade events, the ACD model assumes an exponential distribution for $y_t$ with a time-varying conditional mean, $\mu_{t}$:
\begin{equation}
\begin{array}{rcl}
     y_t  & \sim & \operatorname{Exp}(\mu_t)  \label{eq:tGAS} \\
     \mu_t & = & \omega + \beta \mu_{t-1} + \alpha y_{t-1}.
\end{array}
\end{equation}
\cite{creal2013generalized} show that the ACD model is also a special case of a GAS model, allowing us to consider it in our study of tree- and forest-based extensions of GAS models. See \citet{BauwensHautsch09} for a review of ACD and related models.

For our empirical analysis in this section, we use high-frequency data on SPY, an exchange traded fund tracking the S\&P 500 index, between January 1$^{st}$ 2021 and December 31$^{st}$ 2021. We study the time taken for 10,000 shares of SPY to be transacted, leading to 5,100 durations during this sample period, and corresponding to an average duration of 14.9 minutes. 


Table \ref{tab:acd_oos} presents the out-of-sample forecast performance of the baseline ACD model as well as the competing models considered in previous sections: the distributional random forest (DRF), the ACD tree using only lagged durations as a state variable (Small ACD tree), the ACD using all 13 state variables, and the ACD forest model. We firstly observe that the DRF model for durations, which is pure machine learning tool, is beaten by all other models including the benchmark with $t$-statistics all less than $-2.4$. 

\begin{table}[t!]
    \centering
     \caption{\textbf{Out-of-sample performance of ACD models using negative log-likelihood loss.} This table presents $t$-statistics from Diebold-Mariano tests of out-of-sample forecast performance (top four rows) and average out-of-sample negative $\log \mathcal{L}$ losses (bottom row). A negative $t$-statistic indicates that the model in the column had lower average loss (i.e., a higher out-of-sample log-likelihood) than the model in the row, while a positive $t$-statistic indicates the opposite. \vspace{0.5cm}}
    \begin{tabular}{lccccc}
    \toprule
        &  &  & Small & ACD & ACD  \\
                  & ACD & DRF & ACD Tree & Tree & Forest  \\
         \cmidrule(lr){2-6}
         & \phantom{ACD Tree} & \phantom{ACD Tree} & \phantom{ACD Tree}  & \phantom{ACD Tree} &  \\
         DRF           &  6.170  &         &   &   &   \\
         Small ACD Tree    &  -2.448 &  -5.778 &   &   &   \\
         ACD Tree      &  -1.004 &  -3.988 &  -0.458 &   &   \\
         ACD Forest    &  -2.293 &  -9.358 &  -0.830  & 0.083   &   \\ \midrule
         & & & & & \\
         Avg Loss      &  7.412 & 7.494  & 7.398 & 7.388 & 7.389  \\
         \bottomrule
    \end{tabular}
    \label{tab:acd_oos}
    \vspace{0.75cm}
\end{table}

The baseline ACD model has higher loss in the out-of-sample compared with to tree- and forest-based extensions. Interestingly, the ``small ACD tree'' model, which only uses lagged duration as a state variable, significantly beats the baseline ACD model, while the ``ACD tree'' model, which considers 13 state variables \textit{including} lagged duration, does not significantly beat the baseline model. This reveals the value in imposing some structure (namely, reducing the number of potential state variables) on the tree-based extension in this application. The ``ACD forest'' model significantly beats the baseline ACD model, revealing the forecast gains available from averaging forecasts from randomly formed trees, consistent with \cite{breiman2001random} in a linear regression setting.

\begin{figure}[t!]
    \centering
    \caption{\textbf{Leave-one-out variable importance for the ACD forest.} This figure plots the change in out-of-sample average negative log-likelihood between using the original ACD forest and a ACD forest with a  state variable (listed on the $y$-axis) omitted. Positive values indicate a worsening of forecast performance, and thus that the omitted state variable is an important component of the original model. The horizontal lines represent 95\% confidence intervals for the difference in average log-likelihoods.}
    \includegraphics[scale=0.4]{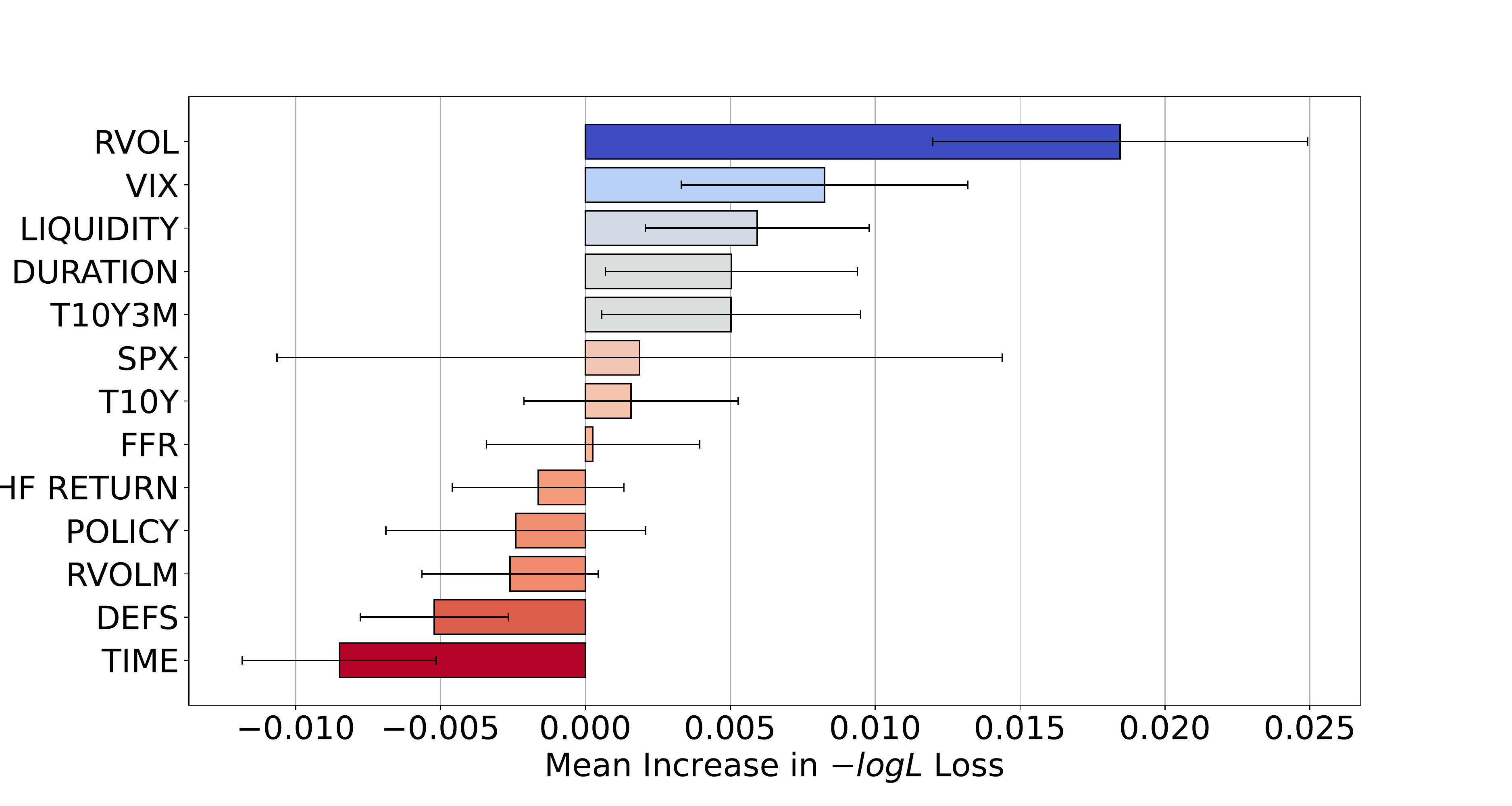}
    \label{fig:acd_vi}
\end{figure}


To understand how forecast accuracy is improved in the ACD forest model, we calculate the variable importance measure for each of the state variables, introduced in the previous section, and present the results in Figure \ref{fig:acd_vi}. The horizontal bars show the  increase in average loss function from omitting a state variable, and a positive value indicates that that state variable is important for forecasting. The lines refer to 95\% confidence intervals computed from Diebold-Mariano tests. We find that the volatility variables RVOL and VIX are two most important state variables, despite the fact that these are measured only daily, and so are constant within a trade day. The next two most important state variables are both high-frequency variables: \citet{amihud2002illiquidity} liquidity, and duration. Interestingly, we find that omitting default spread and time from the set of potential state variables actually improves forecast accuracy, indicating these are harmful when used in a forest-based ACD model.\footnote{Figure \ref{fig:acd_tree} in the supplemental appendix presents the optimal tree structure for the ACD tree. We find three terminal nodes in this structure, all reflecting the direction of the stock and bond markets. We find one state when the stock market is up (representing 55\% of the sample), another when the stock market is down and the bond market is not strongly up (representing 92\% of the remaining sample), and a small third state when the stock market is down and the bond market is strongly up (representing just 3\% of the total sample).}

\begin{figure}[t!]
    \centering
    \caption{\textbf{Parameter estimates as a function of realized volatility (RVOL) for the ACD forest model.} This figure plots the average, across bootstrap samples, values of $\omega/(1-\alpha-\beta)$ (upper-left), $\alpha+\beta$ (upper-right), and $\alpha$ (lower-left) from the ACD model in equation \eqref{eq:tvp}, as well as the average predicted duraction, $\mu_t$ (lower-right) from that model. The state variable is discretized into bins based on 1\% quantiles. The values for each of these quantities from the benchmark ACD model are plotted in horizontal dashed lines. The solid lines are local quadratic polynomials fitted to the grey dots. \vspace{0.5cm}}
    \begin{tabular}{cc}
         \includegraphics[scale=0.45]{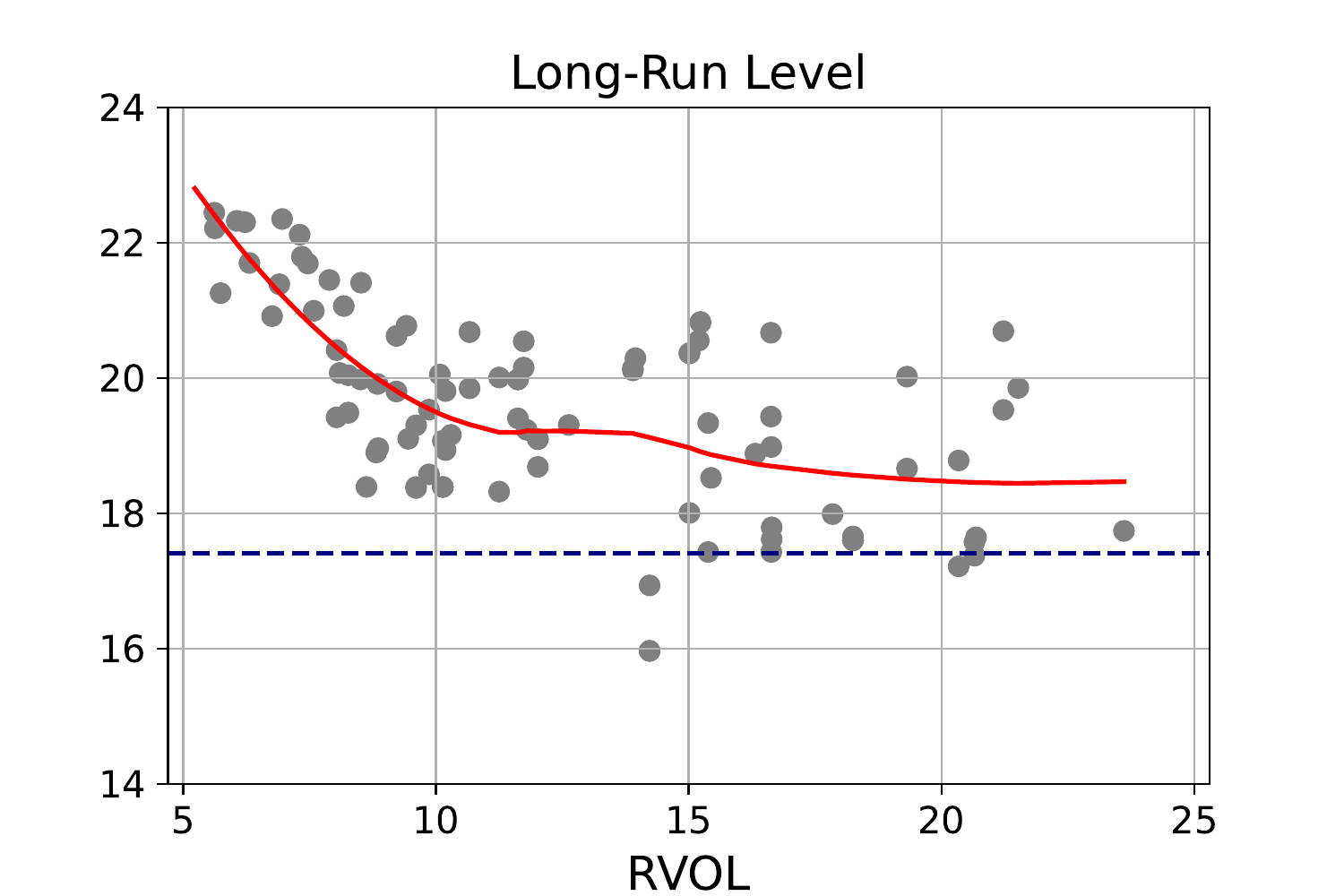} &  
         \includegraphics[scale=0.45]{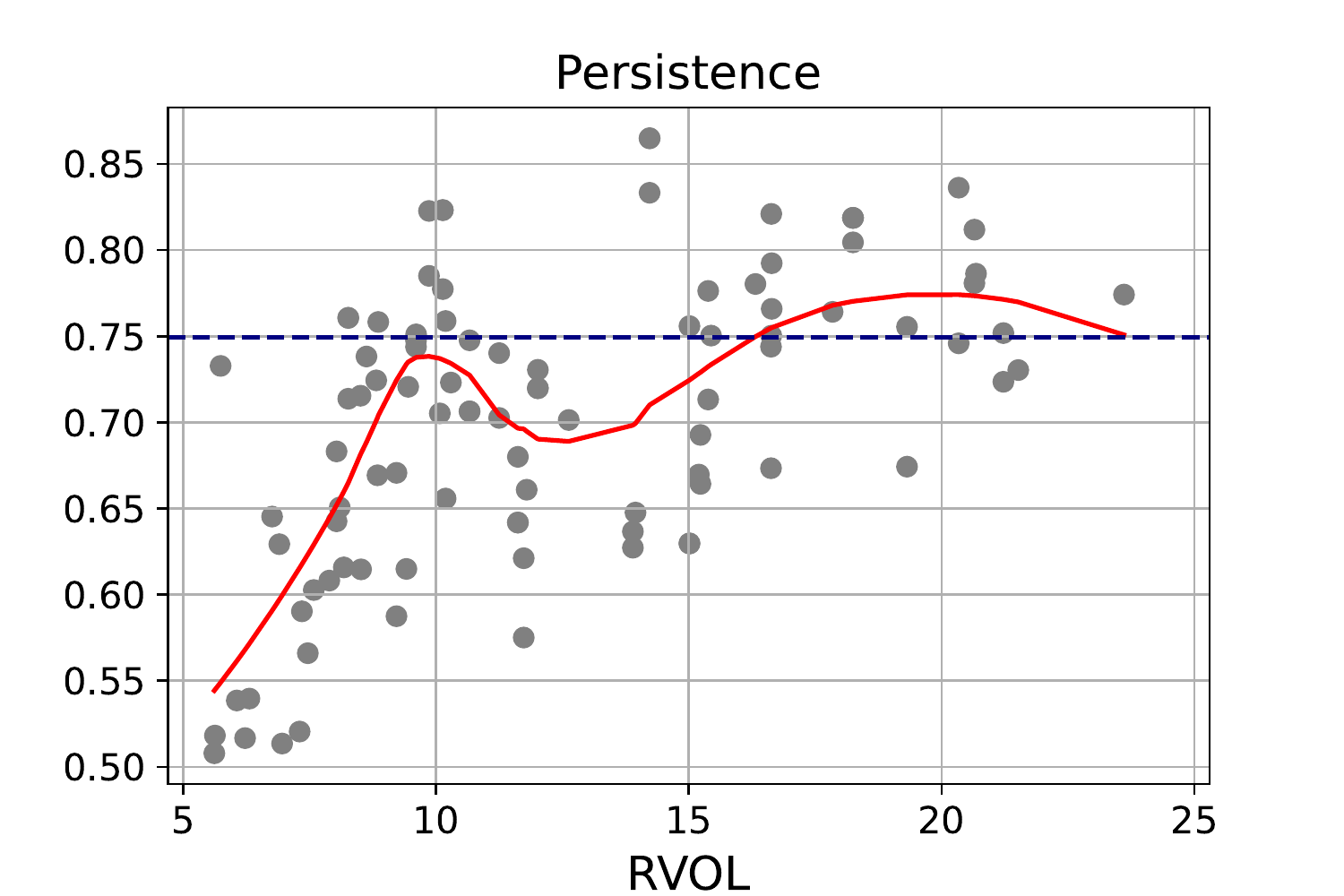} \\
         \includegraphics[scale=0.45]{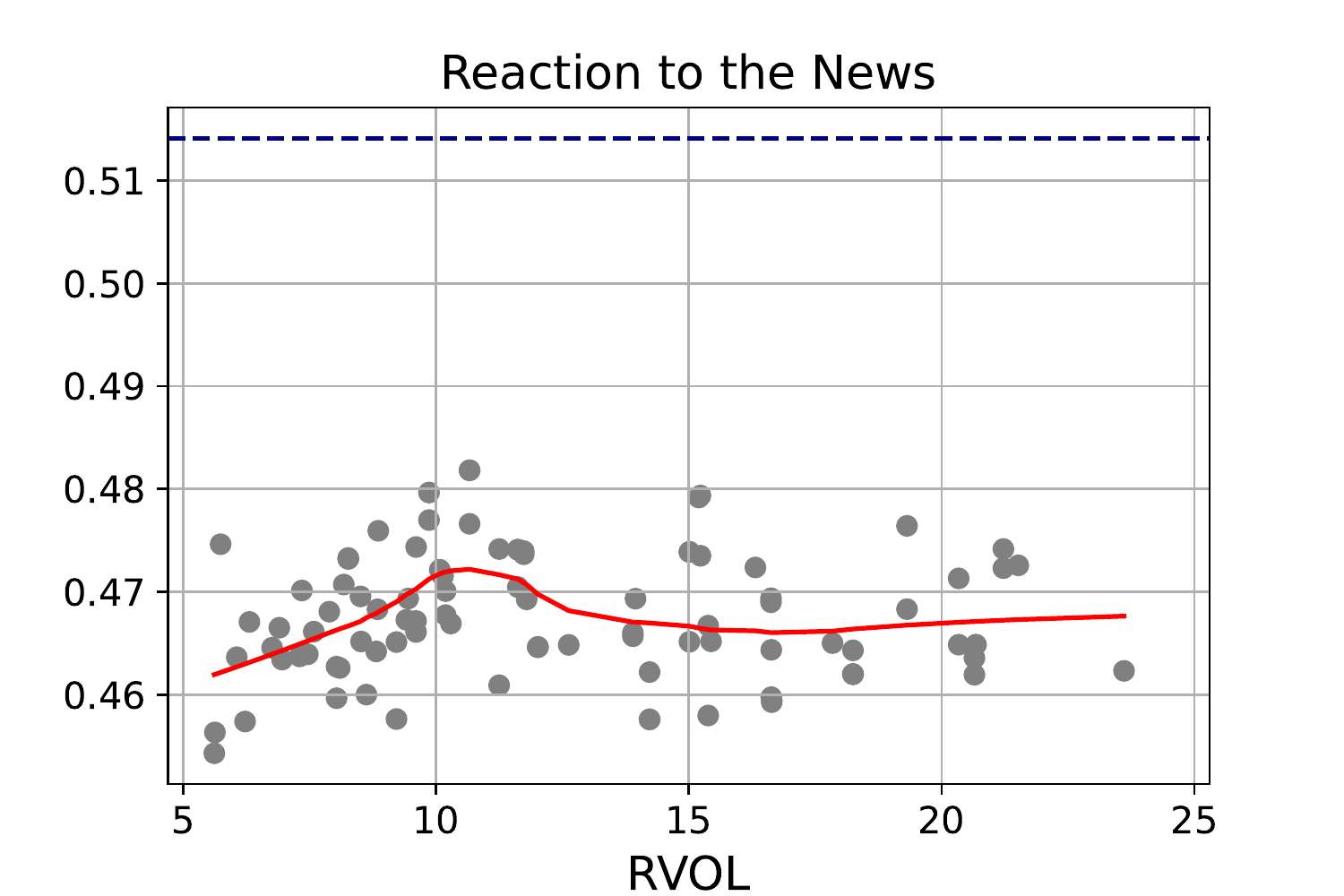} & 
         \includegraphics[scale=0.45]{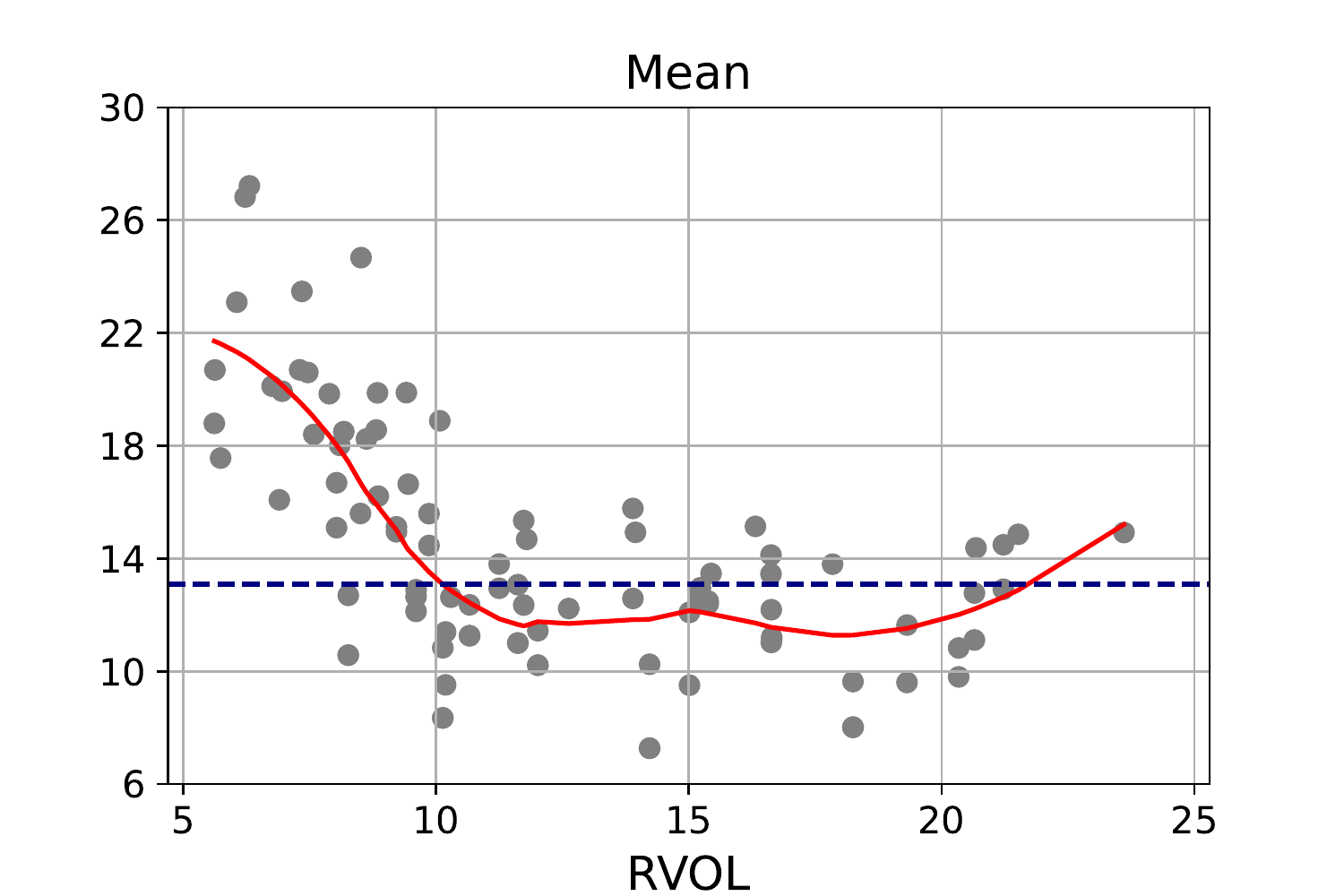}
    \end{tabular}
    \label{fig:ACD_RVOL}
\end{figure}

In Figure \ref{fig:ACD_RVOL} we plot the parameters of the forest ACD model as a function of the most important state variable, realized volatility (RVOL). We see that the long-run average duration implied by the model is highest when volatility is low, at around 23 minutes, and it steeply declines as volatility increases to about 10\%, at around 19 minutes. Persistence, measured by $\alpha+\beta$ in the ACD model, is lowest when volatility is low, and it increases sharply with volatility to around 10\% and is approximately flat beyond that. The parameter governing the reaction of the model to news, $\alpha$, is essentially flat as a function of volatility. The pattern for the forecasts from the ACD forest model, in the lower-right panel, shows that predicted durations are around 12 minutes when volatility is above 10\%, while they are around double that when volatility is low. This is consistent with the positive volume-volatility relationship (see, e.g.,  \citealp{Karpoff87}): volume and durations are negatively correlated (longer durations correspond to lower volumes, and vice versa) and so in periods of lower volatility average trade durations tend to be longer.

\section{Conclusion} \label{sec:conclusion}

Since its publication a decade ago, the class of generalized autoregressive score (GAS) models of \cite{creal2013generalized} and \cite{harvey2013dynamic} has proven to be a popular, parsimonious way to capture time variation in the parameter(s) of a given model. Its parsimonious nature, however, means that some important exogenous information or nonlinearities may be neglected, and how to best incorporate such additional features is difficult to determine \textit{ex ante}. We propose adapting methods from machine learning to search across a wide range of exogeneous variables and capture various forms of nonlinearity: the ``GAS tree'' combines the parsimonious structure of the GAS model with the flexibility of decision trees (\citealp{breiman1984classification,breiman2017classification}), and the ``GAS forest,'' analogous to the random forests of \cite{breiman2001random}, averages the forecasts from many GAS trees each produced on a bootstrap sample of the original data. Our GAS tree and GAS forest models can be applied whenever a GAS model is considered, and require from the researcher only a set of exogenous variables that are thought to be possibly useful.

We apply the proposed GAS tree and GAS forest models in four diverse applications: forecasting stock return volatility, the distribution of stock returns, the joint distribution of stock and bond returns, and high-frequency trade durations. We find that the proposed extensions lead to significantly improved forecasts in all four applications. We moreover uncover economic explanations for the \textit{sources} of these forecast gains. Through inspections of the optimal GAS tree structures, and variable importance and parameter sensitivity analyses for GAS forest forecasts, we find that the best-performing GAS tree and forest models are those that incorporate well-known empirical regularities, such as the leverage effect in volatility, the flight-to-quality effect in stock-bond correlations, and the volume-volatility relationship in trade durations. 

Faced with evolving data generating processes and the resulting ``small'' data sets, the success of machine learning methods in economics and finance relies on good, parsimonious benchmark models as reference points, an observation made nicely in \cite{IsraelKellyMoskowitz2020}. We used GAS models for this purpose; future work may consider augmenting a different class of forecasting models with  machine learning methods. 

\appendix
\newpage
\section{Derivation of the score function for the t-GAS copula} \label{app:derivation}
In this section we present the score function for the t-copula analysis discussed in Section \ref{sec:emp_app}. 
We refer to \cite{creal2013generalized} for the details of the univariate applications (the GARCH and t-GAS models).

\subsection{Notation}
We adopt the notation of \cite{creal2011dynamic} for ease of comparability with that article. The Kronecker product is denoted by $A \otimes B$ for any matrices $A$ and $B$. $A_{\otimes}$ stands for $A\otimes A$. The function $vec(A)$ vectorizes matrix $A$ into a column vector, and $vech(A)$ vectorizes just the lower triangle of $A$, which eliminates duplicates in the case that $A$ is symmetric. The duplication matrix is implicitly defined as the solution to $\mathcal{D}\operatorname{vech}(A)=\operatorname{vec}(A)$. Finally, $\mathbb{E}_{t-1}$ denotes the expectation conditional on the information available up to period $t-1$.    

\subsection{The probability density function of $t$ copula}
We adopt Student's t copula specification in our empirical analysis and its probability density function is given by
\begin{equation}
c(\mathbf{u}_t ; \Sigma_t, \nu)=\frac{\Gamma\left(\frac{\nu+2}{2}\right)\Gamma\left(\frac{\nu}{2}\right)}{\sqrt{|\Sigma_t|}\left[\Gamma\left(\frac{\nu+1}{2}\right)\right]^2}\left(1+\frac{\mathbf{x}_t^{\prime} \Sigma_t^{-1} \mathbf{x}_t}{\nu}\right)^{-\frac{\nu+2}{2}} \prod_{i=1}^2\left(1+\frac{x_{i,t}^2}{\nu}\right)^{\frac{\nu+1}{2}}
\end{equation}
where $\mathbf{x}_t=[x_{1,t},x_{2,t}]=[T^{-1}_{\nu}(u_{1,t}),T^{-1}_{\nu}(u_{2,t})]'$ obtained by applying the inverse of the univariate t distribution with $\nu$ degrees of freedom, $\Gamma(\cdot)$ is gamma function and $\Sigma_t$ is 2-by-2 correlation matrix. We denote the off-diagonal element of $\Sigma_t$ with $\rho_t$ which is the variable of interest:
\begin{equation}
    \Sigma_t = \begin{bmatrix}
        1 & \rho_t \\
        \rho_t & 1
    \end{bmatrix}
\end{equation}

\subsection{The score and information matrix}
We use inverse information matrix of the score function as a scaling factor in all applications. Given the complex structure of the Student's t copula, derivation of the information matrix requires tedious calculations, but \cite{creal2011dynamic} provide a closed-form formula of both score and information matrix. 
Based on their results, we can write
\begin{equation} \label{eq:score_info}
\begin{array}{rcl}
\nabla_t & = &\frac{\partial \log c_t\left(y_t \mid \Sigma_t ; \nu\right)}{\partial f_t} \\
& = & \frac{1}{2} (\mathcal{D} \Psi_t )^{\prime} \Sigma_{t \otimes}^{-1}\left[w_t \mathbf{x}_{t \otimes}-\operatorname{vec}\left(\Sigma_t\right)\right] \\
\mathcal{I}_{t \mid t-1} & = & \mathbb{E}_{t-1}\left[\nabla_t \nabla_t^{\prime}\right] \\
& = &\frac{1}{4} (\mathcal{D} \Psi_t )^{\prime} J_{t \otimes}^{\prime}\left[g G-\operatorname{vec}(\mathrm{I}) \operatorname{vec}(\mathrm{I})^{\prime}\right] J_{t \otimes} \mathcal{D} \Psi_t
\end{array}
\end{equation}
where $\Psi_t \equiv \frac{\partial \operatorname{vech(\Sigma_t)}}{\partial \rho_t}$, $J_t$ is such that $\Sigma_t^{-1}=J_t' J_t$, $w_t \equiv \frac{\nu+2}{\nu-2+\mathbf{x}_t^{\prime} \Sigma_t^{-1} \mathbf{x}_t}$, $g \equiv \frac{v+2}{v+4}$, and the explicit form of matrix $G$ is 
\begin{equation}
    G = \begin{bmatrix} 
    3 & 0 & 0 & 1 \\
    0 & 1 & 1 & 0 \\
    0 & 1 & 1 & 0 \\
    1 & 0 & 0 & 3
    \end{bmatrix}.
\end{equation}
We define the scaled score functions as $s_t = \mathcal{I}_{t \mid t-1}^{-1} \nabla_t$. We apply an additional transformation to $\rho_t$ which ensures that it lies between $-1$ and $1$. Specifically, we assume that $\rho_t = \frac{1-\operatorname{exp}(-\tilde{\rho}_t)}{1+\operatorname{exp}(-\tilde{\rho}_t)}$. In order to obtain scaled score function for transformed $\tilde{\rho}_t$, we multiply the original scaled score with the derivative of transformation function: $\tilde{s}_t = \frac{\partial \tilde{\rho}_t}{\partial \rho_t} s_t$. When we use the explicit form of each component in equation (\ref{eq:score_info}), we obtain the following formula of the scaled score function $t$ copula:
\begin{equation}
    s_t = \left( \frac{2}{1-\rho_t^2}\right) \left(\frac{1+\rho_t^2}{g + (2g-1)\rho_t^2}\right) \left(w_t(x_{1,t}x_{2,t}-\rho_t) - \frac{\rho_t}{1+\rho_t^2}(w_tx_{1,t}^2+w_tx_{2,t}^2-2)\right).
\end{equation}
Note that $(g,w_t) \rightarrow (1,1)$ as $\nu \rightarrow \infty$, we also obtain the scaled score function for Gaussian copula:
\begin{equation}
    s_t = \left( \frac{2}{1-\rho_t^2}\right) \left(x_{1,t}x_{2,t}-\rho_t - \frac{\rho_t}{1+\rho_t^2}(x_{1,t}^2+x_{2,t}^2-2)\right).
\end{equation}

\newpage
\onehalfspacing
\bibliographystyle{apalike}
\bibliography{ref.bib}

\newpage
\doublespacing
\setcounter{page}{1}
\setcounter{footnote}{0}
\label{supp_app}
\begin{center}
	SUPPLEMENTARY MATERIAL FOR   \vspace{10pt} \\
{\Large\bf {Generalized Autoregressive Score \\Trees and Forests} \vspace{10pt} }\\
Andrew J. Patton and Yasin Simsek \\
	This Version: \today \\ 
\end{center}

\bigskip
\bigskip

\renewcommand{\thesection}{S.\arabic{section}}   
\renewcommand{\thepage}{S.\arabic{page}}  
\renewcommand{\thetable}{S.\arabic{table}}   
\renewcommand{\thefigure}{S.\arabic{figure}}

\setcounter{section}{0}
\setcounter{table}{0}
\setcounter{figure}{0}

\begin{table}[h!]
    \centering
    \caption{Out-of-sample performance of t-GAS models using QLIKE loss}
    \begin{tabular}{lccccc}
    \toprule
         & Benchmark & DRF & Tiny Tree & Tree & Forest \\
         \cmidrule(lr){2-6}
         Benchmark &  &  &   &  &  \\
         DRF       & -1.332 &  &   &  &  \\
         Tiny Tree & -2.006 & -0.789 &   &  &  \\
         Tree      & -5.277 & -3.856 & -5.594 &  &  \\
         Forest    & -3.652 & -2.599 & -0.871 & 3.154  &  \vspace{0.2cm} \\
         QLIKE     & 0.403  & 0.382 & 0.369 & 0.324  & 0.358  \vspace{0.2cm} \\
         \bottomrule
    \end{tabular}
    \label{tab:qlike_vol}
\end{table}

\begin{table}[h!]
    \centering
    \caption{Out-of-sample performance of GARCH models using -logL loss}
    \begin{tabular}{lccccc}
    \toprule
         & Benchmark & DRF & Tiny Tree & Tree & Forest \\
         \cmidrule(lr){2-6}
         Benchmark &  &  &   &  &  \\
         DRF       & -3.025 &  &   &  &  \\
         Tiny Tree & -2.300 & 0.350 &   &  &  \\
         Tree      & -5.777 & -2.356 & -5.818  &  &  \\
         Forest    & -6.703 & -2.392 & -2.183  & 1.594  &  \vspace{0.2cm} \\
         -logL     & 1.205 & 1.177  & 1.182 & 1.148 & 1.162  \vspace{0.2cm} \\
         \bottomrule
    \end{tabular}
    \label{tab:logL_vol}
\end{table}

\begin{figure}
    \centering
    \caption{Parameter estimates as a function of T10Y state variable for $t$ Copula forest-based GAS}
    \begin{tabular}{cc}
         \includegraphics[scale=0.45]{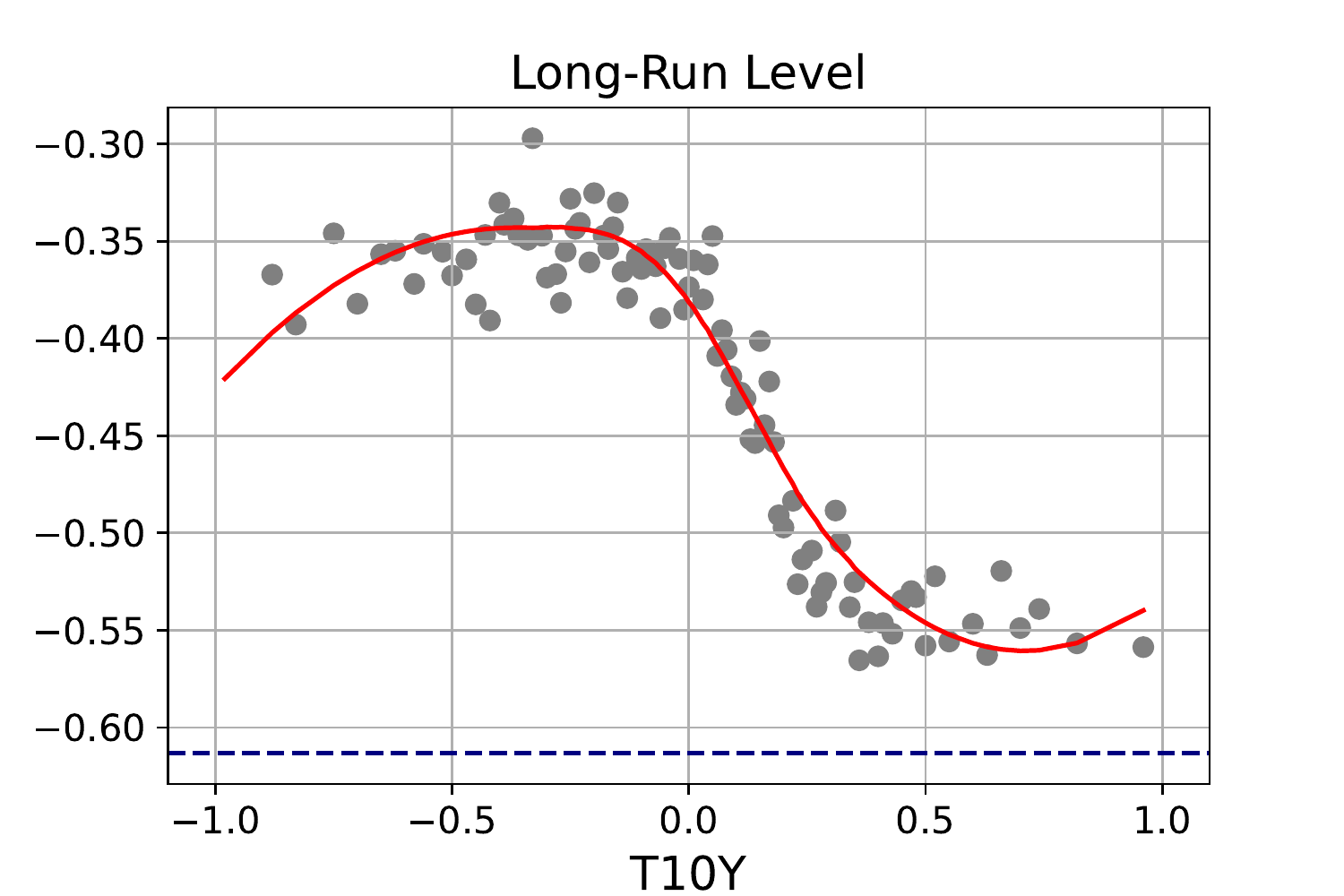} &  
         \includegraphics[scale=0.45]{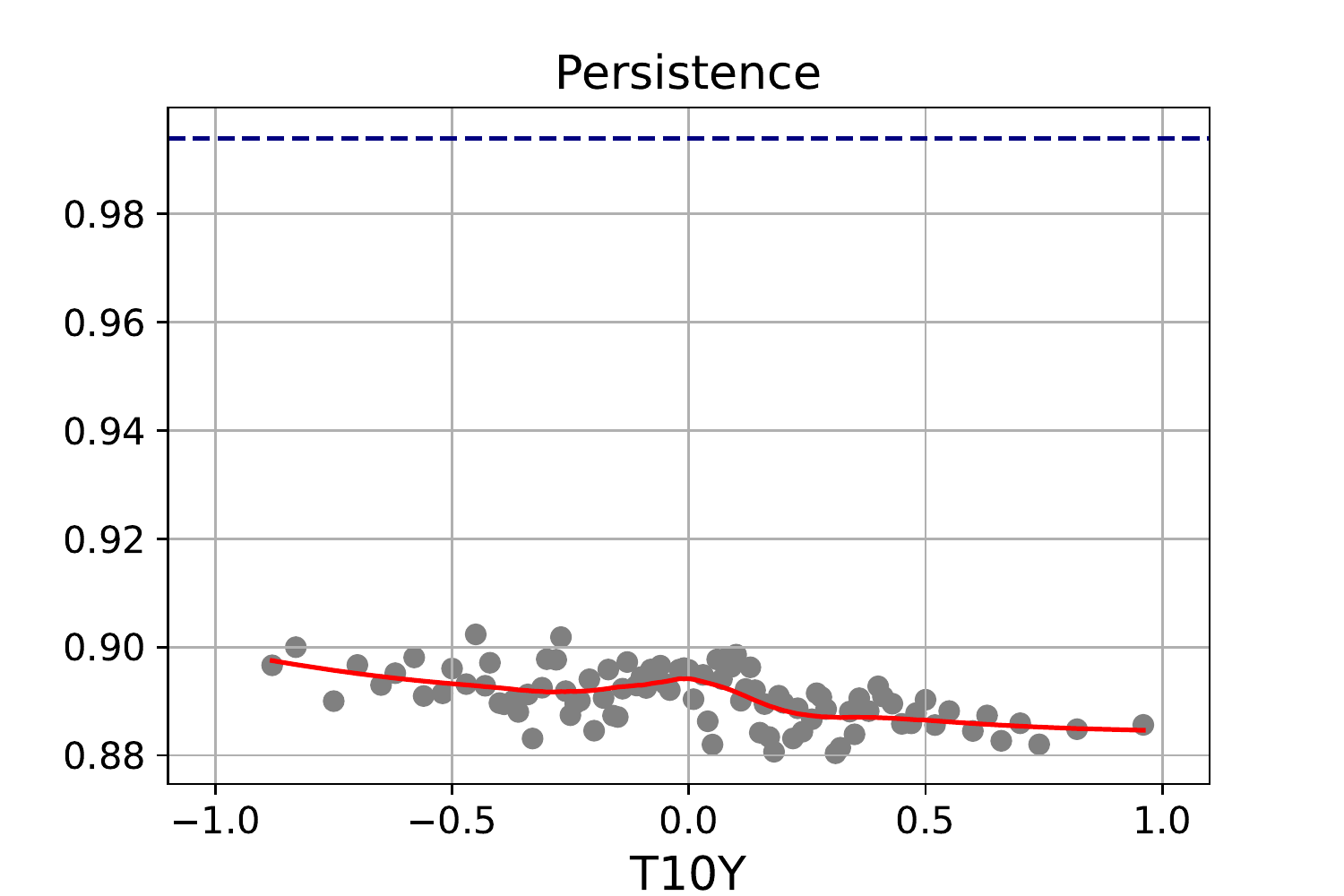} \\
         \includegraphics[scale=0.45]{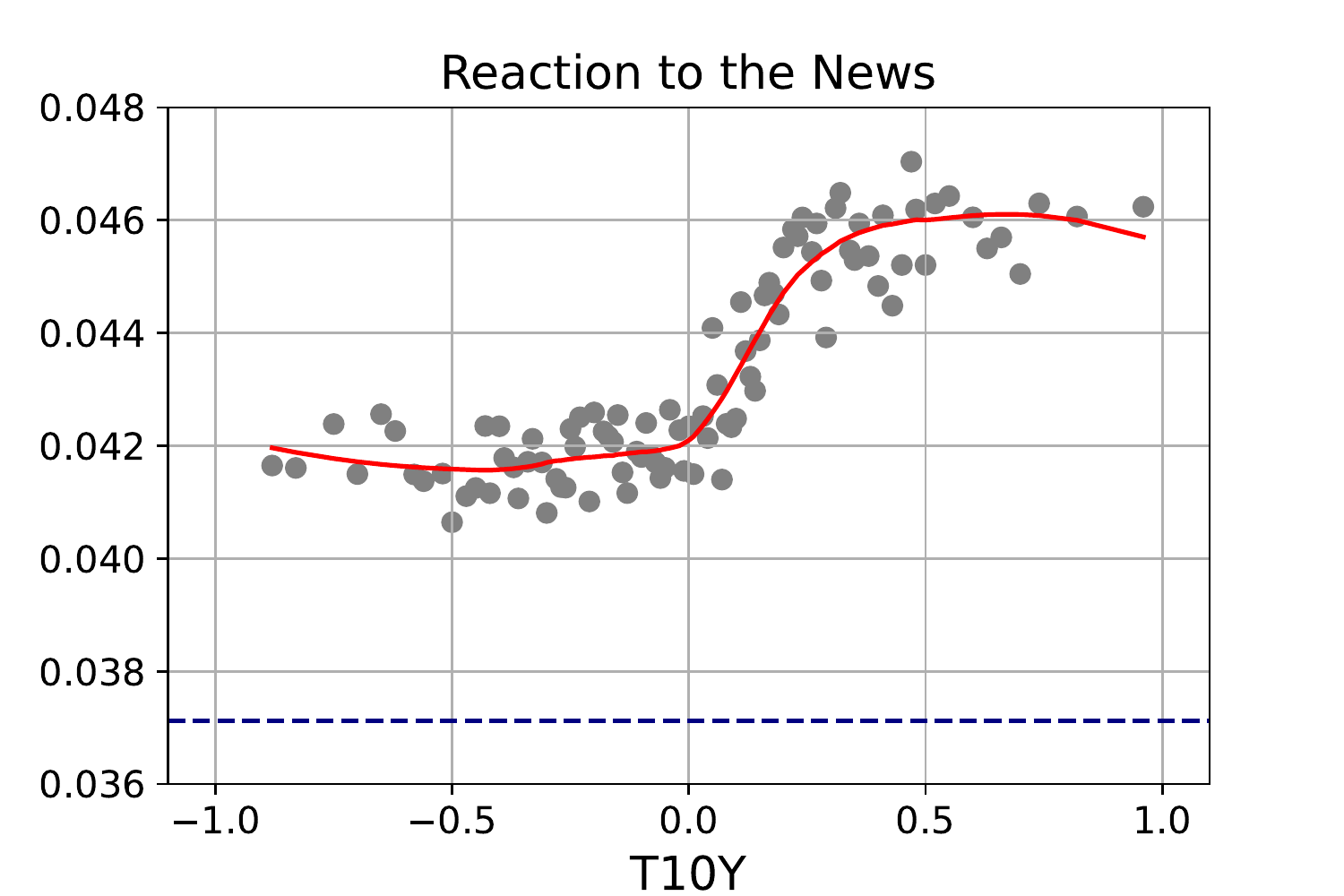} & 
         \includegraphics[scale=0.45]{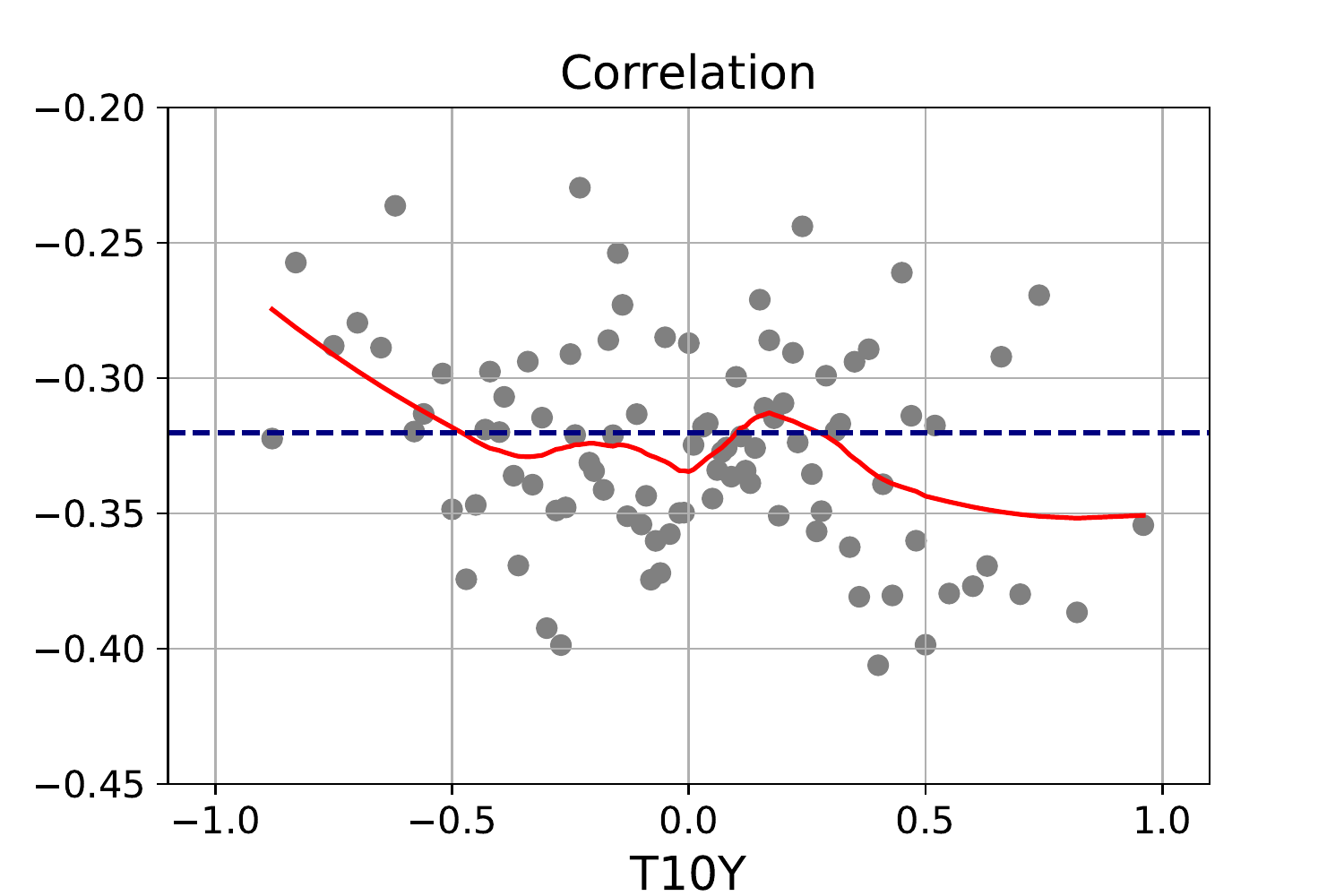}
    \end{tabular}
    \label{fig:tcop_T10Y}
\end{figure}

\begin{figure}
    \centering
    \caption{Parameter estimates as a function of T10Y3M state variable for $t$ Copula forest-based GAS}
    \begin{tabular}{cc}
         \includegraphics[scale=0.45]{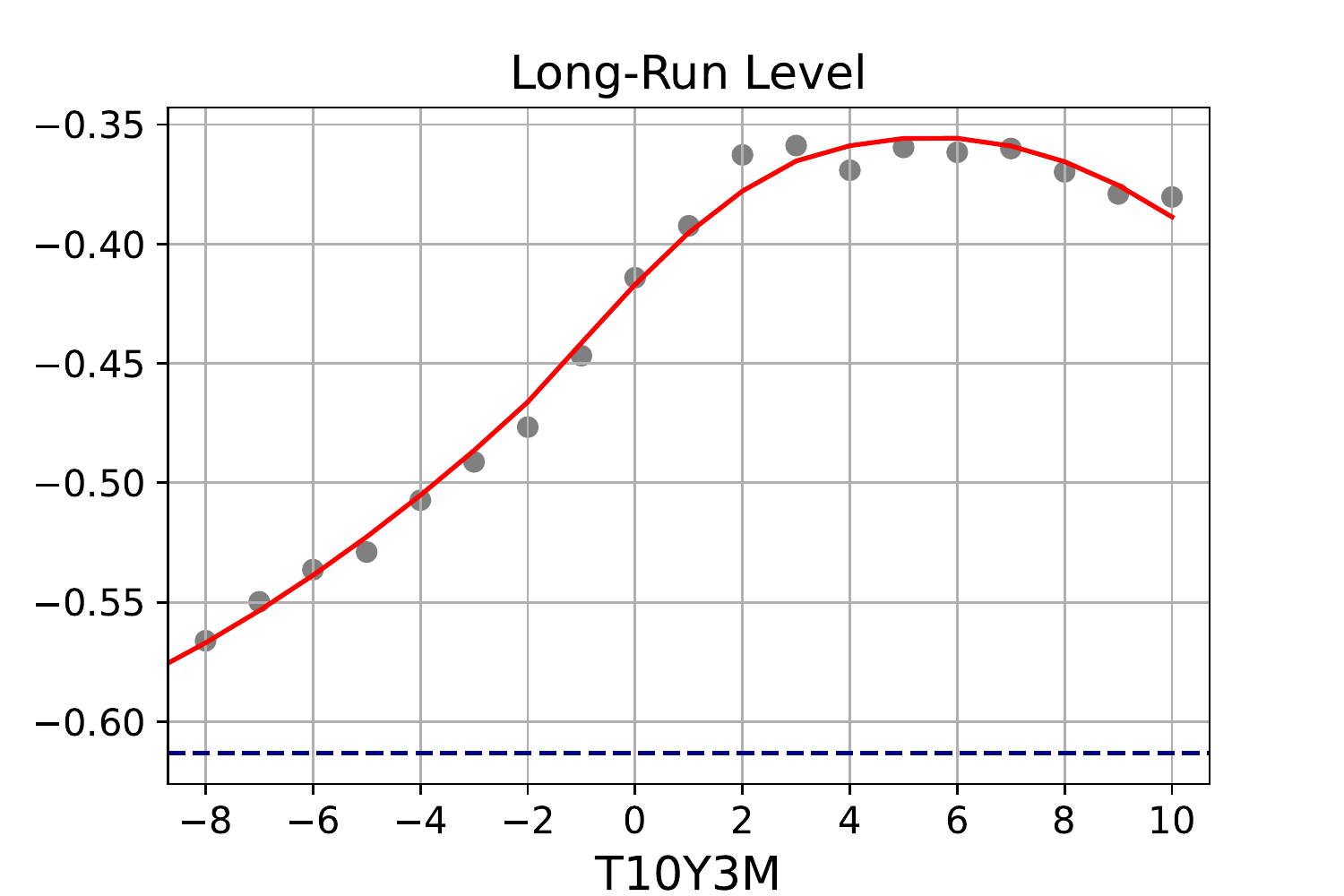} &  
         \includegraphics[scale=0.45]{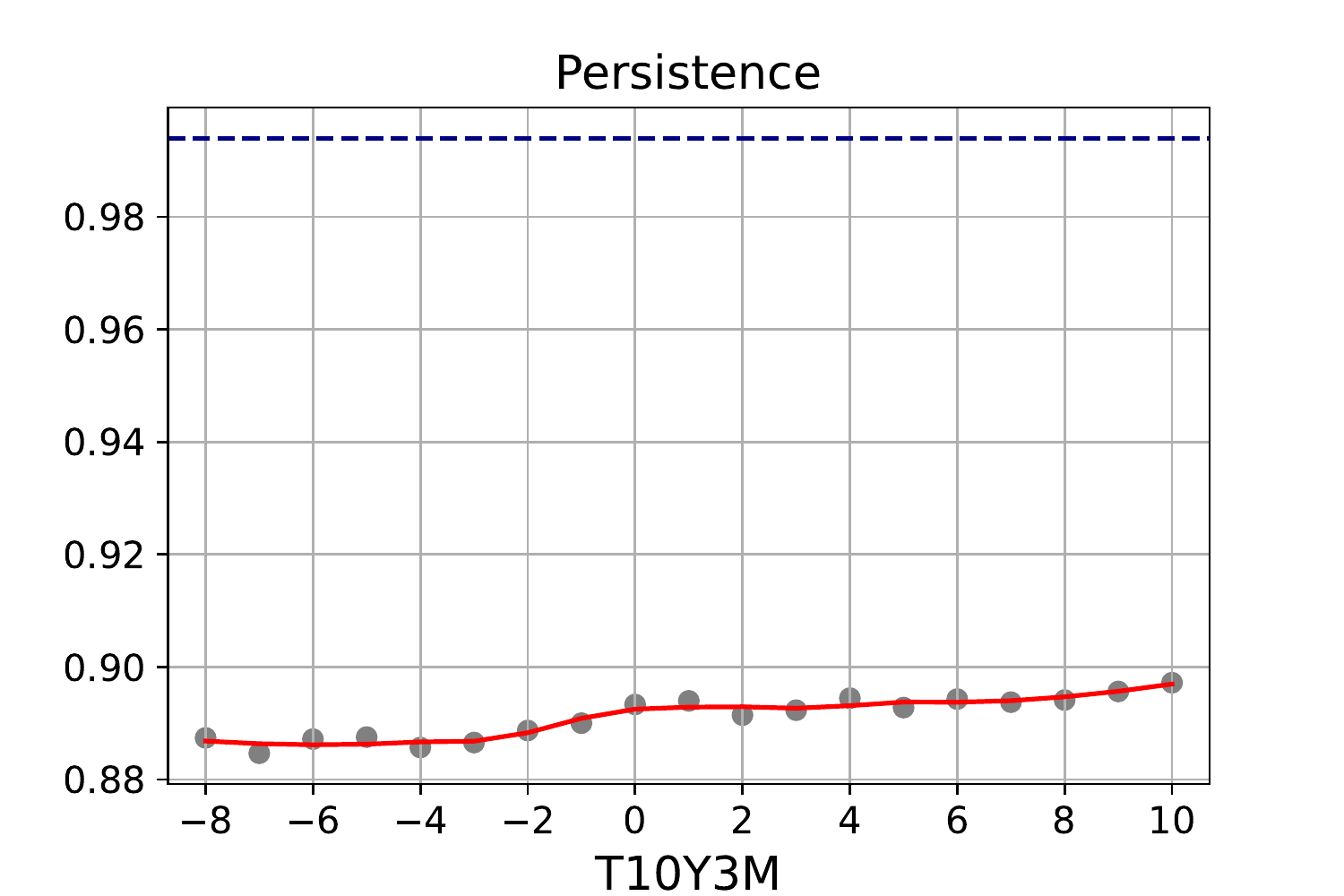} \\
         \includegraphics[scale=0.45]{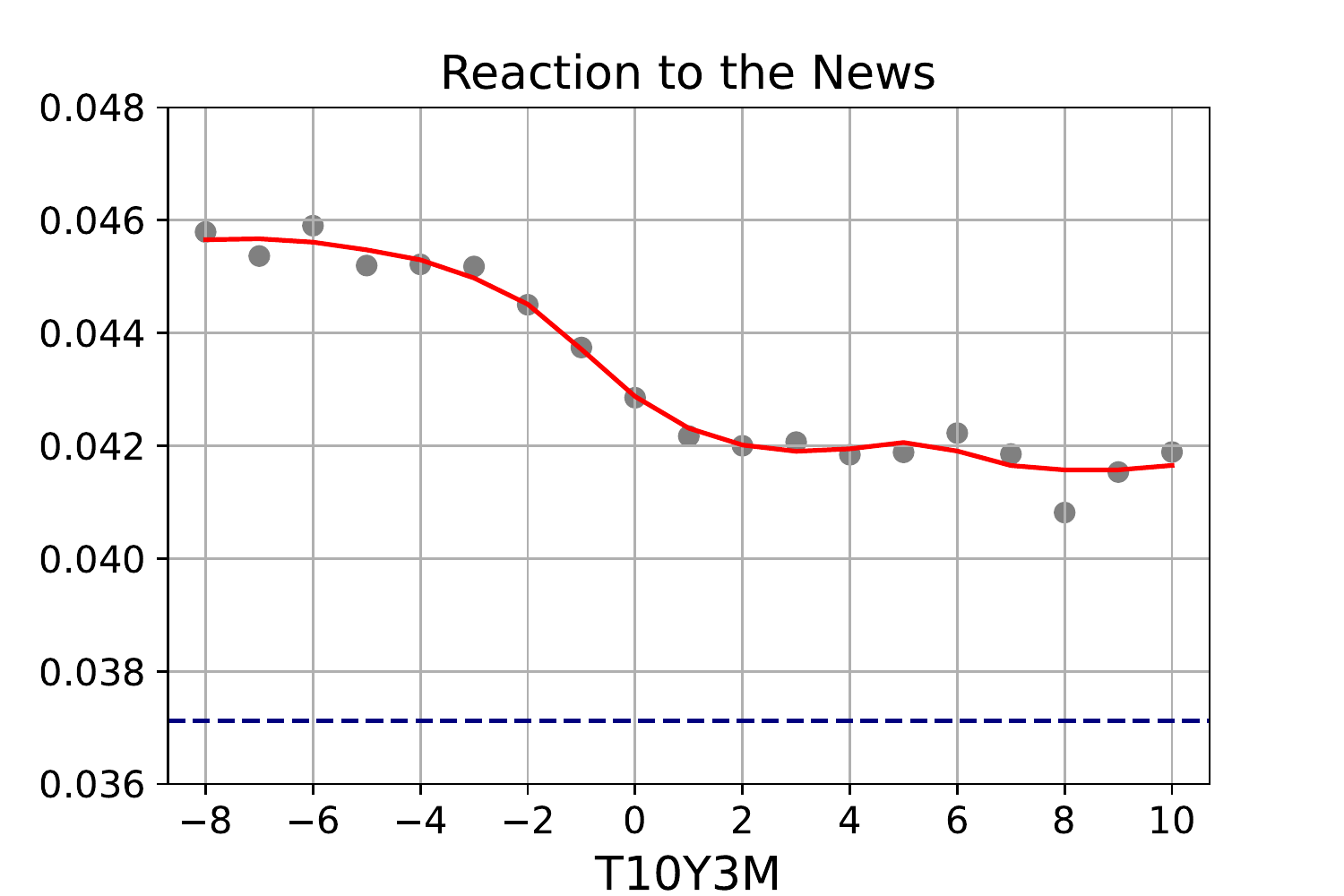} & 
         \includegraphics[scale=0.45]{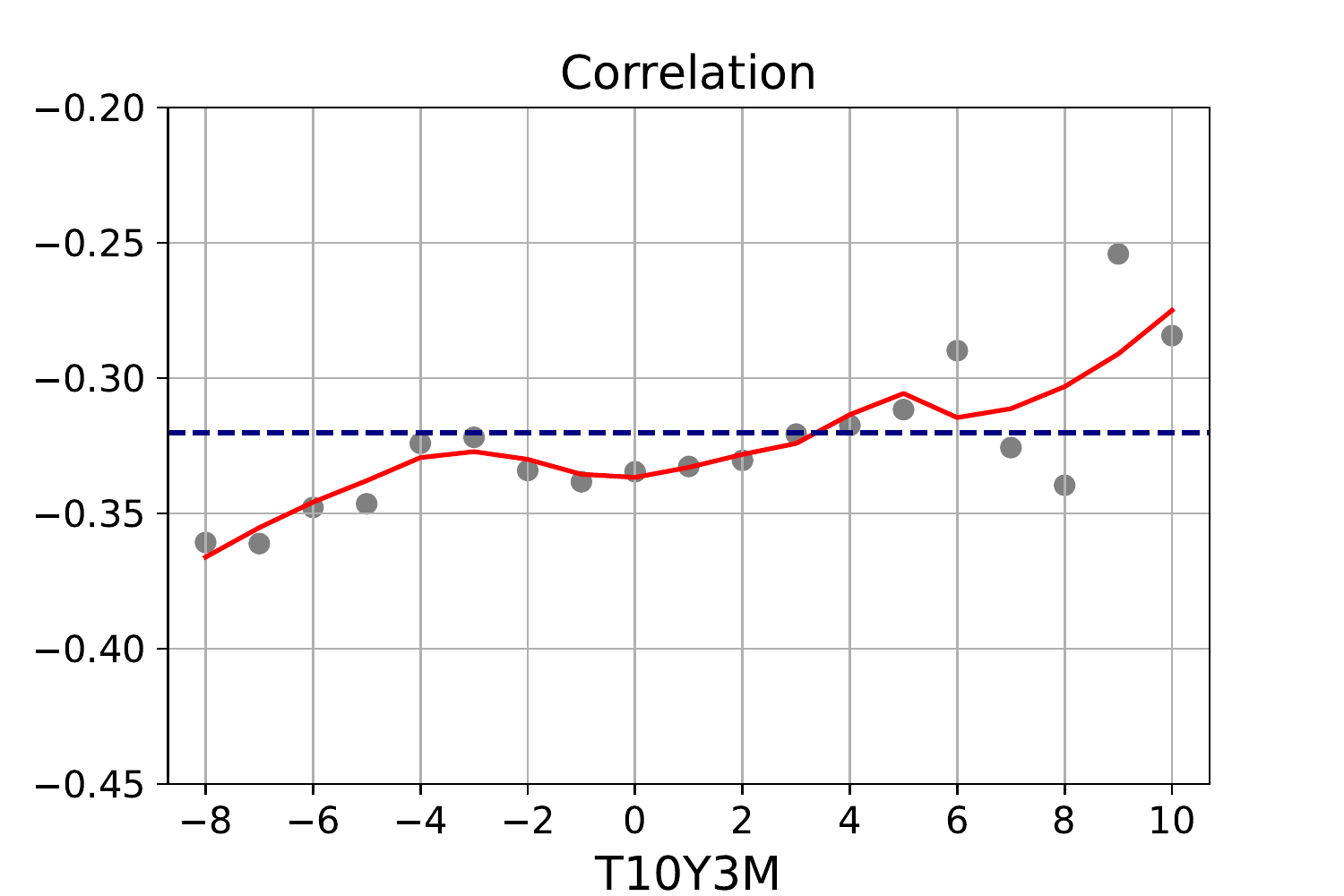}
    \end{tabular}
    \label{fig:tcop_T10Y3M}
\end{figure}

\begin{figure}
    \centering
    \caption{Parameter estimates as a function of VIX state variable for $t$ Copula forest-based GAS}
    \begin{tabular}{cc}
         \includegraphics[scale=0.45]{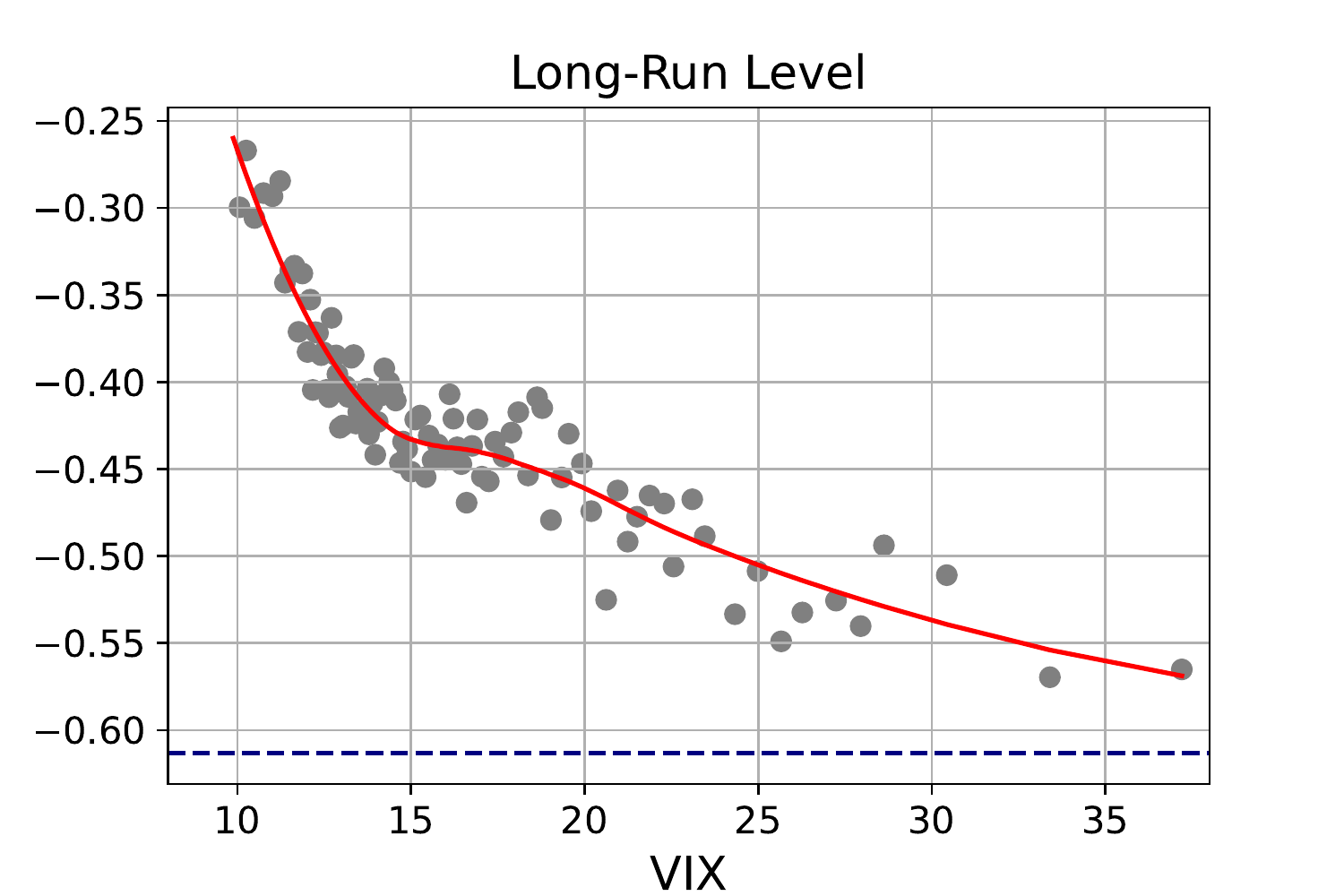} &  
         \includegraphics[scale=0.45]{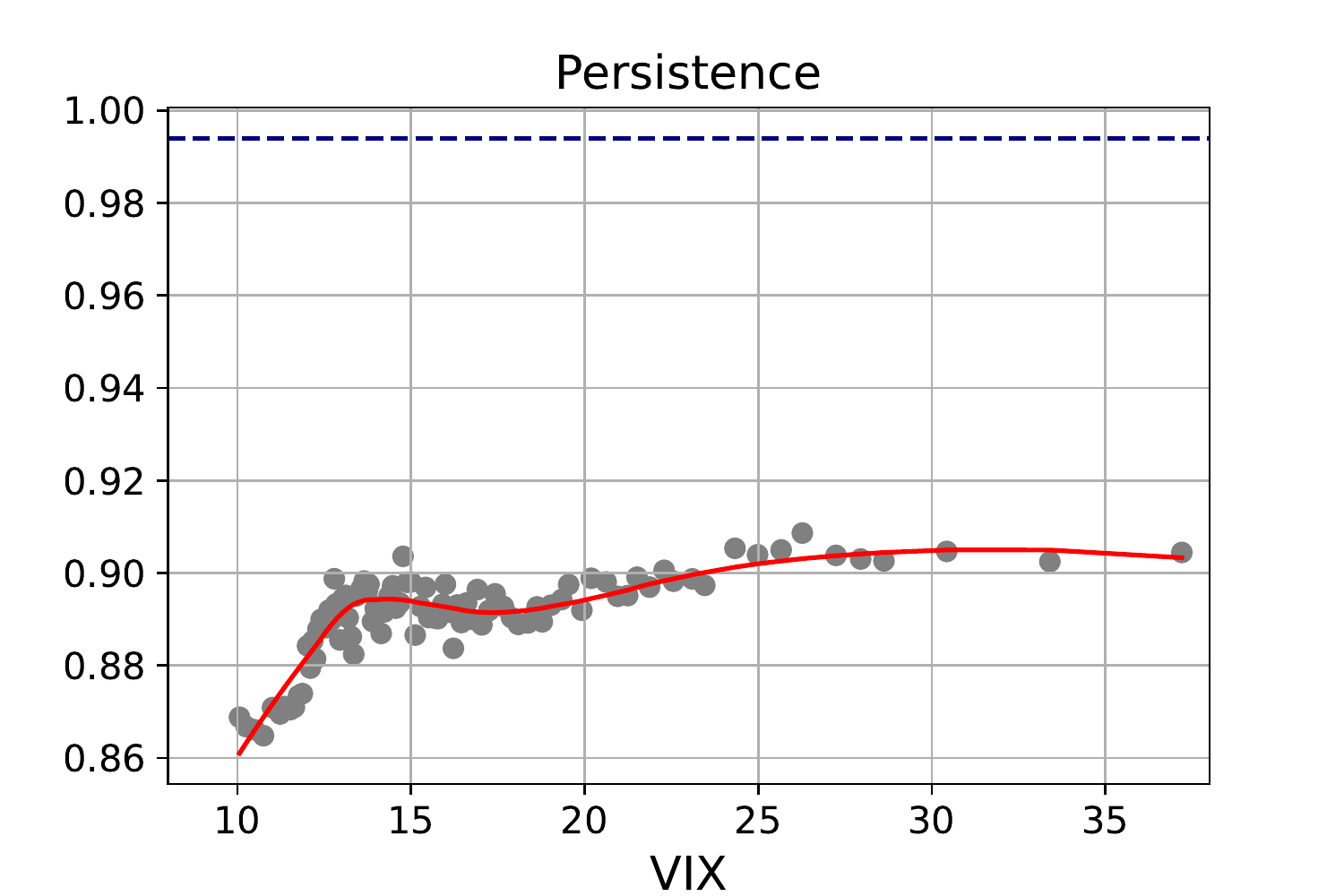} \\
         \includegraphics[scale=0.45]{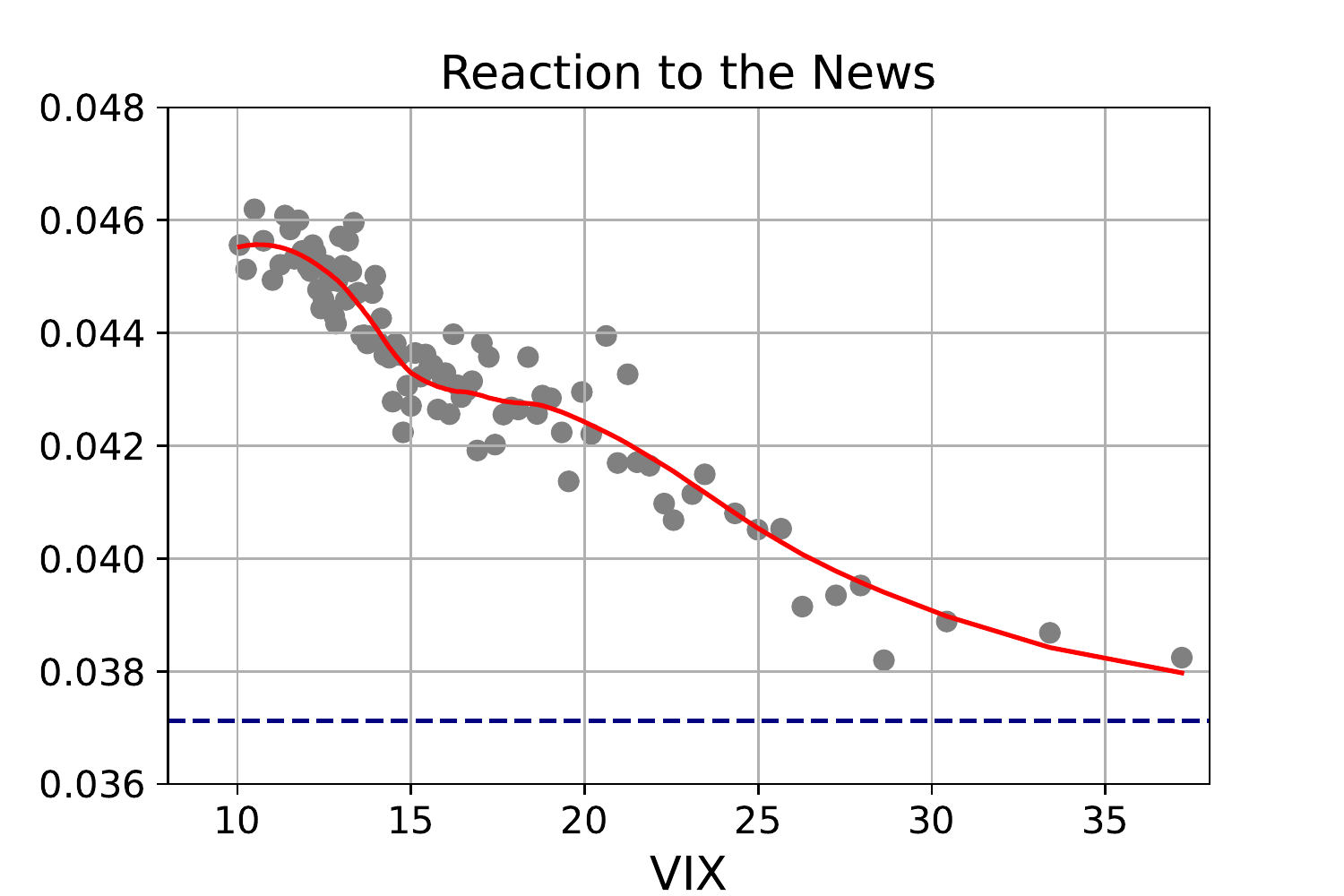} & 
         \includegraphics[scale=0.45]{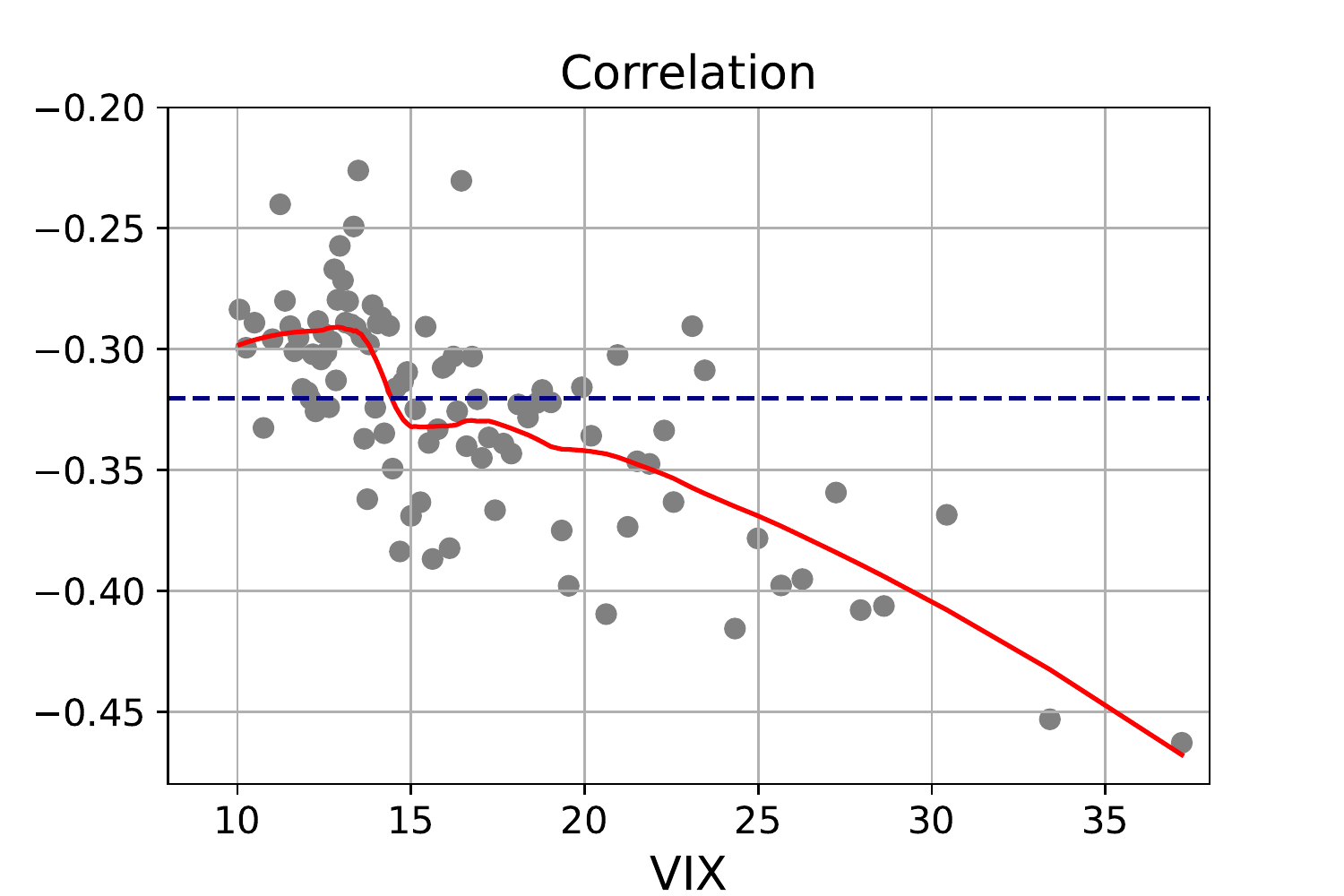}
    \end{tabular}
    \label{fig:tcop_VIX}
\end{figure}

\begin{figure}[t!]
    \centering
    \caption{\textbf{The estimated ACD tree model.} This figure depicts the tree structure for the ACD model. The tree's splits are based on SPX and T10Y, which refer to the S\&P 500 return and 10 year bond return respectively. \vspace{0.0cm}}
    \includegraphics[scale=0.6]{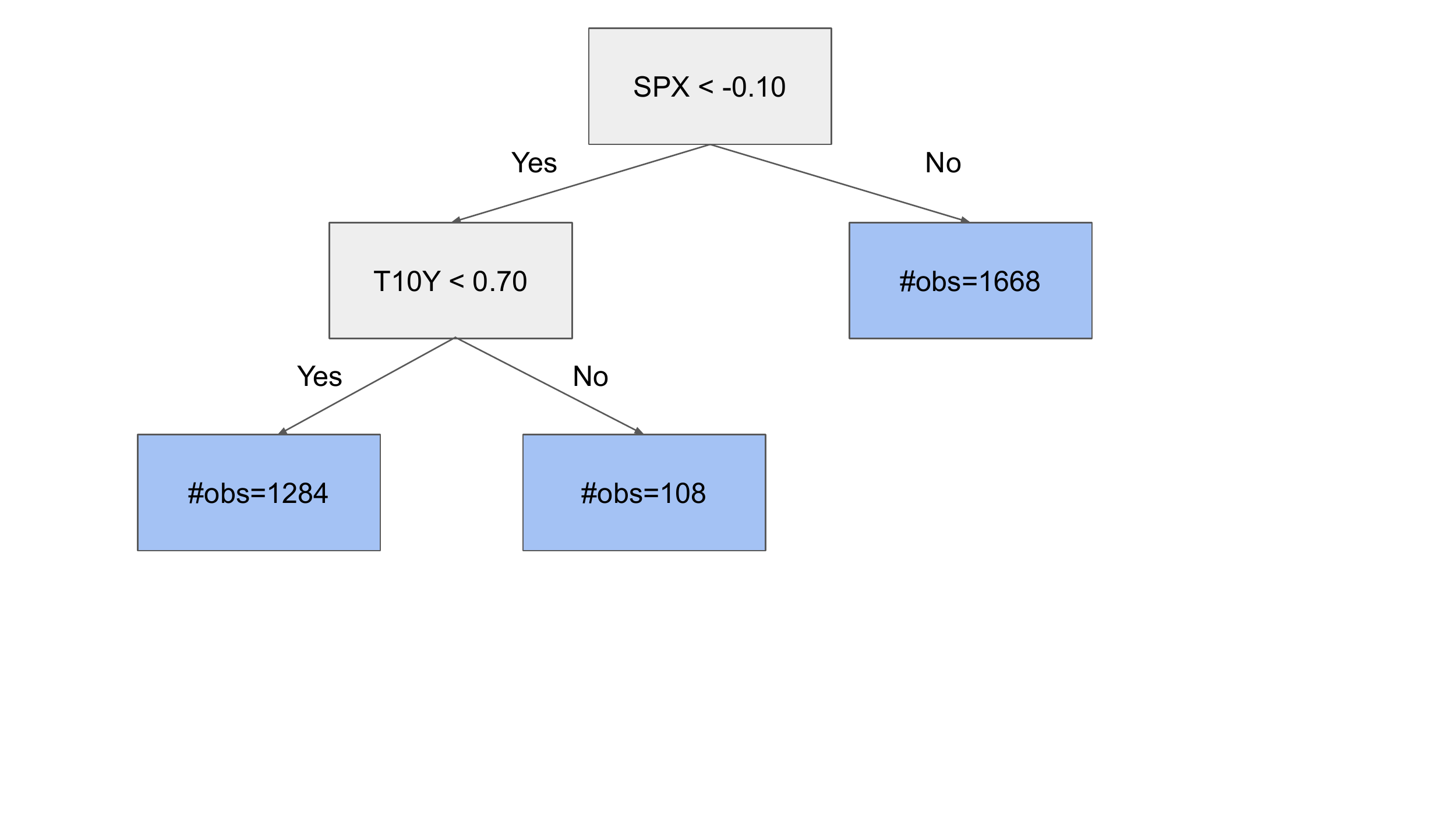}
    \label{fig:acd_tree}
\end{figure}

\begin{figure}[t!]
    \centering
    \caption{Parameter estimates as a function of VIX state variable for ACD forest-based}
    \begin{tabular}{cc}
        \includegraphics[scale=0.45]{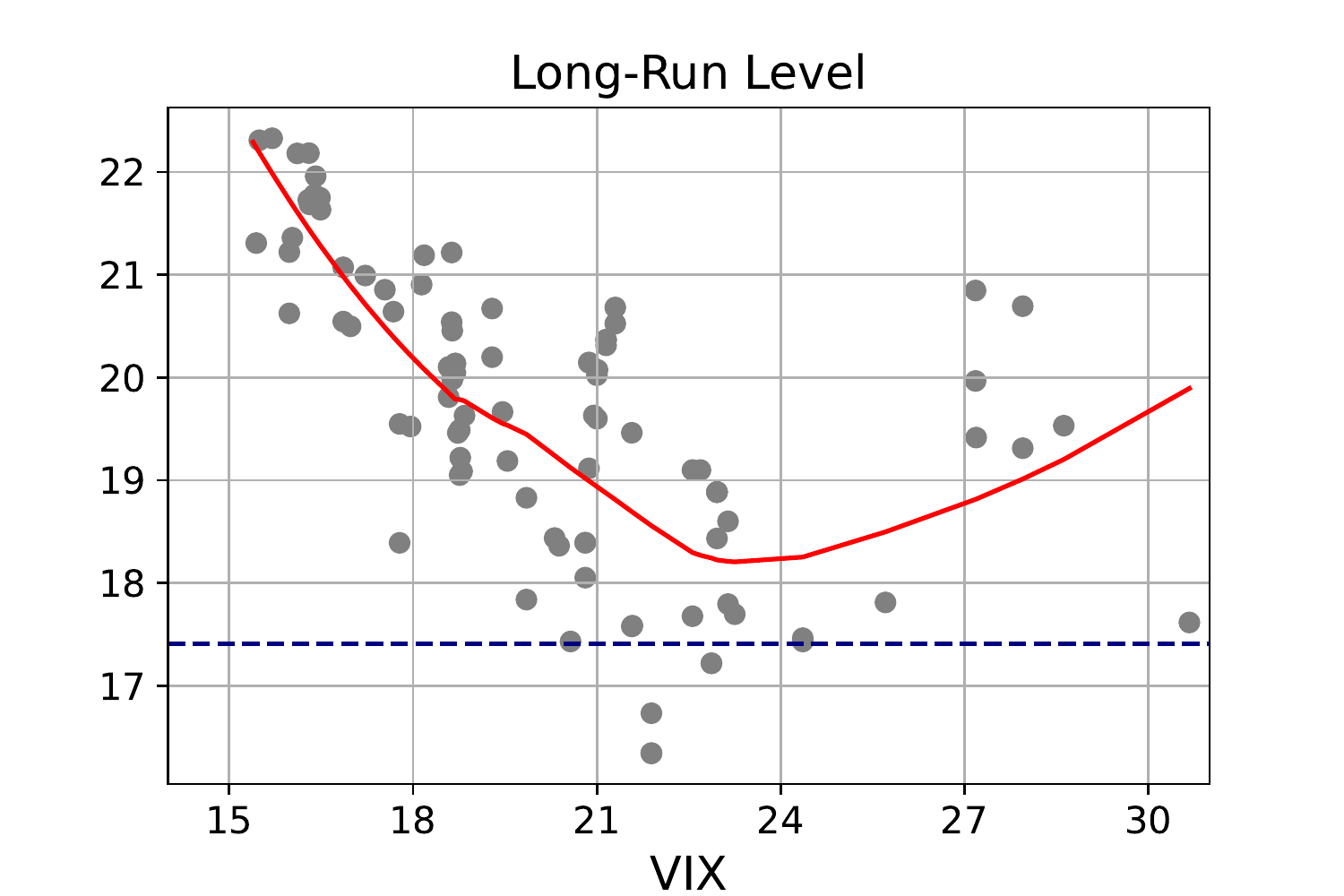} &
         \includegraphics[scale=0.45]{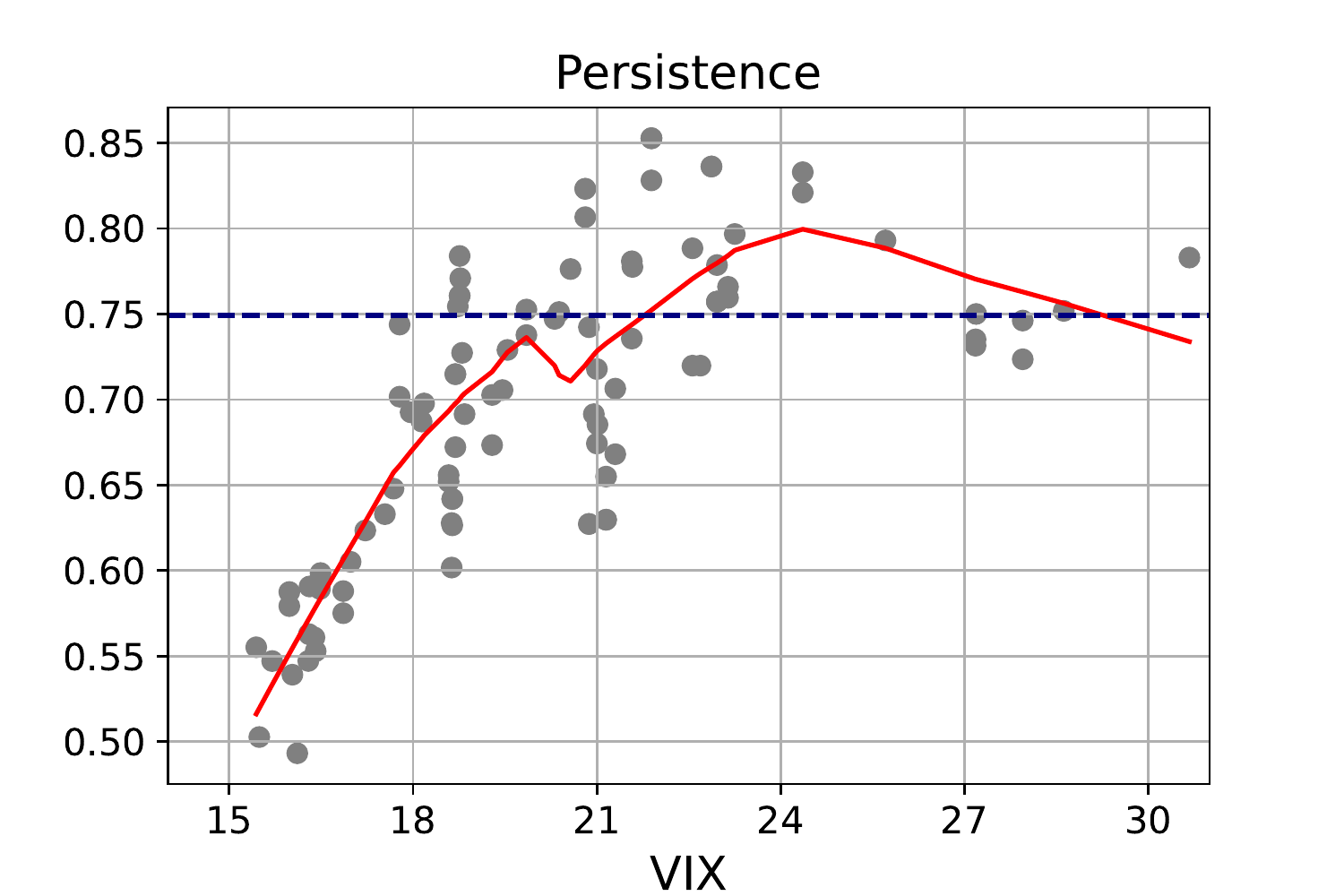} \\  
         \includegraphics[scale=0.45]{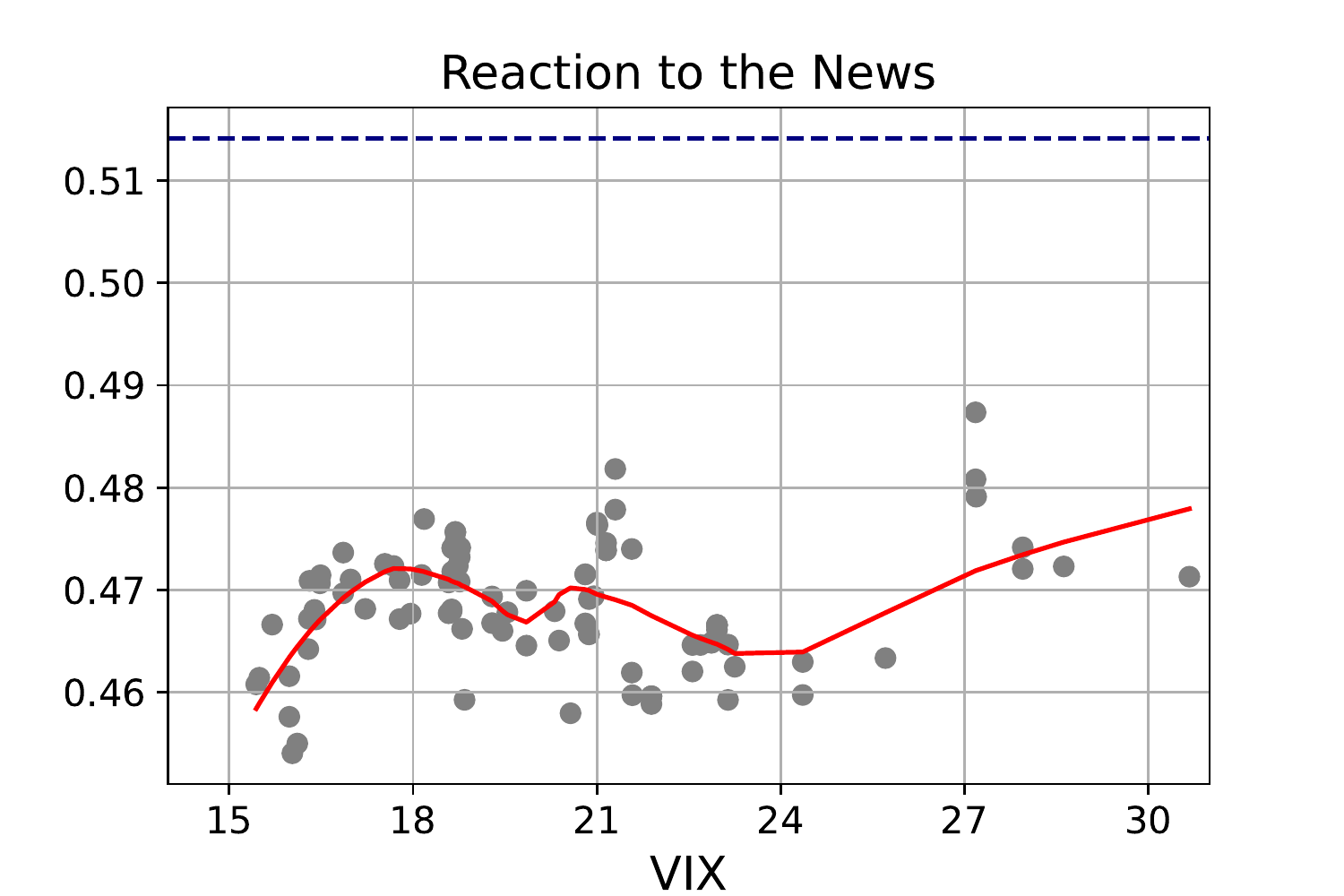} &
         \includegraphics[scale=0.45]{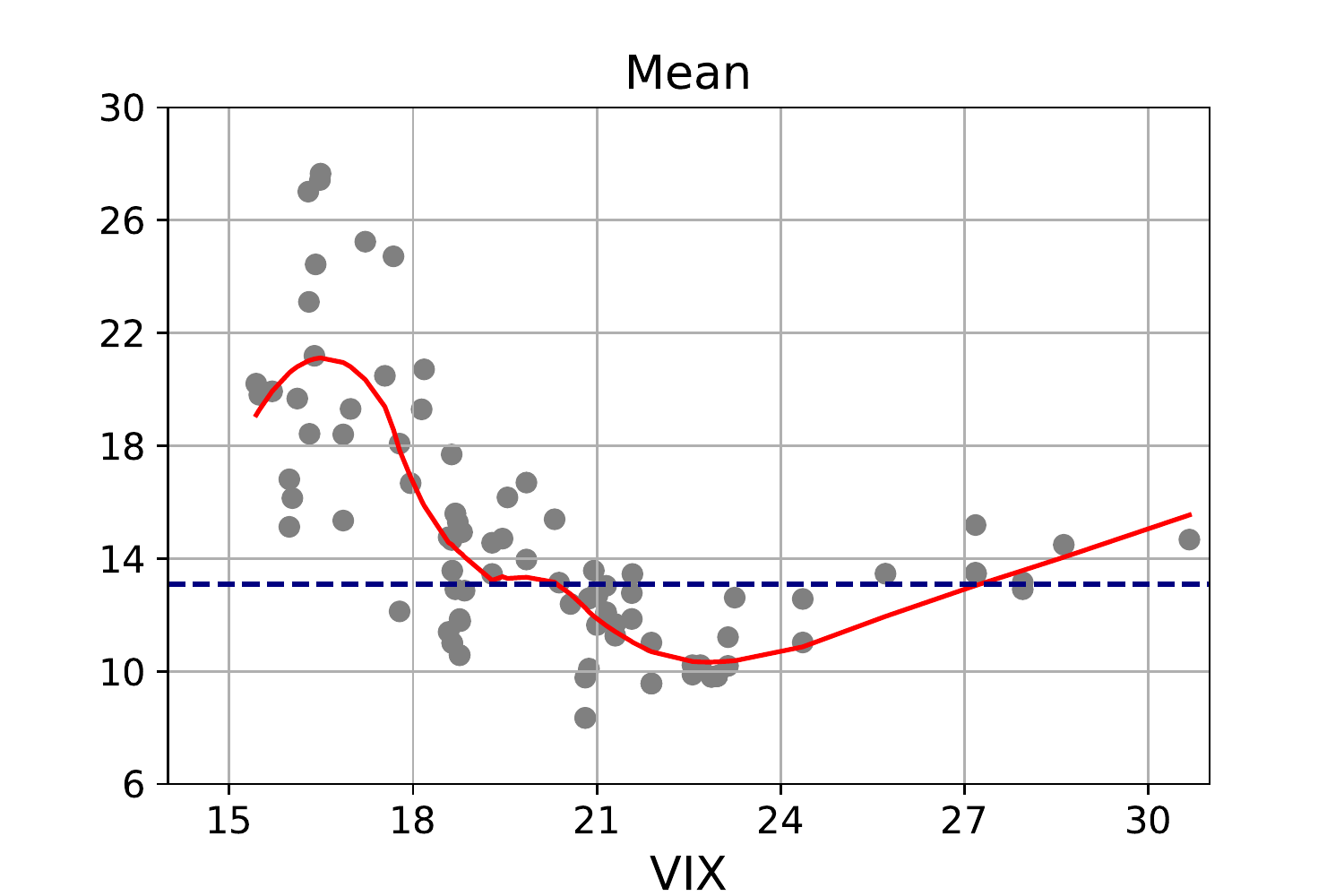}
    \end{tabular}
    \label{fig:ACD_VIX} \vspace{0.5cm}
\end{figure}

\begin{figure}
    \centering
    \caption{Parameter estimates as a function of T10Y3M state variable for ACD forest-based}
    \begin{tabular}{cc}
         \includegraphics[scale=0.45]{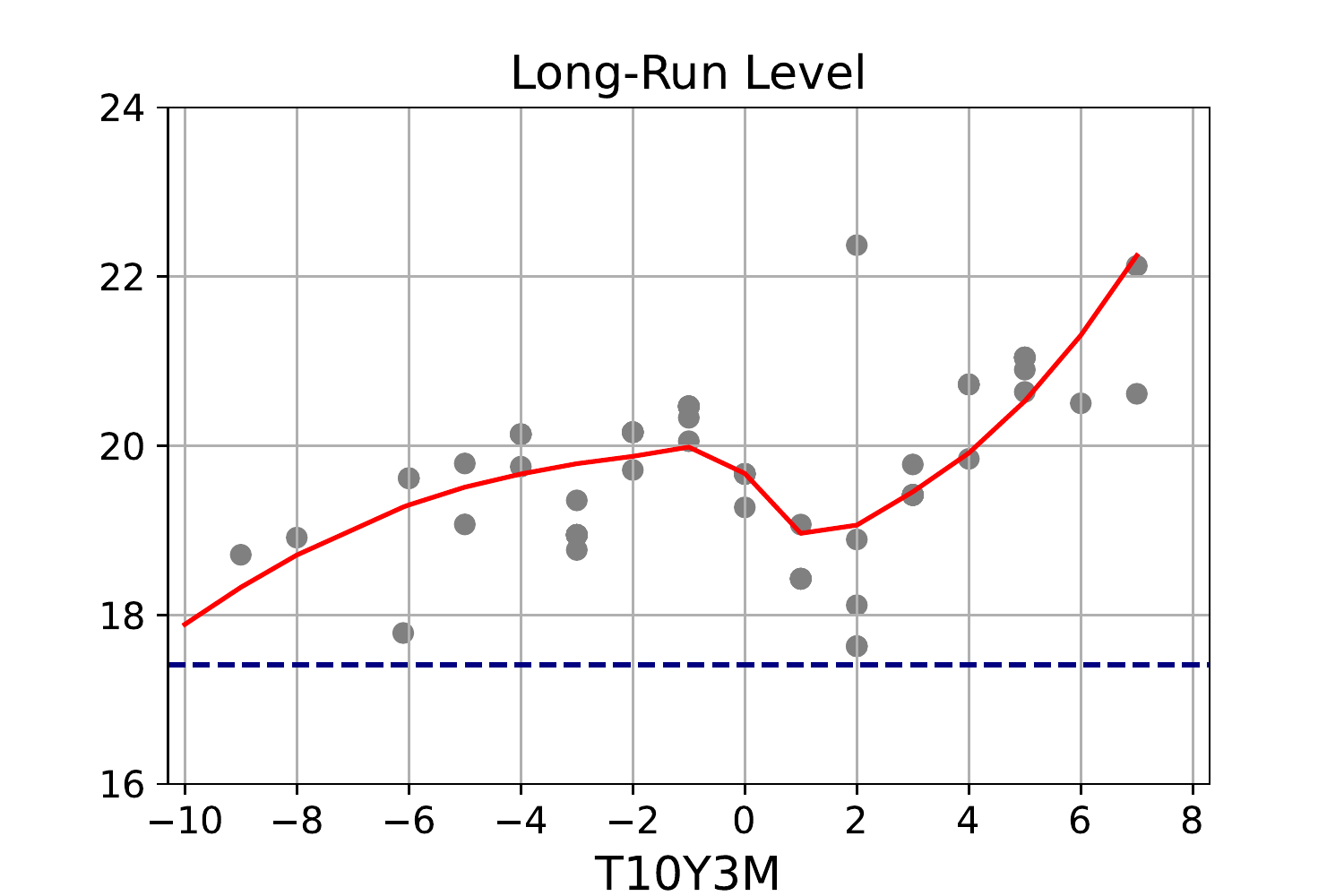} &  
         \includegraphics[scale=0.45]{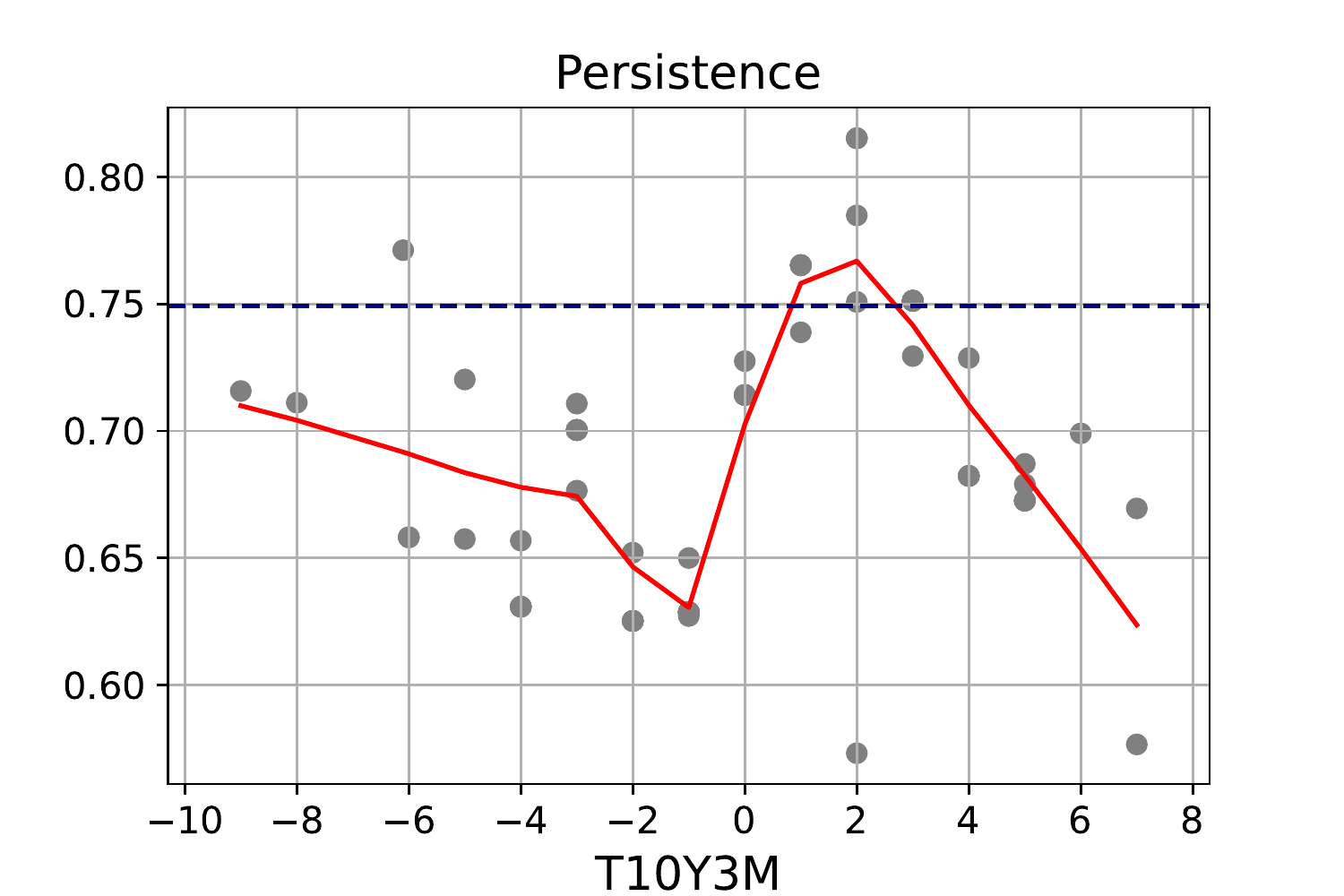} \\
         \includegraphics[scale=0.45]{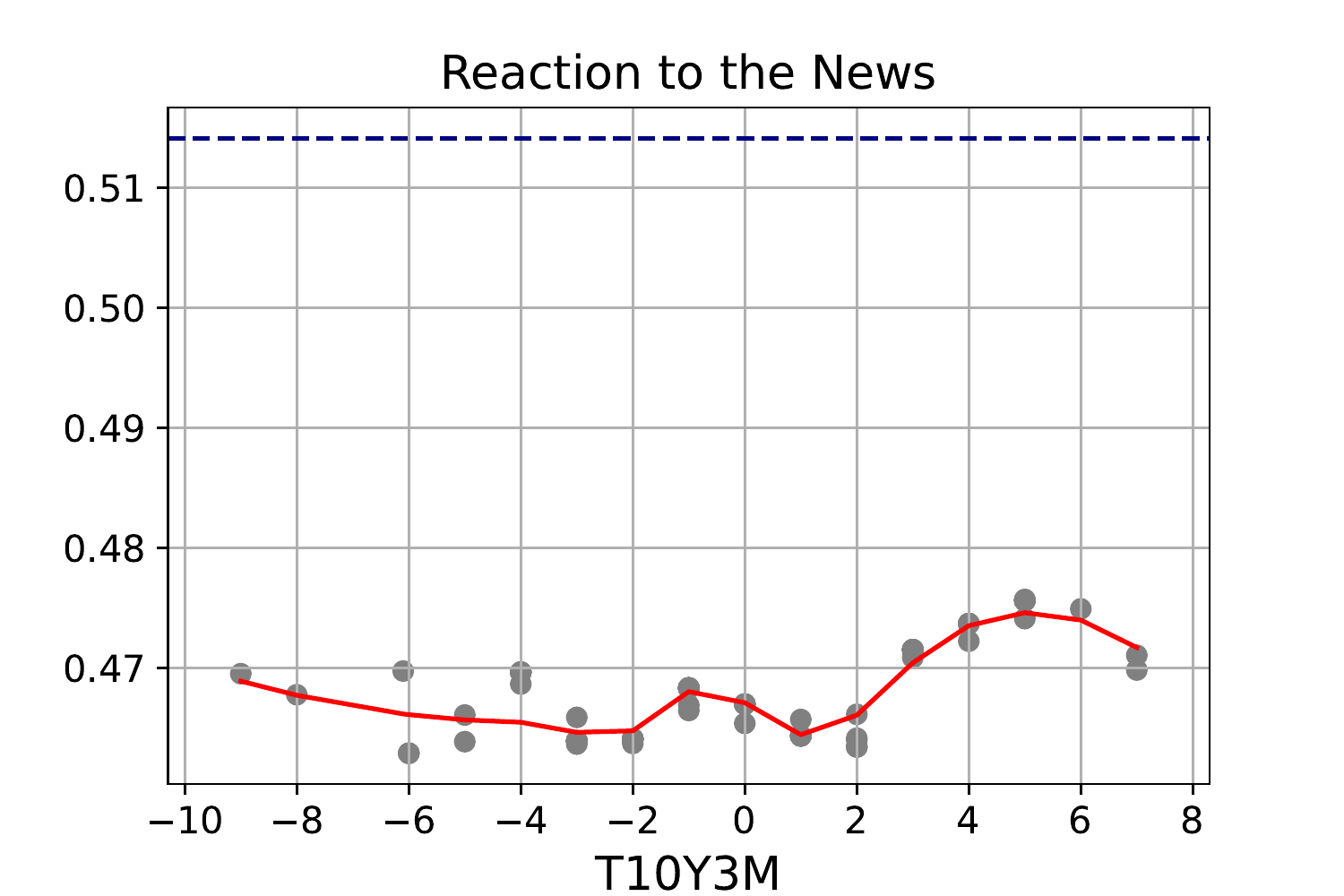} & 
         \includegraphics[scale=0.45]{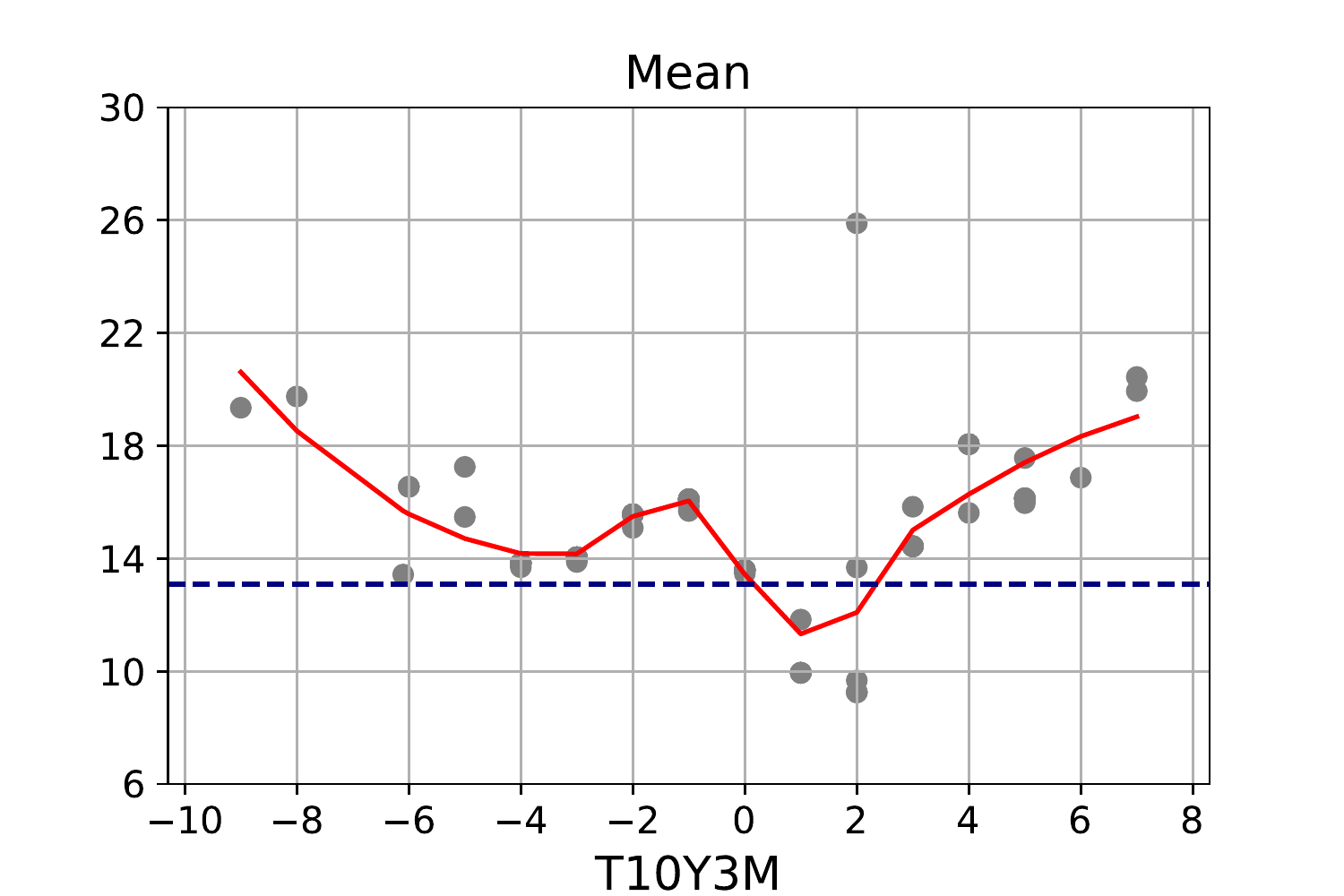}
    \end{tabular}
    \label{fig:ACD_T10Y3M}
\end{figure}

\begin{figure}
    \centering
    \caption{Parameter estimates as a function of DURATION state variable for ACD forest-based}
    \begin{tabular}{cc}
         \includegraphics[scale=0.45]{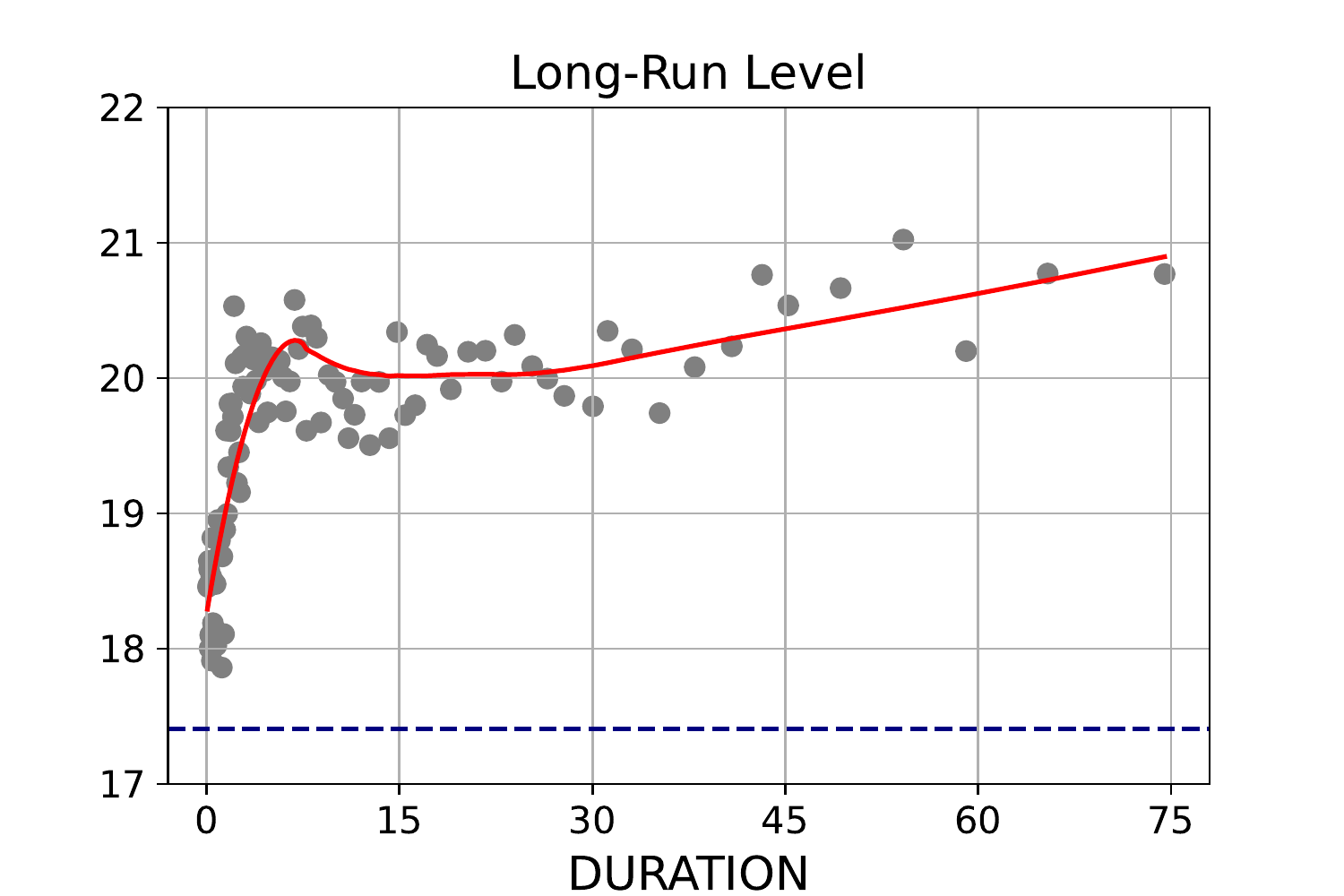} &  
         \includegraphics[scale=0.45]{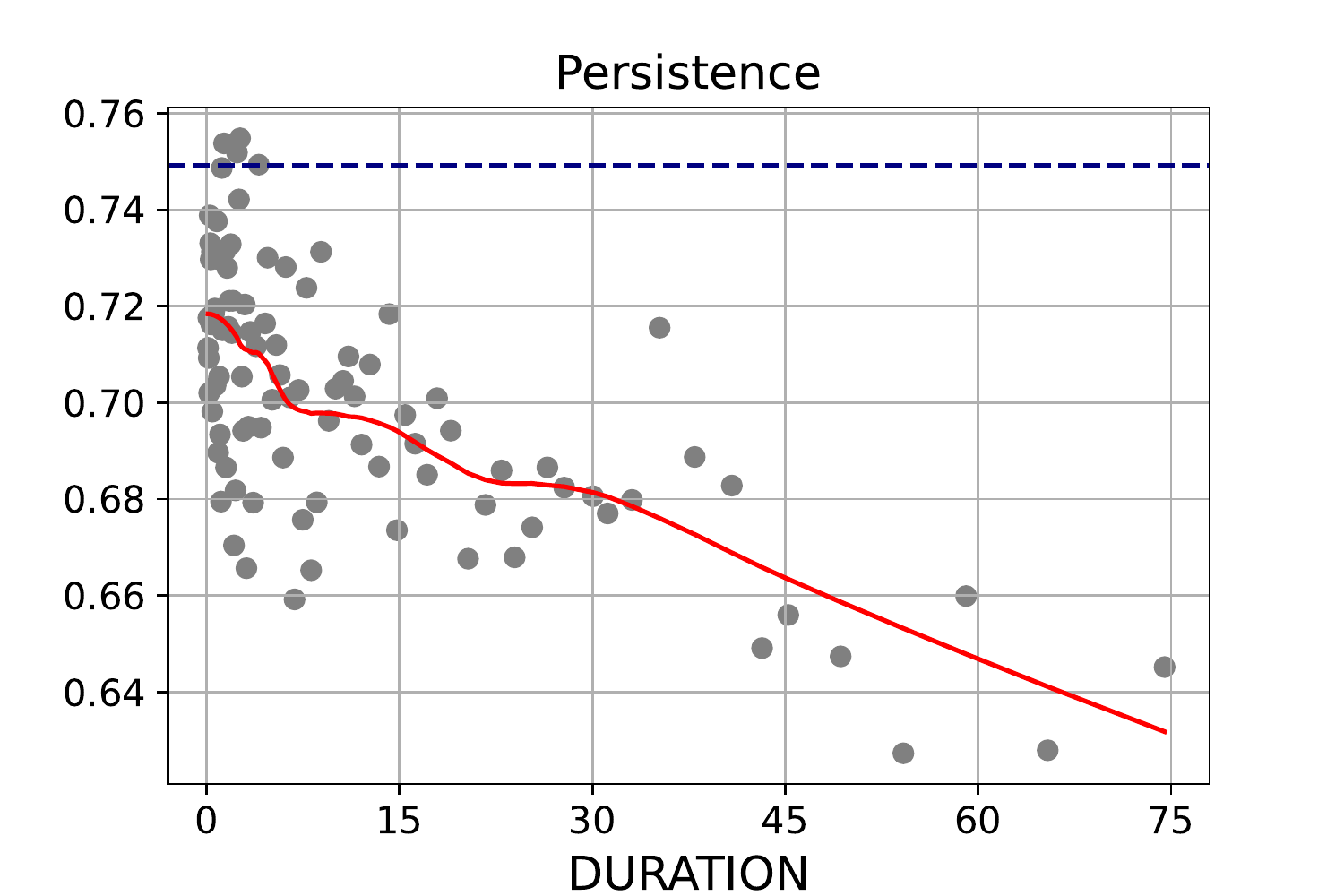} \\
         \includegraphics[scale=0.45]{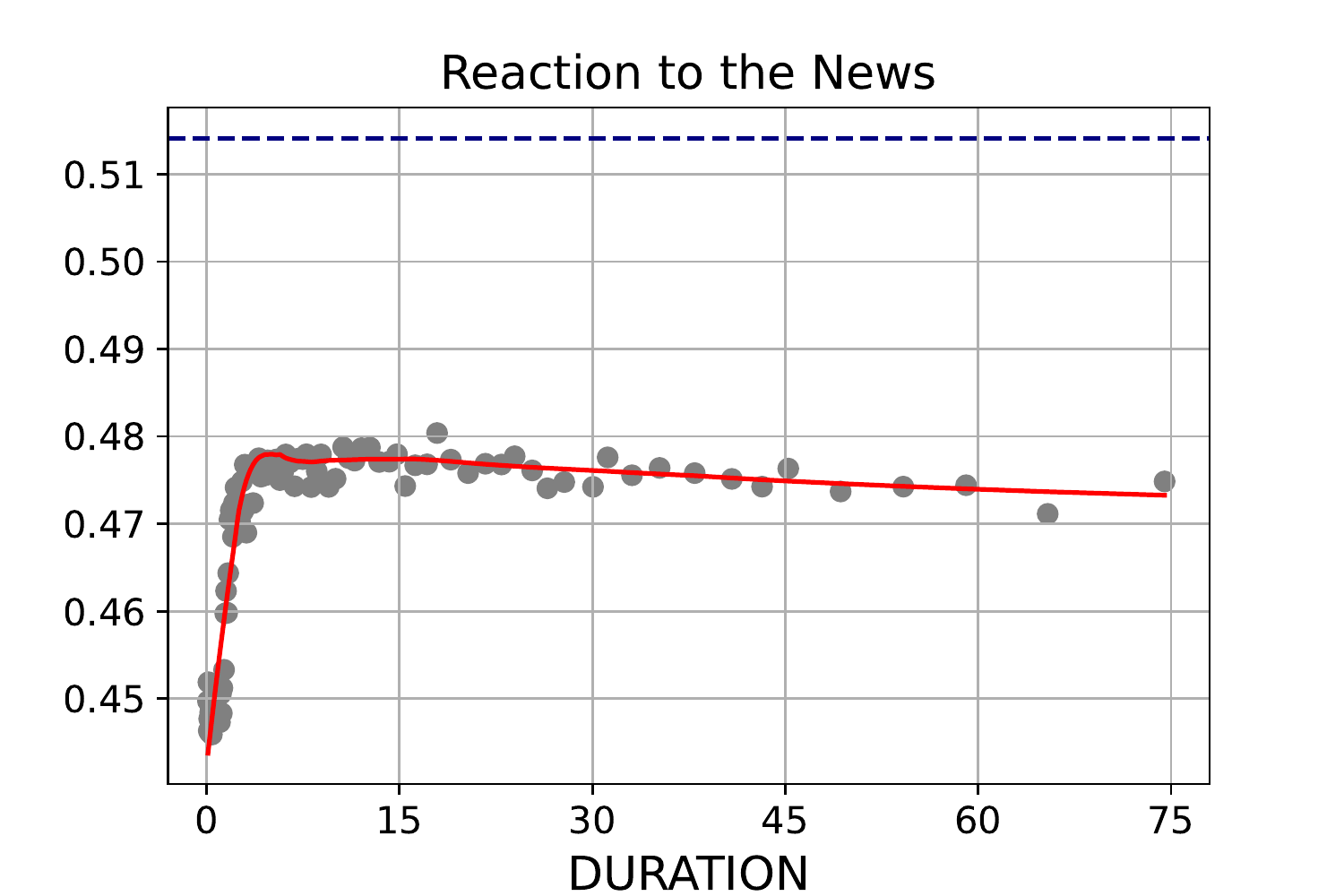} & 
         \includegraphics[scale=0.45]{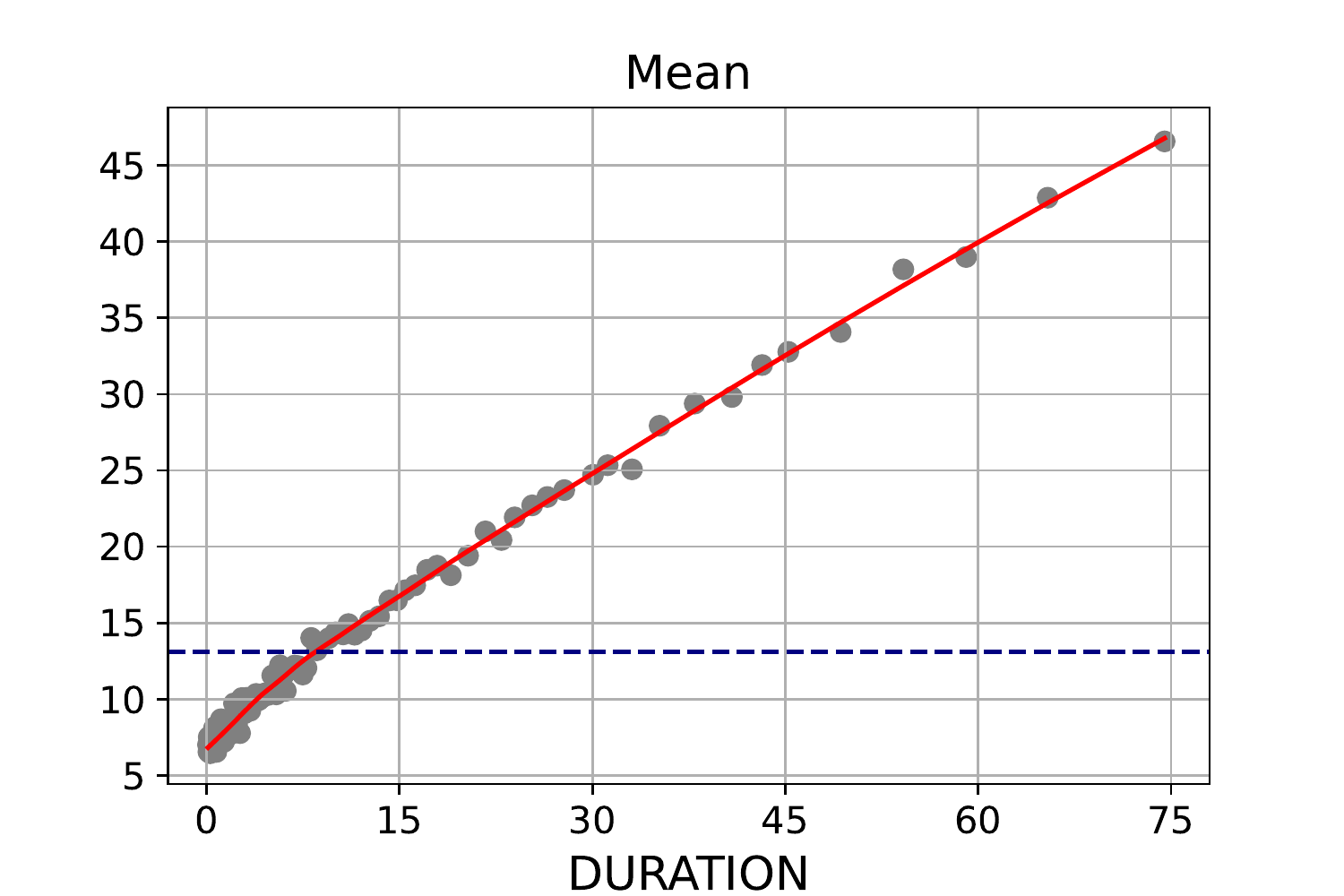}
    \end{tabular}
    \label{fig:ACD_DURATION}
\end{figure}

\begin{figure}
    \centering
    \caption{Parameter estimates as a function of LIQUIDITY state variable for ACD forest-based}
    \begin{tabular}{cc}
         \includegraphics[scale=0.45]{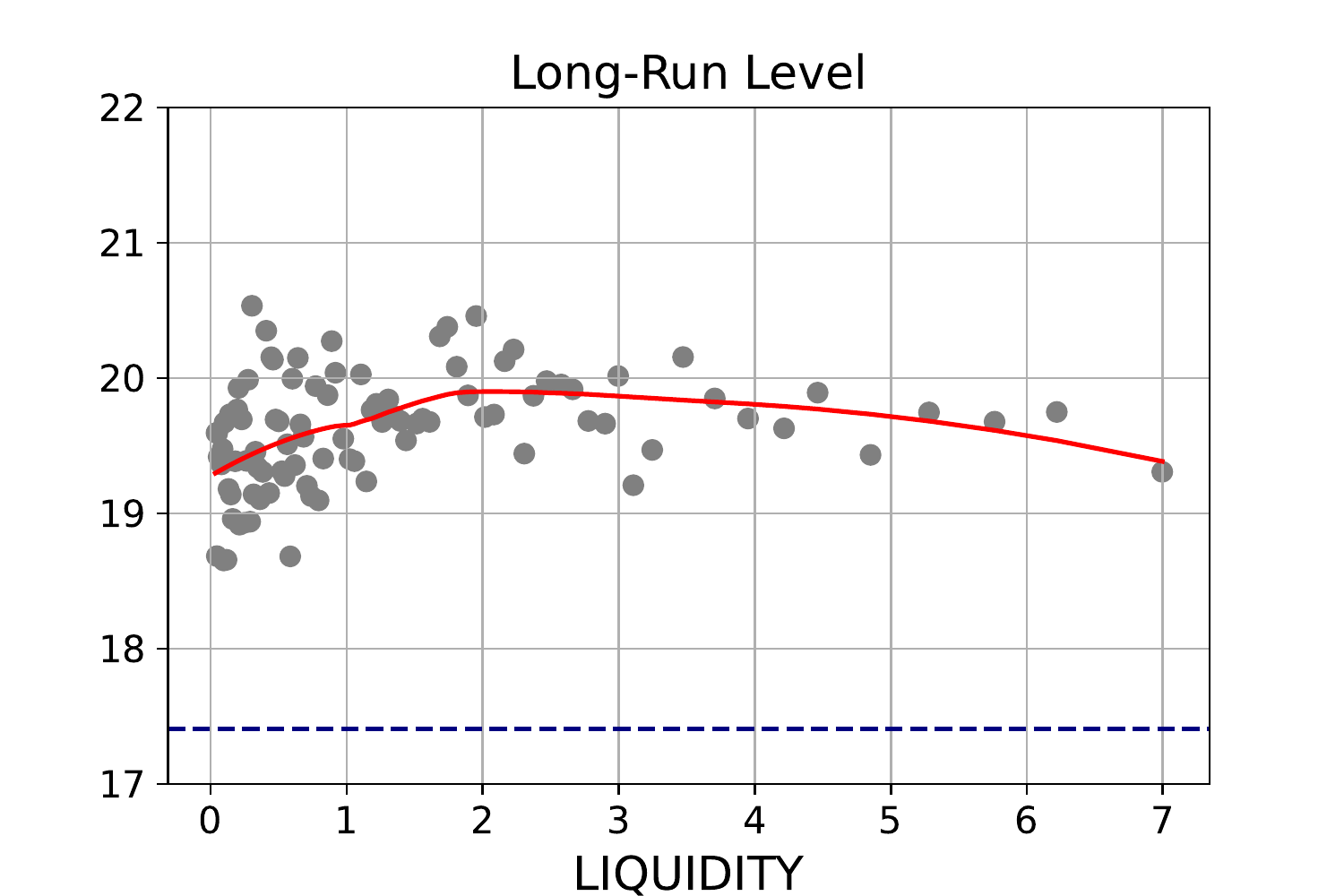} &  
         \includegraphics[scale=0.45]{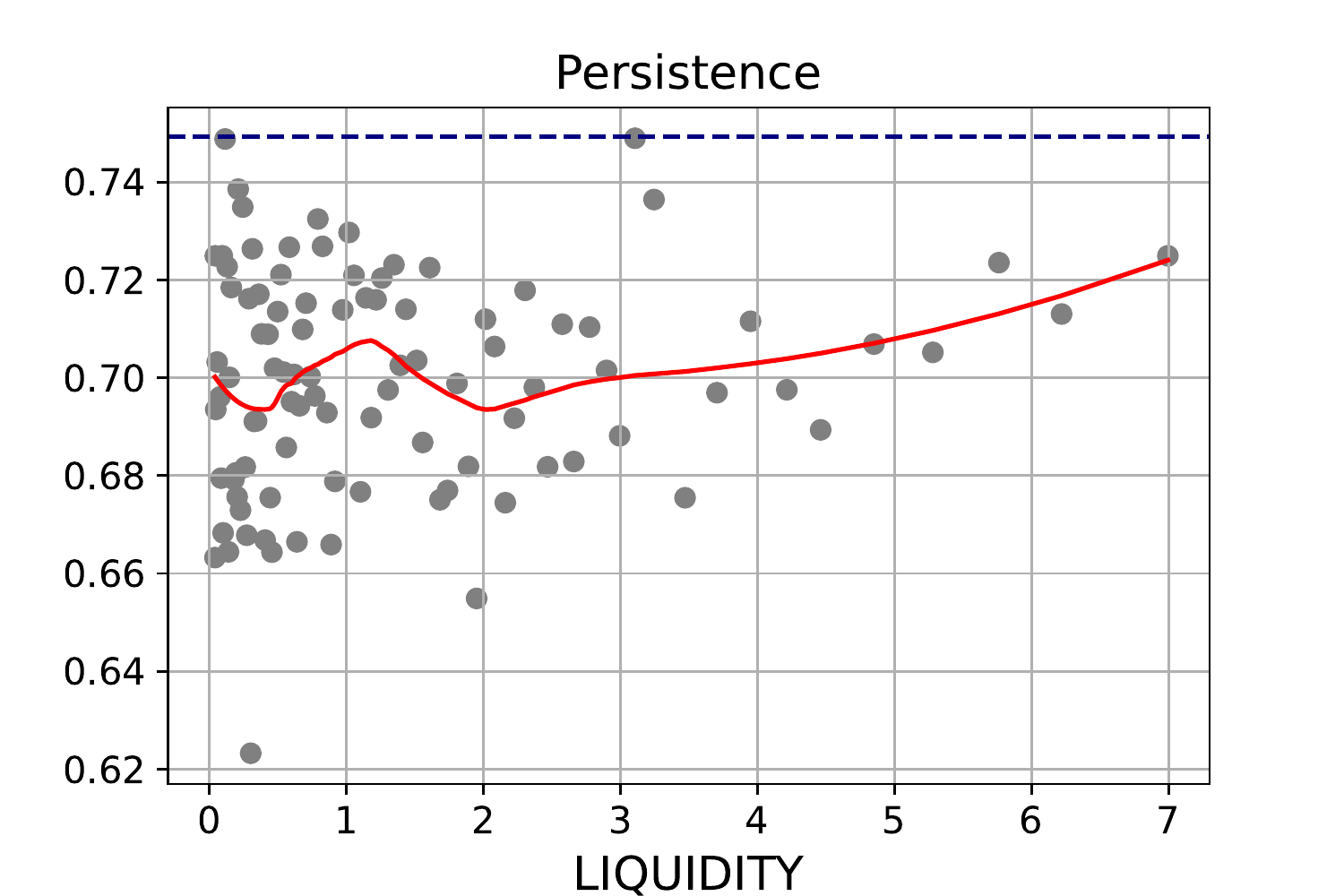} \\
         \includegraphics[scale=0.45]{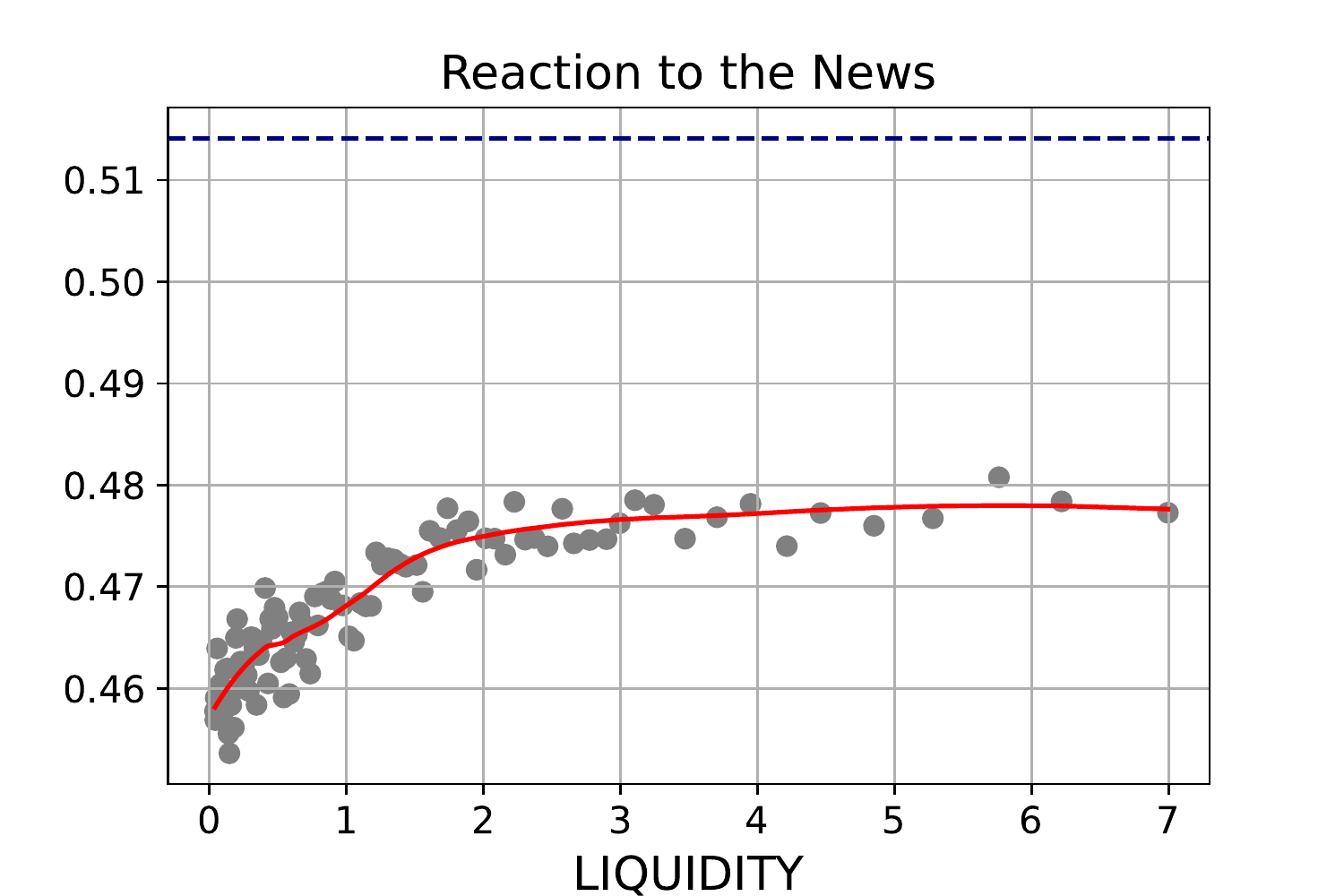} & 
         \includegraphics[scale=0.45]{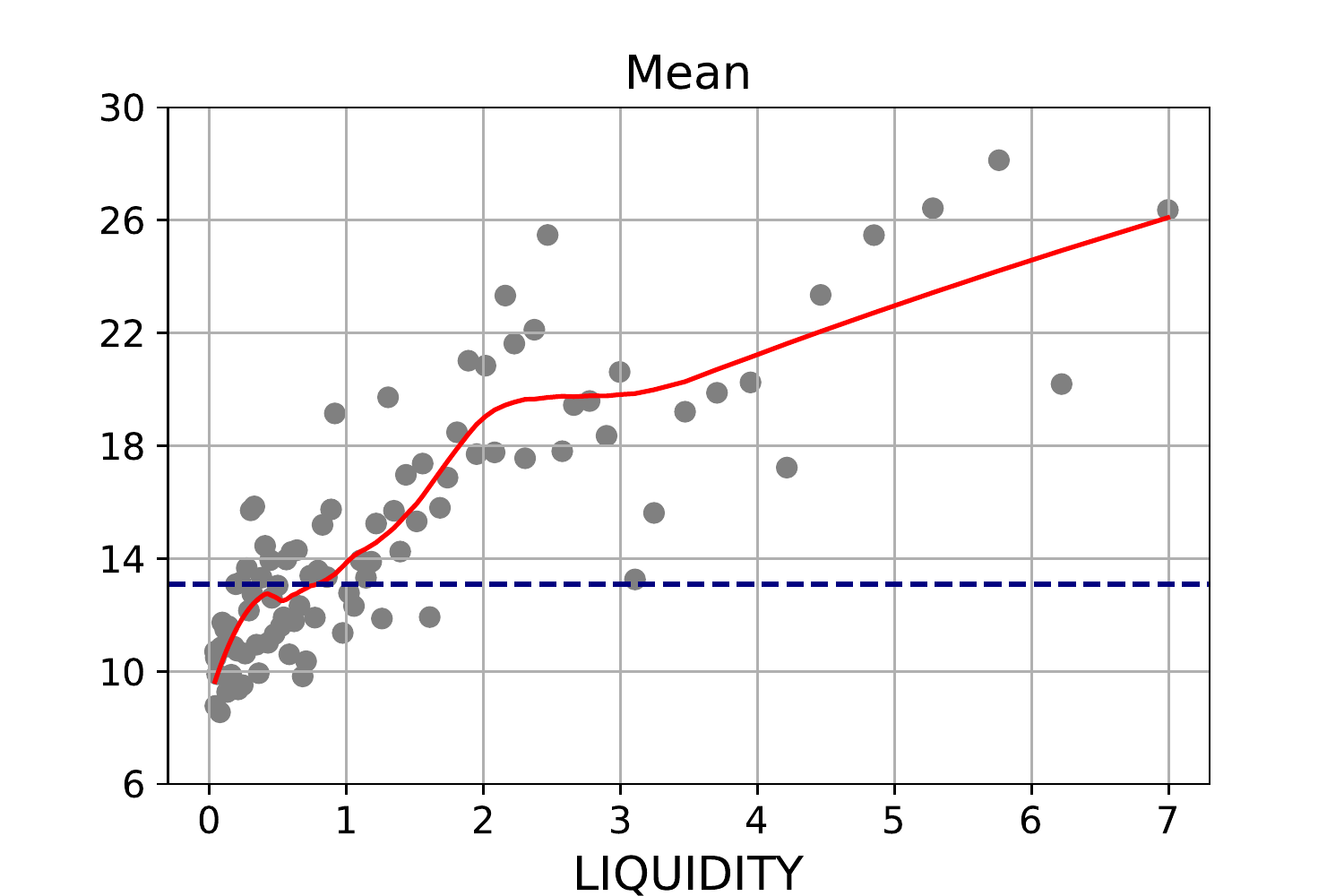}
    \end{tabular}
    \label{fig:ACD_LIQUIDITY}
\end{figure}

\end{document}